\documentclass[10pt,journal]{IEEEtran}
\usepackage[utf8]{inputenc} 
\usepackage[T1]{fontenc}
\usepackage{url}
\usepackage{ifthen}
\usepackage{cite}
\usepackage[cmex10]{amsmath} % Use the [cmex10] option to ensure complicance
\usepackage{amsmath,amsfonts}

\usepackage{array}
\usepackage{textcomp}
\usepackage{stfloats}
\usepackage{url}
\usepackage{verbatim}
\usepackage{graphicx}
\usepackage{cite}
\usepackage{algorithmicx,algorithm}
\usepackage{multicol}

\usepackage{algpseudocode}
\usepackage{bm}
\usepackage{verbatim}
\usepackage{booktabs}
\usepackage{amsmath,amssymb,amsfonts}
\usepackage{threeparttable}
\usepackage{multirow}
\usepackage{bbding}
\usepackage{pifont}
\usepackage{diagbox}
\usepackage{array}
\usepackage{tablefootnote}

\usepackage{stackengine}

\usepackage[caption=false,font=normalsize,labelfont=sf,textfont=sf]{subfig}

\usepackage{hyperref}
\hypersetup{colorlinks,%				% Reference color setup	
	linkcolor=blue,%
	citecolor=blue}

\begin{document}

\title{Deep Joint Source-Channel Coding for Adaptive Image Transmission over MIMO Channels
%\thanks{Identify applicable funding agency here. If none, delete this.}
}

%%% Many authors with many affiliations:
\author{
Haotian Wu,~\IEEEmembership{Graduate Student Member,~IEEE},
Yulin~Shao,~\IEEEmembership{Member,~IEEE},
Chenghong Bian, \\
Krystian Mikolajczyk,~\IEEEmembership{Senior Member,~IEEE},
Deniz~G\"und\"uz,~\IEEEmembership{Fellow,~IEEE}

\thanks{H. Wu, C. Bian, K. Mikolajczyk, and D. G\"und\"uz are with the Department of Electrical and Electronic Engineering, Imperial College London, London SW7 2AZ, U.K. (e-mails: \{haotian.wu17, c.bian22, k.mikolajczyk, d.gunduz\}@imperial.ac.uk). Y. Shao is with the State Key Laboratory of Internet of Things for Smart City and the Department of Electrical and Computer Engineering, University of Macau, Macau S.A.R (e-mail: ylshao@um.edu.mo). He is also a visiting researcher at the Department of Electrical and Electronic Engineering, Imperial College London, London SW7 2AZ, U.K.
}

\thanks{This work received funding from the UKRI for the projects AIR (ERC-Consolidator Grant, EP/X030806/1), SONATA (EPSRC-EP/W035960/1), and the SNS JU project 6G-GOALS under the EU’s Horizon program (grant agreement No. 101139232). The work of Y. Shao was supported in part by the Start-up Research Grant under Grant SRG2023-00038-IOTSC, and in part by the Science and Technology Development Fund, Macao SAR, under Grant 0068/2023/RIB3.}

\thanks{For the purpose of open access, the authors have applied a Creative Commons Attribution (CCBY) license to any Author Accepted Manuscript version arising from this submission.}

\thanks{This paper was presented in part at the IEEE International Conference on Communications (ICC) in May 2023.}

}

\maketitle

%%%%%%
%% Abstract: 
%% If your paper is eligible for the student paper award, please add
%% the comment "THIS PAPER IS ELIGIBLE FOR THE STUDENT PAPER
%% AWARD." as a first line in the abstract. 
%% For the final version of the accepted paper, please do not forget
%% to remove this comment!
%%
%%%interaction-baseded. semantic paramlim.
%%fb csi, beamforming...
%%hybrid contextual...
%%find a optimzation scheme....in this framework....
%%%try to find a picture, describe the future....least fb for best improvement...
%%
%joint communication and sensing...
%make some compression comparation....
%add some description over the caption...
\begin{abstract}
We introduce a vision transformer (ViT)-based deep joint source and channel coding (DeepJSCC) scheme for wireless image transmission over multiple-input multiple-output (MIMO) channels, called DeepJSCC-MIMO. We employ DeepJSCC-MIMO in both open-loop and closed-loop MIMO systems. The novel DeepJSCC-MIMO architecture surpasses the classical separation-based benchmarks, exhibiting robustness to channel estimation errors, and flexibility in adapting to diverse channel conditions and antenna configurations without requiring retraining. Specifically, by harnessing the self-attention mechanism of the ViT, DeepJSCC-MIMO intelligently learns feature mapping and power allocation strategies tailored to the unique characteristics of the source image and prevailing channel conditions. Extensive numerical experiments validate the significant improvements in both distortion quality and perceptual quality achieved by DeepJSCC-MIMO for both open-loop and closed-loop MIMO systems across a wide range of scenarios. Moreover, DeepJSCC-MIMO exhibits robustness to varying channel conditions, channel estimation errors, and different antenna numbers, making it an appealing technology for emerging semantic communication systems.
\end{abstract}


\section{Introduction}
The exponential growth of {wireless multimedia applications, encompassing augmented reality, virtual reality, real-time streaming, and edge intelligence}, has significantly heightened the demand for efficient wireless transmission of image/video signals, particularly under strict delay constraints. Consequently, this has sparked a notable surge in research on developing optimized image communication systems over wireless channels \cite{bourtsoulatze2019deep, 9714510, xu2021wireless, 9791398, kurka2021bandwidth, bian2024process,yang2022deep, yoo2023role, erdemir2022generative,bian2022deep}.

The conventional solution, motivated by Shannon's separation theorem, is to independently design source and channel coding, which is optimal in the asymptotic limit of infinite block length for ergodic source and channel distributions \cite{shannon1948mathematical}. \textcolor{black}{However, the separation-based approach is known to be sub-optimal in the practical finite block length regime, which is becoming increasingly relevant in emerging applications involving Internet-of-things and edge intelligence.} Despite its known benefits, designing practical joint source-channel coding (JSCC) schemes has been an ongoing challenge for many decades. Most existing studies on JSCC either focus on theoretical scenarios, e.g., Gaussian \textcolor{black}{or Gauss-Markov} sources\cite{hekland2009shannon, vazquez2014analog, aguerri2015joint,vaezi2014distributed}, or jointly optimize parameters of separate source and channel codes \cite{Peng:JSAC:00, Pu:TIP:07, Fresia:SPM:10, Golmohammadi:TCOM:22}, \textcolor{black}{as there is a lack of practical JSCC schemes and theoretical performance bounds on the achievable mean squared error distortion for lossy JSCC in finite block lengths \cite{9954060}, particularly when dealing with practical sources such as images.} On the other hand, significant progress has been made in the recent years in designing deep learning-based JSCC (DeepJSCC) schemes thanks to the introduction of deep neural networks (DNNs) \cite{bourtsoulatze2019deep, 9714510, xu2021wireless, 9791398}, which allow to directly map the input source signal to channel symbols, and vice versa at the decoder. This approach removes the necessity of explicit source compression or error correction, or the usage of bits as a common unit of representation of signals. \textcolor{black}{Instead, the DeepJSCC approach is optimized to directly generate channel symbols with the goal of enhancing the overall transmission quality across diverse source and channel distributions. This is achieved through an end-to-end training methodology that refrains from imposing constraints on the distribution of the source or the channel.} 
% Notably, it demonstrates significant effectiveness in handling complex distributed sources, such as images and videos.}

The first DeepJSCC scheme for image transmission was presented in \cite{bourtsoulatze2019deep}, which was shown to outperform the concatenation of the better portable graphics (BPG) image compression codec with low-density parity-check (LDPC) codes. It was later extended to adaptive channel bandwidth scenario in \cite{kurka2021bandwidth} and \cite{yang2022deep}, and to the transmission over multi-path fading channels in \cite{9714510} and \cite{wu2022channel}. \textcolor{black}{Its feasibility in practical systems has also been evaluated, including building prototypes \cite{tung:ICASSP:20, yoo2023role}, or its implementation with constrained channel inputs \cite{tung2021deepjscc}, or under peak-to-average power ratio constraints \cite{DTAT}.} However, existing DeepJSCC schemes solely consider single-antenna transmitters and receivers. While there is a growing literature successfully employing DNNs for various multiple-input multiple-output (MIMO)-related tasks in wireless communications, such as detection, channel estimation, or beamforming \cite{samuel2019learning, mashhadi2021pruning, he2020model}, no previous work has so far applied DeepJSCC to the more challenging MIMO scenario, with the exception of our initial results in \cite{wu2022vision}. %MIMO systems are known to boost the throughput and spectral efficiency of wireless communications, showing significant improvements in the capacity and reliability\cite{paulraj2004overview}. 
JSCC over MIMO channels was studied in \cite{gunduz2008joint} from a theoretical perspective, \textcolor{black}{providing theoretical bounds for the distortion exponent as a function of the bandwidth ratio considering Gaussian sources.} It is challenging to design a practical JSCC scheme for MIMO channels, where the model needs to learn how to map the input signal to multiple antennas, and then to recover the input from multiple channel outputs, where each input-output antenna pair  experiences a distinct channel condition.  

Advantages of DeepJSCC and end-to-end MIMO schemes in practical MIMO transmission scenarios have prompted the investigation of DeepJSCC over MIMO systems. The first deep autoencoder (DAE) based end-to-end MIMO communication method was introduced in\cite{o2017deep}. In the subsequent work \cite{song2020benchmarking}, the authors systematically establish the symbol error rate benchmarks for MIMO channels by evaluating several DAE-based models with channel state information (CSI). A singular-value decomposition (SVD) based DAE is proposed in \cite{zhang2022svd} to \textcolor{black}{achieve competitive performance for the closed-loop MIMO system with CSI at the transmitter (CSIT).} These works indicate the potential of end-to-end MIMO communication schemes to enhance transmission quality when coupled with advanced deep learning (DL) technologies. However, we remark that the aforementioned works focus on the transmission of \textcolor{black}{bits/one-hot vectors} for \textcolor{black}{specific} signal-to-noise ratio (SNR) values, and do not consider the contextual features of the input signals, also referred to as `semantics' in the literature \cite{Gunduz:JSAC:23,lo2023collaborative}, and the model's adaptability to varying channel conditions, \textcolor{black}{as outlined in Table \ref{table:previous_work}.} Two recent studies that are more relevant to this work are \cite{shi2023excess} and VST\cite{yao2022versatile}, where \cite{shi2023excess} theoretically analyzes the excess distortion exponent for semantic-aware MIMO systems, while the authors in \cite{yao2022versatile} propose a semantically coded transmission scheme utilizing MIMO, assuming that CSI is available at the receiver (CSIR). 
 \begin{table*}[t]
    \centering
    %\caption{\textcolor{black}{Overview of the input source, generalizability, adaptability, channel robustness, and high-level conclusions for potential end-to-end MIMO pipelines}}
    \caption{\textcolor{black}{Overview of alternative end-to-end MIMO communication pipelines.}}
    \label{table:previous_work}
    \begin{tabular}{|p{0.142\textwidth}<{\centering}|p{0.13\textwidth}<{\centering}|p{0.144\textwidth}<{\centering}|p{0.125\textwidth}<{\centering}|p{0.13\textwidth}<{\centering}|p{0.15\textwidth}<{\centering}|}%
    \hline
    {\textbf{Potential Schemes}}& End-to-end AE\cite{o2017deep} & DAE benchmarks\cite{song2020benchmarking}&  SVD-DAE\cite{zhang2022svd}&VST\cite{yao2022versatile}& Our work\\
    \hline
     \textbf{Input source}  & One-hot input & One-hot input & Bits& Images & Images\\
     \hline
     \textbf{CSIR scenarios}  & \checkmark & $\checkmark$ &  \ding{55}& $\checkmark$ & $\checkmark$\\
         \hline
    \textbf{CSIT scenarios}  & $\checkmark$  & $\checkmark$& $\checkmark$& \ding{55} & $\checkmark$\\
        \hline
    \textbf{Antenna flexibility}  &  \ding{55} & \ding{55} & \ding{55} & \ding{55}  & \checkmark\\
    \hline
    \textbf{Channel adaptability}  &  \ding{55} &\ding{55} & \checkmark &  \checkmark & \checkmark\\
    \hline
    \textbf{Channel robustness}  & \checkmark &\ding{55} &\ding{55} & \ding{55} & $\checkmark$\\
    \hline
    \hline
     \textbf{Metrics} & Bit error rate & Block error rate&Bit error rate & PSNR & PSNR $\&$ LPIPS\\ 
     \hline
    \multirow{2}{*}{\textbf{Main baseline}} & SVD+QPSK+ & STBC$^*$/SVD + GS+ & \multirow{2}{*}{Plain-DAE} &\multirow{2}{*}{BPG+LDPC}&{BPG/DL-based} \\ 
    & equal power & bit/power loading& &&compression+Capacity\\ 
    \hline
    \multirow{2}{*}{\textbf{Results}} 
     & \multicolumn{3}{c|}{DAE outperforms/matches the baseline for CSIR; }& Method outperforms & Method outperforms  \\ 
      & \multicolumn{3}{c|}{DAE outperforms baseline at some specific SNRs for CSIT.$^{\dag}$}& the baseline & the baseline \\ 
    \hline
     \multirow{2}{*}{\textbf{Main insights}} & Pioneered the DAE-based MIMO & Benchmark end-to-end MIMO systems&Evaluate the model-driven DAE &  {Semantic coding over CQI $^{\ddag}$}& {First unified DeepJSCC scheme over MIMO} \\ 
\hline
    \end{tabular}
    \begin{tablenotes}
      \item $*$ STBC represents space-time block codes; GS represents geometrically-shaped signal constellations; QPSK is quadrature phase shift keying modulation; SVD is singular value decomposition precoding operation.
      \item $\dag$ For the CSIT, DAE-based models can outperform baseline at some specific SNRs, but cannot perform better than properly chosen benchmarks\cite{song2020benchmarking}.
      \item $\ddag$ CQI represents channel quality indicators, e.g., the SNR value, which needs to be assumed accessible for the transmitter in the open-loop MIMO system.
      
   \end{tablenotes}
\end{table*}

To the best of our knowledge, prior research has not delved into the specific exploration of adaptive JSCC MIMO systems designed for wireless image transmission, encompassing both scenarios with open-loop MIMO with only receiver-side CSI (CSIR) and closed-loop MIMO with both transmitter and receiver side CSI (CSIT). Such a system would be capable of leveraging the semantics of the source signal and channel conditions simultaneously, and efficiently managing both fluctuating channel conditions and potential channel estimation errors, all within a cohesive framework.

\textcolor{black}{The differences from earlier studies and the ensuing challenges we aim to address in this paper can be outlined across four key points: performance improvement, model generalizability, channel adaptability, and channel robustness. These facets become particularly important to ensure the framework's effectiveness and practicality.}
\begin{itemize}
    \item \textcolor{black}{Performance improvement: 
    Prior DL-based designs for MIMO systems \cite{song2020benchmarking, zhang2022svd, o2017deep},  have mostly focused on symbol detection in a digital communication framework. Moreover, even the state-of-art DL-based solution \cite{song2020benchmarking} cannot outperform a properly selected traditional model-driven solution, which underscores the inherent difficulty of learning effective communication strategies in MIMO systems. Another important aspect of the performance is the memory and computational complexity, which further motivates competitive and practical end-to-end architectures.%Subsequent challenges arise from the retrieval of coupled signals and practicality problem in the implementation of a JSCC scheme over MIMO channels, motivating for our study into a model-driven-based DeepJSCC approach.
    } 
    
    \item \textcolor{black}{Model generalizability: Most of the proposed schemes \cite{zhang2022svd, yao2022versatile} have been designed specifically for open-loop or closed-loop systems. Moreover, they require retraining to be used in different antenna configurations. The impracticality of customizing distinct models tailored for specific scenarios is apparent, prompting our search for a unified DeepJSCC solution. Specifically, our ambition is to design a unique DeepJSCC scheme which can generalize to both CSIR and CSIT scenarios, and varying antenna configurations.}

    \item \textcolor{black}{Channel adaptability: The adaptability to diverse channel conditions constitutes a pivotal attribute of DL-based communication schemes \cite{xu2021wireless, zhang2022svd}. The separate training of neural networks for different channel conditions necessitates massive storage space for multiple networks corresponding to distinct channel conditions. Previous investigations of channel adaptability in DeepJSCC within single input single output (SISO) system focus on incorporating the SNR as an additional network input \cite{xu2021wireless}. However, transferring these approaches to MIMO channels remains challenging, where the model needs to simultaneously account for the coupled channel inputs and the channel matrix. Consequently, we employ model-driven techniques to disentangle the channels and introduce an innovative paradigm, integrating the self-attention mechanism and CSI heatmap, to learn and adapt to the channel conditions  effectively.}

    \item \textcolor{black}{Channel Robustness: CSIR and CSIT are typically acquired through channel estimation and feedback, respectively, which can introduce estimation errors that can lead to reduction in the reliability of the transmitted information \cite{baltersee2001achievable}. Authors in \cite{yoo2006capacity} demonstrated that, in the presence of imperfect CSIT and minimal noise, MIMO performance can be enhanced through power allocation compared to systems without CSIT. However, an increase in noise level can significantly deteriorate the performance. Meanwhile, the joint design of the system with channel estimation errors enables model robustness against channel estimation errors \cite{he2020model, o2017deep}. This observation motivates us to incorporate imperfect CSI into the proposed DeepJSCC framework.} 
\end{itemize}

This paper presents an innovative DeepJSCC scheme meticulously tailored for MIMO image transmission \textcolor{black}{to address the above practical concerns.} Our approach introduces a unified vision transformer (ViT)-based DeepJSCC scheme, named DeepJSCC-MIMO, which accommodates both scenarios with CSIR and CSIT. Inspired by the attention mechanism in developing flexible communication schemes \cite{wu2022channel, xu2021wireless, shao2022attentioncode, ozfatura2022all,wu2023transformer}, we leverage the self-attention mechanism inherent in ViT for adaptive wireless image transmission. Specifically, we represent the channel conditions with a channel heatmap and adapt the JSCC encoding and decoding parameters according to this heatmap. Our approach can learn global attention parameters between the source image and the channel conditions across all the intermediate layers of the DeepJSCC encoder and decoder. Intuitively, we expect this design to simultaneously learn channel symbol mapping and power allocation, considering different channel conditions, channel estimation errors, and different antenna number configurations. Therefore, unlike most existing literature on DNN-aided MIMO communications, we eliminate the need for training separate DNN models for different transceiver settings and channel conditions. The proposed DeepJSCC-MIMO is a practical and unified scheme with robustness to channel estimation errors. 

Our main contributions can be summarized as follows:

\begin{itemize}
\item \textcolor{black}{This paper presents the first DeepJSCC scheme over the MIMO system for image transmission. The key innovation lies in utilizing a ViT-based model, which adeptly exploits both the semantic features of the image and the CSI through a self-attention mechanism. This design of jointly leveraging attention over both the source and channel aspects can be readily applied to other emerging communication systems concerning semantic sources.}

\item \textcolor{black}{DeepJSCC-MIMO is a comprehensive and versatile solution that can be seamlessly applied to both open-loop and closed-loop MIMO systems. Importantly, DeepJSCC-MIMO demonstrates remarkable flexibility and adaptability by efficiently accommodating diverse channel conditions and antenna number configurations. This is achieved without the necessity for retraining, thanks to the utilization of the channel heatmap and the self-attention mechanism.
%, DeepJSCC-MIMO intelligently captures and adapts to the characteristics of the channel conditions and antenna configurations. 
Furthermore, the DeepJSCC-MIMO scheme exhibits significant resilience to channel estimation errors. }

\item \textcolor{black}{Extensive numerical evaluations validate the superiority of the proposed model, showcasing significant enhancements in both distortion quality and perceptual quality across a wide range of channel conditions and bandwidth ratios when compared to traditional separate source and channel coding schemes.}
\end{itemize}

\section{System Model}
We consider an $M\times M$ MIMO system, where an $M$-antenna transmitter aims to deliver an image $\bm{S}\in \mathbb{R}^{h\times w\times 3}$ to an $M$-antenna receiver ($h$ and $w$ denote the height and width of the image, while $3$ refers to the color channels R, G, and B). The transmitter encodes the image into a vector of channel symbols $\bm{X}\in \mathbb{C}^{M\times k}$, where $k$ \textcolor{black}{denotes the number of channel uses dedicated to transmitting one image.} We define the \textit{bandwidth ratio} as $R\triangleq k/n$, \textcolor{black}{representing the average number of available channel symbols per source dimension}, where $n=3hw$ is the number of source symbols. %Intuitively, $R$ reflects the number of available channel symbols for each source symbol. 
The transmitted signal $\bm{X}$ is subject to a power constraint $P_s$ as:
\begin{equation}
   \frac{1}{Mk}\|\bm{X}\|_F^2\leq P_s,
   \label{power}
\end{equation}
where $\|\cdot\|_\text{F}$ denotes the Frobenius norm, and we set $P_s=1$ without loss of generality.

The channel model can be written as: 
\begin{equation}
   \bm{{Y}}=\bm{H}\bm{X}+\bm{W},
\label{MIMO_channel}
\end{equation}
where $\bm{X}\in \mathbb{C}^{M\times k}$ and $\bm{Y}\in \mathbb{C}^{M\times k}$ denote the channel input and output matrices, respectively, while $\bm{W}\in \mathbb{C}^{M\times k}$ is the additive white Gaussian noise (AWGN) term, \textcolor{black}{whose entries ${W}[i,j]\sim \mathcal{CN}(0,\sigma_w^2)$} follow an independent and identically distributed (i.i.d.) complex Gaussian distribution with zero mean and variance $\sigma_w^2$. The entries of the channel gain matrix $\bm{H}\in \mathbb{C}^{M\times M}$ also follow a complex Gaussian distribution with zero mean and variance $\sigma_h^2$, i.e., $\bm{H}[i,j]\sim \mathcal{CN}(0,\sigma_h^2)$. We consider a block-fading channel model, in which the channel matrix $\bm{H}$ remains constant for $k$ channel uses, corresponding to the transmission of one image, and takes an independent realization in the next block. 

Given the channel output $\bm{Y}$, the receiver reconstructs the  source image as $\bm{\hat{S}}\in \mathbb{R}^{h\times w\times 3}$. \textcolor{black}{The reconstruction quality is quantified through the peak signal-to-noise ratio (PSNR) and the learned perceptual image patch similarity (LPIPS) metrics\cite{8578166}. The PSNR metric serves as an indicator of image distortion at the per-pixel level, defined as}:
\begin{equation}
\text{PSNR} \triangleq 10\log_{10}\frac{\|\bm{S}\|_\infty^2}{\text{MSE}(\bm{S},\bm{\hat{S}})} ~ (\text{dB}),
\end{equation}
\textcolor{black}{where $\|\cdot\|_\infty$ is the infinity norm and  $\text{MSE}(\bm{S},\bm{\hat{S}})\triangleq \frac{1}{3hw} \|\bm{S}-\bm{\hat{S}}\|^2_2$ is the mean squared error (MSE) between $\bm{S}$ and $\bm{\hat{S}}$. LPIPS, on the other hand,  computes the dissimilarity between the feature vectors of the input and reconstructed images, where a lower LPIPS score implies a better perceptual quality, defined as:}
\textcolor{black}{
\begin{equation}
    \text{LPIPS} (\bm{S},\bm{\hat{S}}) \triangleq \sum_L \frac{1}{H_lW_l}\sum_{h,w}\|\bm{w_l}\odot(\bm{{y}^l}-\bm{\hat{y}^l})\|^2_2,
\end{equation}
where $\bm{{y}^l}, \bm{\hat{y}^l}\in \mathbb{R}^{H_l\times W_l\times C_l}$ are the intermediate features derived from the $l$-th layer of the given network (VGG in our case) composed of a total of $L$ layers. $H_l, W_l,$ and $C_l$ are the intermediate feature map's height, width, and channel dimensions, $\bm{w^l}\in \mathbb{R}^{C_l}$ is the weight vector, and $\odot$ is the channel-wise feature multiplication operation \cite{8578166}.}

\begin{comment}
We use the peak signal-to-noise ratio (PSNR) as the distortion metric:
\begin{equation}
\text{PSNR}=10\log_{10}\frac{\|\bm{S}\|_\infty^2}{\text{MSE}(\bm{S},\bm{\hat{S}})} ~ (\text{dB}),
\label{psnr}
\end{equation}
where $\|\cdot\|_\infty$ is the infinity norm and  $\text{MSE}(\bm{S},\bm{\hat{S}})\triangleq \frac{1}{3hw} \|\bm{S}-\bm{\hat{S}}\|^2_2$ is the mean squared error (MSE) between $\bm{S}$ and $\bm{\hat{S}}$.   
\end{comment}

\begin{figure}[t] 
    \centering
    \includegraphics[scale=0.45]{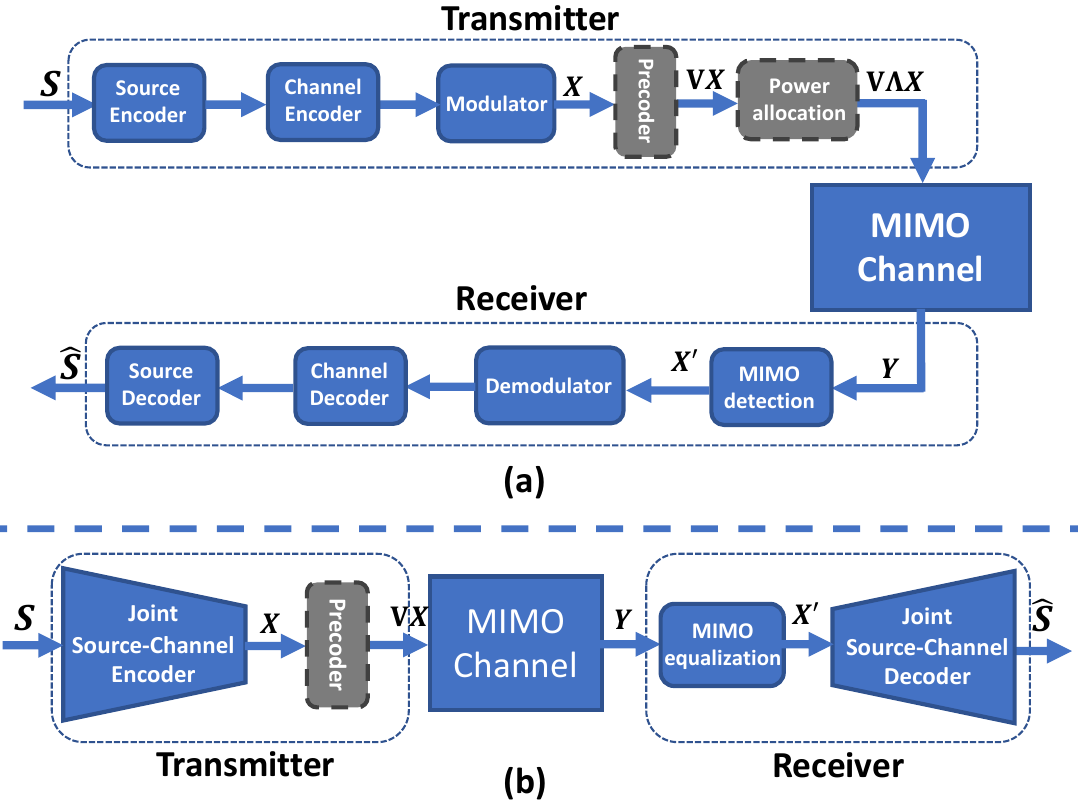}
    \caption{Block diagram of the MIMO image transmission system: (a) conventional separate source-channel coding scheme and (b) DeepJSCC scheme, where the gray blocks with dashed lines are the additional operations for the closed-loop MIMO system.}
    \label{MIMO_pipeline}
\end{figure}

As shown in Fig. \ref{MIMO_pipeline}, there can be two approaches to solve this problem, i.e., the traditional separate source and channel coding scheme and the JSCC scheme. This paper considers both the CSIR (open-loop MIMO) and CSIT (closed-loop MIMO) scenarios for each transmission scheme.

\subsection{Open-loop MIMO with CSIR}
In an open-loop MIMO system, CSI is only available at the receiver,  enabling it to equalize the transmitted signals and subsequently decode the image.

\subsubsection{Separate source and channel coding scheme}
In an open-loop MIMO system with separate source and channel coders, the transmitter follows a sequential process, applying source coding, channel coding, and the modulation in order. {We note that, due to the lack of CSI, the transmitter has to choose the compression and channel coding rates and the constellation independently of the current channel realization.
}

Given the received signal $\bm{Y}$ and the CSI, the receiver decodes the source signal as $\bm{\hat{S}}$ by sequentially performing the MIMO detection, demodulation, channel decoding, and source decoding operations.

%can perform the MIMO detection algorithms to retrieve the multiple transmitted data $\bm{X}'$ from each transmitting antenna. With $\bm{X}'$, the receiver sequentially performs demodulation, channel decoding, and source decoding to reconstruct the source signal denoted as $\bm{\hat{S}}$.

\subsubsection{DeepJSCC scheme} Different from the traditional separation-based scheme, DL-based JSCC scheme uses a joint source-channel encoder at the transmitter, denoted as $f_{\bm{\theta}}$, to directly map the source signal $\bm{S}$ into the channel input matrix $\bm{X}$. \textcolor{black}{The encoder $f_{\bm{\theta}}$ is trained to achieve the best end-to-end performance across diverse channel realizations in an average sense since the transmitter does not have access to CSI.} This process can be represented as:
\begin{equation}
    \bm{X}=f_{\bm{\theta}}(\bm{S}).
\end{equation}
%Without CSIT, the JSCC encoder $f_{\bm{\theta}}$ is optimized to learn the source-to-channel mapping for the best reconstruction quality over the distribution of channel conditions and source images.

At the receiver, a MIMO equalization algorithm, such as zero-forcing (ZF) or minimum mean square error (MMSE), is employed first to exploit the CSI to decouple the entangled signal $\bm{Y}$ as $\bm{X'}$. Subsequently, a JSCC decoder, referred to as $f_{\bm{\phi}}$, is employed to reconstruct the source signal based on the available CSI represented by ($\bm{H}$, $\sigma_w^2$), along with $\bm{X'}$. This can be expressed as: 

\begin{equation}
   \bm{\hat{S}}=f_{\bm{\phi}}(\bm{X'},\bm{H},\sigma_w^2).
\end{equation}

An alternative approach is to directly learn the decoding process using the CSI without explicitly performing channel equalization. This involves training the model to minimize the distortion in an end-to-end manner. However, in order to achieve better transmission quality, we adopt a model-driven approach that explicitly performs channel equalization prior to the decoding process, \textcolor{black}{leveraging the domain knowledge to simplify the learning process, similarly to \cite{he2020model, 9714510}.}

\subsection{Closed-loop MIMO with CSIT}
Within a closed-loop MIMO system, the CSI is accessible to both the transmitter and receiver, which allows them to apply pre-coding and power allocation at the transmitter, and MIMO equalization at the receiver, thereby improving the image transmission quality.

\subsubsection{Separate source and channel coding scheme}
The transmitter sequentially performs source coding, channel coding, and modulation to generate the channel input matrix $\bm{X}$, the elements of which are constellations with average power normalized to 1. Additional operations at the transmitter with CSIT are precoding and power allocation, which can boost the communication rate. Specifically, given the CSI, we first decompose the channel matrix $\bm{H}$ by singular-value decomposition (SVD), yielding $\bm{{H}}=\bm{U\Sigma V^H}$, where $\bm{U}\in \mathbb{C}^{M\times M}$ and $\bm{V}\in \mathbb{C}^{M\times M}$ are unitary matrices, $\bm{V}^{H}$ is the conjugate transpose of $\bm{V}$, and $\bm{\Sigma}$ is a diagonal matrix whose singular values are in descending order. We denote $\bm{\Sigma}$ by $\text{diag}(s_1,s_2,\dots, s_M)$, where $s_1\geq s_2\geq \dots \geq s_M$.

We denote the power allocation matrix by $\bm{\Lambda}$, where $\bm{\Lambda}$ is diagonal, with its diagonal elements being the power allocation weights for the signal streams of separate antennas. $\bm{\Lambda}$ can be derived using the standard water-filling algorithm \cite{scutari2009mimo}. With power allocation and SVD precoding (precoding $\bm{X}$ into $\bm{VX}$), Eqn. \eqref{MIMO_channel} can be rewritten as:
\begin{equation}\label{eq:MIMO1}
   \bm{{Y}}=\bm{H}\bm{V}\bm{\Lambda X}+\bm{W}=\bm{U\Sigma}\bm{\Lambda X}+\bm{W}. 
\end{equation}
Multiplying both sides of Eqn. \eqref{eq:MIMO1} by $\bm{U}^H$ (MIMO detection) gives us:
\begin{equation}
   \bm{X'}=\bm{\Sigma}\bm{\Lambda X}+\bm{U}^{H}\bm{W}.
   \label{svd_MIMO_out}
\end{equation}

SVD-based precoding converts the MIMO channel into a set of parallel subchannels with different SNRs.
In particular, the SNR of the $i$-th subchannel is determined by the $i$-th singular value of $s_i$ and the $i$-th power allocation coefficient of $\bm{\Lambda}$. 

Given $\bm{X'}\in \mathbb{C}^{M\times k}$, the receiver performs demodulation, channel decoding, and source decoding sequentially to reconstruct the source image as $\bm{\hat{S}}$.

\subsubsection{DeepJSCC scheme}
For the DeepJSCC scheme, we exploit DL technologies to parameterize the encoder and decoder functions, which are trained jointly on an image dataset and the channel model as in Eqn. \eqref{MIMO_channel}. 
Let us denote the DeepJSCC encoder and decoder by $f_{\bm{\theta}}$ and $f_{\bm{\phi}}$, respectively, where $\bm{\theta}$ and $\bm{\phi}$ denote the network parameters. We have
\begin{equation}
  \bm{X}=f_{\bm{\theta}}(\bm{S},\bm{H},\sigma_w^2).
   \label{dl_encoder}
\end{equation}
Unlike the separate source-channel coding scheme, the transmitter does away with power allocation. Instead, we leverage the DeepJSCC encoder to perform feature extraction and channel symbol mapping, and power allocation, all at once. Intuitively, the DNN is expected to transmit critical features over subchannels with higher SNRs, thereby improving the transmission performance.

We note here that one option is to train the encoder/decoder networks directly, hoping they will learn to exploit the spatial degrees-of-freedom that the MIMO channel provides. We will instead follow the model-driven approach, where we will exploit the SVD and first convert the MIMO channel into subchannels. The received signal can then be written as
\begin{equation}
\bm{{Y}}=\bm{HVX+W}=\bm{U\Sigma}\bm{X}+\bm{W}.
\label{precode_dl_MIMO}
\end{equation}

To simplify the training, we apply MIMO equalization at the receiver by left multiplying both sides of Eqn. \eqref{precode_dl_MIMO} by $\bm{\Sigma}^{\dagger}\bm{U}^H$ to obtain
\begin{equation}
%\bm{z'}=\bm{U^{H}Y}=\bm{\Sigma X}+\bm{U^HW},
\bm{X'}=\bm{\Sigma}^{\dagger}\bm{U}^{H}\bm{Y}=\bm{X}+\bm{W'},
\label{MIMO_detec}
\end{equation}
where $\bm{W}'\triangleq \bm{\Sigma^{\dagger}U^HW}\in \mathbb{C}^{M\times k}$ is the equivalent noise term, $\bm{\Sigma^{\dagger}}$ is the Moore-Penrose inverse of matrix $\bm{\Sigma}$, and $\bm{U}^{H}$ is the conjugate transpose of matrix $\bm{U}$.

Finally, we feed both $\bm{X'}$ and the CSI into the DeepJSCC decoder to recover the image as
\begin{equation}
 \bm{\hat{S}}=f_{\bm{\phi}}(\bm{X'},\bm{H},\sigma_w^2).
   \label{dl_decoder}
\end{equation}

\textcolor{black}{In summary, our DeepJSCC framework integrates some well-established algorithms, such as SVD-based precoding and equalization operations, from \textcolor{black}{previous} model-driven MIMO solutions. The objective is to improve the end-to-end performance of DL-based JSCC approach by exploiting domain knowledge from conventional MIMO design.}

\section{Proposed method}
\begin{figure}[t] 
    \centering
    \includegraphics[scale=0.55]{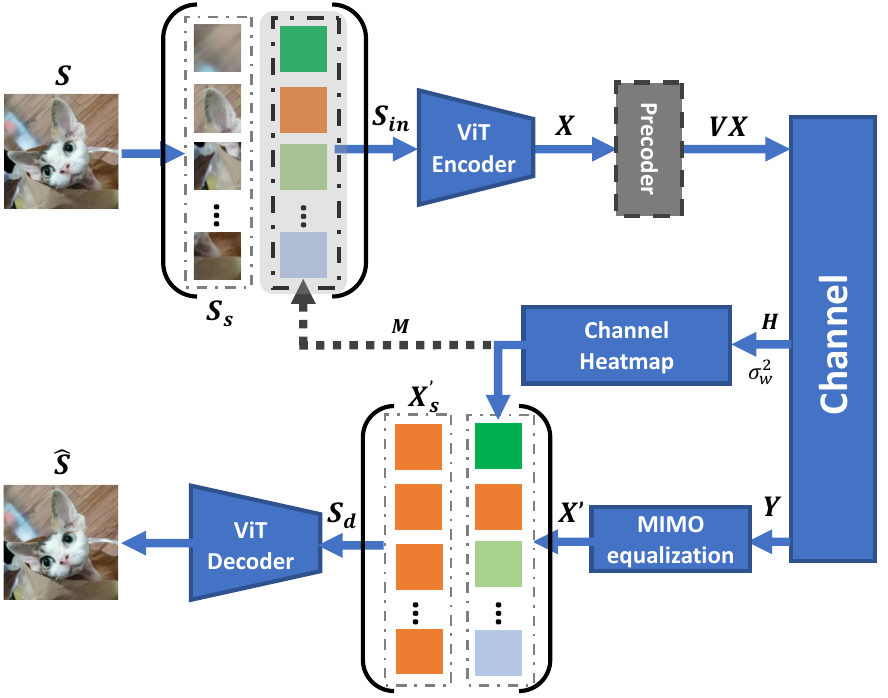}
    \caption{The pipeline of the DeepJSCC-MIMO scheme, where the source image $\bm{S}$ is encoded by a ViT-encoder and reconstructed by a ViT-decoder as $\bm{\hat{S}}$. Precoding operation in dashed line will be performed if the CSI is available at the transmitter, and the CSI ($\bm{H},\sigma_w^2$) is fed to the encoder, if available, and decoder in the form of a ``heatmap'' $\bm{M}$ to facilitate the JSCC encoding/decoding process.}
    \label{Secom_pipeline}
\end{figure}

\begin{figure*}[tb] 
    \centering
    \includegraphics[scale=0.7]{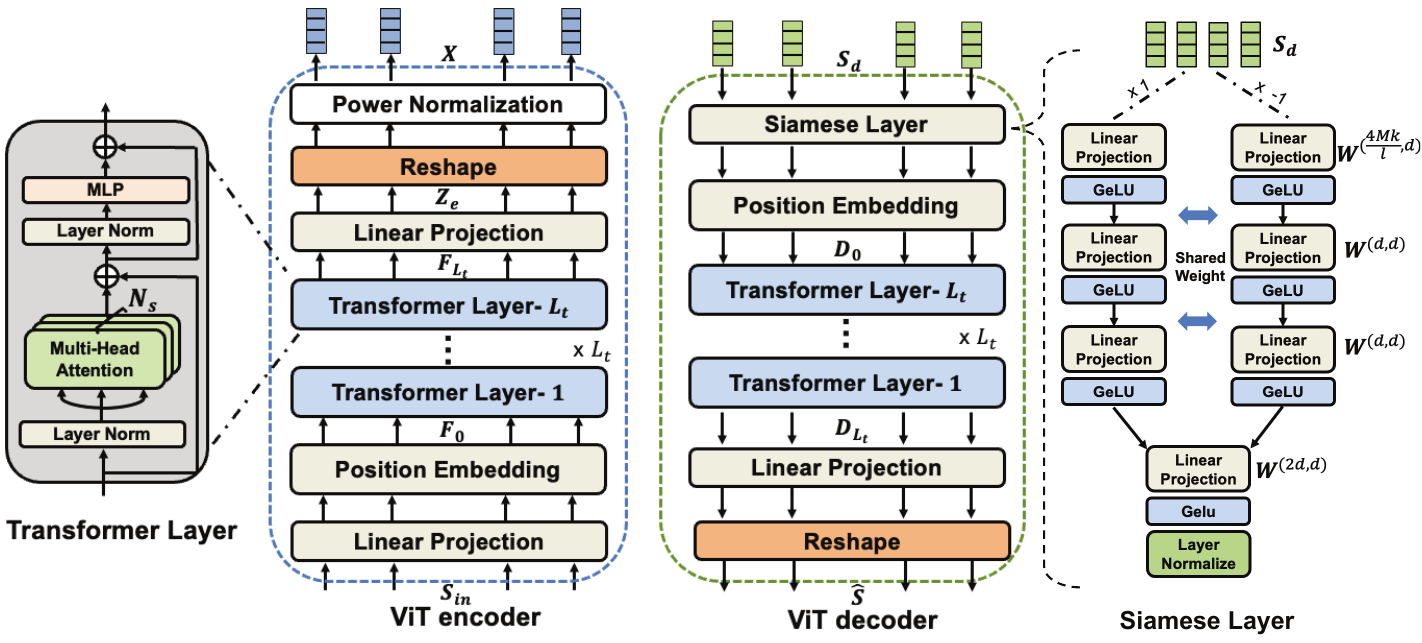}
    \caption{The architecture of the ViT-based encoder and decoder, where both encoder and decoder comprise linear projection layers, a positional embedding layer, and multiple transformer layers.
    }
    \label{ViT_structure}
\end{figure*}

This section presents a novel DeepJSCC architecture called DeepJSCC-MIMO, which aims to enable efficient image transmission in MIMO systems. %The DeepJSCC-MIMO scheme is mainly based on the vision transformer and incorporates self-attention mechanisms to exploit the CSI. 
The pipeline of our DeepJSCC-MIMO scheme is \textcolor{black}{depicted} in Fig. \ref{Secom_pipeline} and \textcolor{black}{outlined} in Algorithm \ref{MIMO_algorithm}. \textcolor{black}{In this scheme,} DeepJSCC-MIMO employs a pair of ViT-based encoder $f_{\bm{\theta}}$ and decoder $f_{\bm{\phi}}$ \textcolor{black}{to encode and decode channel symbols, utilizing the inherent self-attention mechanism over both source features and channel conditions.} Further elaborations on the inner structures of these ViTs are provided in Fig. \ref{ViT_structure}. In the following, we detail the pipeline of DeepJSCC-MIMO in five main steps: image-to-sequence transformation, channel heatmap construction, ViT encoding, ViT decoding, and the loss function.

\subsection{Image-to-sequence transformation}
To construct the input of DeepJSCC-MIMO, we first convert the three-dimensional input image $\bm{S}$ into a sequence of vectors, denoted by $\bm{S_s}=\textit{Seq}(\bm{S})$. Specifically, given a source image $\bm{S}\in\mathbb{R}^{h\times w\times 3}$, we divide $\bm{S}$ into a grid of $p\times p$ patches, and {reshape each patch into a vector of dimension $\mathbb{R}^{\frac{3hw}{p^2}}$}. In this way, $\bm{S}$ is converted to $\bm{S_s}\in\mathbb{R}^{l\times c}$, where $l={p^2}$ is the sequence length and $c\triangleq\frac{3hw}{p^2}$ is the dimension of each vector in the sequence.

\subsection{Channel heatmap construction}
To enable efficient training, we construct a channel heatmap from CSI, indicating the effective noise variance faced by each channel symbol generated by ViT. Let us define $\bm{P_n}\in \mathbb{R}^{M\times k}$ as the average power of the equivalent additive noise term $\bm{W'}$. Specifically, for different scenarios, $\bm{{P}_{n}}$ is defined as:
\begin{equation}
   \bm{{P}_{n}}\triangleq \begin{cases}
\sigma_w^2 (\bm{H_w}\odot\bm{\bar{H}_w})
\bm{J}_{M\times k}, & \text{MIMO with CSIR}\\
{\sigma}_w^2\bm{\Sigma}^{\dagger}\bm{U}^H\bm{J}_{M\times k}, & \text{MIMO with CSIT},\\
\end{cases}
\label{forward_link}
\end{equation}
{where $\odot$ denotes the Hadamard product and $\bm{\bar{H}_w}$ is the element-wise conjugate of matrix $\bm{{H}_w}$, which is obtained from the zero-forcing operation as: $\bm{H_w}\triangleq\bm{(H^HH)^{-1}H^H}$}, and $\bm{J}$ is the matrix of ones.

Then the heatmap $\bm{M}\in\mathcal{R}^{l\times \frac{2Mk}{l}}$ is constructed as follows:
\begin{equation}
   \bm{{M}}\triangleq \texttt{reshape} \left(\texttt{concat}\left(\frac{1}{2}\bm{P_n},\frac{1}{2}\bm{P_n}\right)\right),
   \label{heatmap_open}
\end{equation}
where $\texttt{concat}(\cdot)$ and $\texttt{reshape}(\cdot)$ denote the concatenation and reshape operations, respectively. We concatenate two matrices $\frac{1}{2}\bm{P_n}$ to get the shape of $\mathbb{R}^{M\times 2k}$, and reshape it into $\bm{M}\in\mathcal{R}^{l\times \frac{2Mk}{l}}$, which represents the equivalent noise term faced by each real encoder output element.

We expect that feeding the CSI in the form of a heatmap will simplify the training process as the model only needs to focus on the `additive' noise power faced by each channel symbol. As we will show later, this design enables our model to be used under various antenna numbers without retraining.

\subsection{{ViT encoding}}
For the two scenarios, the input sequences to our ViT encoder $\bm{{S}_{in}}\in \mathbb{R}^{l\times c_{in}}$ are given by:
\begin{equation}
   \bm{{S}_{in}}= \begin{cases}
\bm{S_s}, & \text{MIMO with CSIR}\\
\texttt{concat}(\bm{S_s},\bm{M}), & \text{MIMO with CSIT},\\
\end{cases}
\label{ViT_input}
\end{equation}
where the dimensions of the vectors in these sequences are $c_{in}=c$ and $c_{in}=c+\frac{2Mk}{l}$ for the MIMO with CSIR and MIMO with CSIT scenarios, separately.

The architecture of our encoder is shown in Fig.  \ref{ViT_structure}, which mainly consists of linear projection layers, a positional embedding layer, and several transformer layers. 

\textbf{Linear projection and positional embedding:}
$\bm{S_{in}}$ is first linearly projected by $\bm{W_{0}}\in \mathbb{R}^{c_{in}\times d}$ followed by a positional embedding operation $P_{e}(\cdot)$ to get the initial input $\bm{F_{0}}\in \mathbb{R}^{l\times d}$ for the following transformer layers:
\begin{equation}
\bm{F_{0}}=\bm{S_{in}}\bm{W_{0}}+P_{e}(\bm{p}),
\label{pos_embed}
\end{equation}
where $d$ is the output dimension of the hidden projection layer. {Positional embedding $P_{e}(\cdot)$ serves the purpose of representing the spatial arrangement of sequences for better performance, %ensuring that distinct positions are allocated with unique encoded representations, 
where two typical methods are dense layer-based position embedding (DPE) \cite{zheng2021rethinking} and conditional position embedding (CPE) \cite{chu2021twins}. In specific, our $P_{e}(\cdot)$ is implemented with a DPE that employs a dense layer to embed the index vector $\bm{p}$ of each patch into a $d$ dimensional vector.}

\textbf{Transformer Layer:}
As shown in Fig. \ref{ViT_structure}, the intermediate feature map $\bm{F_{i}}$ is generated by the $i\textit{-}th$ transformer layer by a multi-head self-attention (MSA) block and a multi-layer perceptron (MLP) layer as:
\begin{equation}
\bm{F_{i}}=MSA(\bm{F_{i-1}})+MLP(MSA(\bm{F_{i-1}})),
\label{trans_layer_eqn}
\end{equation}
where $\bm{F_{i}}\in \mathbb{R}^{l\times d}$ is the output sequence of the $i$-th transformer layer, GeLU activation and layer normalization operations are applied before each MSA and MLP block. 

Each MSA block consists of $N_{s}$ self-attention (SA) modules with a residual skip, which can be formulated as:
\begin{equation}
MSA(\bm{F_{i}})=\bm{F_{i}}+[SA_1(\bm{F_{i}}),\cdots,SA_{N_{s}}(\bm{F_{i}})]\bm{W_i},
\label{msa}
\end{equation}
where the output of all SA modules $SA(\bm{F_{i}})\in \mathbb{R}^{l\times d_{s}}$ are concatenated for a linear projection $\bm{W_i}\in \mathbb{R}^{d_{s}N_{s}\times d}$, $d_s=d/N_{s}$ is the output dimension of each SA operation.

For each SA module, the operations are formulated as:
\begin{equation}
SA(\bm{F_{l-1}})=softmax\left( \frac{\bm{qk}^{T}}{\sqrt{d}}\right)\bm{v},
\label{SA_eqn}
\end{equation}
where the $\bm{q},\bm{k},\bm{v}\in \mathbb{R}^{l\times d_s}$ are the query, key, and value vectors generated through three linear projection layers $\bm{W_q,W_k,W_v}\in \mathbb{R}^{d\times d_s}$ as:
\begin{equation}
    \bm{q}=\bm{F_{l-1}W_q},~~\bm{k}=\bm{F_{l-1}W_k},~~\bm{v}=\bm{F_{l-1}W_v}.
    \label{sa}
\end{equation}

\textbf{Linear projection and power normalization:}
After $L_t$ transformer layers, we apply a linear projection $\bm{W_c}\in \mathbb{R}^{d\times \frac{2Mk}{l}}$ to map the output of the transformer layers $\bm{F_{L_t}}$ to the channel symbols as: 
\begin{equation}
\bm{Z_e}=\bm{F_{L_t}}\bm{W_c},
\label{last_dense}
\end{equation}
where $\bm{Z_e}\in \mathbb{R}^{l\times \frac{2Mk}{l}}$ is then reshaped and normalized to satisfy the power constraints to form the complex channel input symbols $\bm{X}\in \mathbb{C}^{M\times k}$.

\subsection{ViT decoding}
\textbf{Channel equalization:} To simplify the decoding process, we perform an equalization operation on the channel output $\bm{Y}$ to decouple the transmitted data of each antenna as $\bm{X'}\in \mathbb{C}^{M\times k}$ before feeding the signal into the ViT decoder. We apply different equalization methods for the MIMO systems with CSIR and with CSIT:
\begin{equation}
   \bm{{X}'}= \begin{cases}
\bm{H_w}\bm{Y}+f_{res}(\bm{Y},\bm{H}), & \text{MIMO with CSIR}\\
\bm{\Sigma^{\dagger}U^HY}, & \text{MIMO with CSIT}.\\
\end{cases}
\label{equalization}
\end{equation}
 
\begin{figure}[t]
     \centering
    \includegraphics[scale=0.65]{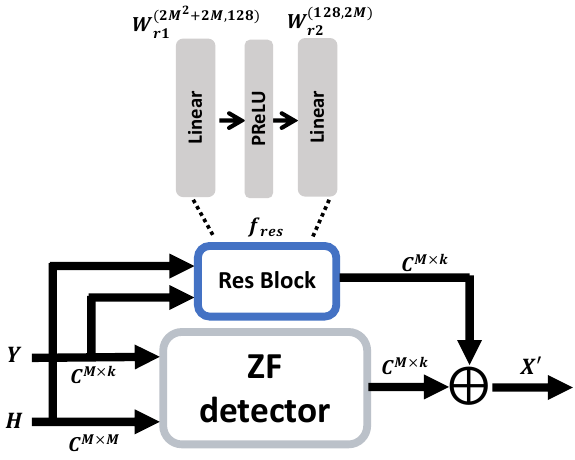}
     \caption{Structure of DL-aided channel equalization method, where the input at the $i$-th channel use is the concatenation of all elements within $\bm{H}$ and the channel output}
     \label{MIMO_dl_detection}
 \end{figure}

To perform MIMO equalization in an open-loop MIMO system with CSIR, we designed a DL-aided equalization method to retrieve the transmitted signal $\bm{X'}$ from $\bm{Y}$, as detailed in Fig. \ref{MIMO_dl_detection}. 

Specifically, we add a residual block, denoted as $f_{res}(\cdot)$, to learn the compensation after the zero-forcing (ZF) channel equalization operation based on the received signals and CSI, given as:
\begin{align}
 \bm{{X'}} & = \bm{H_wY}+f_{res}(\bm{Y},\bm{H})\\ 
  & =\bm{X}+\bm{H_wW}+f_{res}(\bm{Y},\bm{H}),
\end{align}
where $\bm{{W'}}\triangleq \bm{H_wW}\in \mathbb{C}^{M\times {k}}$ is the equivalent additive noise term from the channel. The residual operation $f_{res}(\cdot)$ encompasses two linear layers, parameterized with $\bm{W_{r1}}\in \mathbb{R}^{(2M^2+2M)\times 128}$ and $\bm{W_{r2}}\in \mathbb{R}^{128\times 2M}$, and a PReLU activation function, \textcolor{black}{formulated as:
\begin{equation}
f_{res}(\bm{H},\bm{Y})\triangleq \text{PReLU}(\bm{f_{res_{in}}}\bm{W_{r1}})\bm{W_{r2}}
\end{equation}
 %The input of the $f_{res}(\cdot)$ at the $i$-th channel use is the concatenation of all elements within $\bm{H}$ and $\bm{Y}[:,i]$}. 
where the elements of $\bm{H}$ and $\bm{Y}$ in each time slot are concatenated to form an input $\bm{f_{res_{in}}}\triangleq\texttt{concat}(\bm{H,\bm{Y}})\in \mathbb{R}^{2M^2+2M}$ for the residual block.}
\textcolor{black}{We jointly optimize this residual block with the whole pipeline and expect it to compensate for ZF equalization when faced with channel estimation errors.} Note that we can also apply other MIMO equalization methods. However, we have observed that a simple ZF estimation method, together with a learned compensation, is sufficiently effective. %Compared with traditional separation schemes equipped with a better MIMO detection method, such as the sphere decoding algorithm, we use a simple ZF algorithm to evaluate the powerful performance of our DeepJSCC-MIMO scheme. 
Moreover, the ZF method converts the MIMO channel into equivalent parallel sub-channels, which suits our approach of applying the self-attention mechanism over different sub-channels and source signals. 
%After equalization, $\bm{{X'}}$ is then fed into a ViT decoder to reconstruct the source image. 

To perform MIMO equalization in a closed-loop MIMO system with CSIT, we perform SVD decomposition and precoding as in Eqn. \eqref{precode_dl_MIMO}. At the receiver, we perform equalization as in Eqn. \eqref{MIMO_detec} to get: 
\begin{equation} \bm{{X'}}=\bm{\Sigma^{\dagger}U^HY}=\bm{X}+\bm{\Sigma^{\dagger}U^HW},
\end{equation}
which converts the MIMO channel into a set of parallel subchannels. 

\textcolor{black}{In summary, the proposed DeepJSCC-MIMO scheme obtains disentangled channel output $\bm{X}'$ via channel equalization procedures, which is shown to improve decoding performance, as demonstrated in \cite{wu2023deep_arxiv}.} 

Given the equalized signal $\bm{X'}$, which is reshaped into $\bm{X^{'}_{s}}\in \mathbb{R}^{l\times \frac{2Mk}{l}}$, and the noise heatmap $\bm{M}$, a ViT-based decoder $f_{\bm{\phi}}$ is designed to recover the source image as $\bm{\hat{S}}=f_{\bm{\phi}}(\bm{X^{'}_{s}},\bm{M})$. Our ViT-based decoder consists of a Siamese layer, positional embedding layer, transformer layer, and linear projection layer, elaborated upon below:

\begin{algorithm}[t] 
    \caption{Algorithm of the DeepJSCC-MIMO scheme}
    \textbf{For the transmitter:} \\
        \textbf{Input:} $\bm{S}\in \mathbb{R}^{h\times w\times 3}$, $\bm{M}\in \mathbb{R}^{l\times \frac{2Mk}{l}}$ (MIMO with CSIT)\\
        \textbf{Output:} $\bm{X}\in \mathbb{C}^{M\times k}$\\
    \vspace{-15pt}
    \begin{algorithmic}[1]
        \State \texttt{$\bm{S_s}={Seq}(\bm{S})\in\mathbb{R}^{l\times c}$} 
     \Comment{Image Sequentialization}
       \If{\textit{MIMO with CSIR}}
        \State \texttt{$\bm{{S_{in}}}=\bm{S_s}$}\;
        \State \texttt{$\bm{X}=f_{\theta}(\bm{S_{in}})\in \mathbb{C}^{M\times k}$}     
        \Comment{ViT encoding}
        \ElsIf{\textit{MIMO with CSIT}}
        \State \texttt{$\bm{{S_{in}}}={concat}(\bm{S_s},\bm{M})$}\;
        \State \texttt{$\bm{X}=f_{\theta}(\bm{S_{in}})\in \mathbb{C}^{M\times k}$}     
        \Comment{ViT encoding}
        \State \texttt{$\bm{{H}}=\bm{U\Sigma V^H}$}       
        \Comment{SVD decomposition}
        \State \texttt{$\bm{X}=\bm{VX}$}     
        \Comment{Precoding}
        \EndIf
    \end{algorithmic}
    
    %\vspace{2pt}
    \textbf{Channel model: }
    \texttt{$\bm{Y}=\bm{HX+W}$} \\
    \textbf{For the receiver:} \\
        \textbf{Input:} $\bm{Y}\in \mathbb{C}^{M\times {k}}$, $\bm{H=U\Sigma V^H}$, $\bm{M}$\\
        \textbf{Output:} $\bm{\hat{S}}\in \mathbb{R}^{h\times w\times 3}$\\
    \vspace{-15pt}
    \begin{algorithmic}[1]
    %\State \texttt{$\bm{H=U\Sigma V^H}$}\
    \If{\textit{MIMO with CSIR}}
    \State \texttt{$\bm{{X'}}=\bm{H_wY}+f_{res}(\bm{Y,H})$}
        %\State \texttt{$\bm{{X'}}=\bm{(H^HH)^{-1}H^HY}+f_{res}(\bm{Y},\bm{H})$}\\
        \Comment{DL-aided equalization}    
    %    \Else
    %        \State \texttt{$\bm{{X'}}=\bm{(H^HH)^{-1}H^HY}$}\\
    %        \Comment{ZF-based equalization}    
    %    \EndIf
    \ElsIf{\textit{MIMO with CSIT}}
        \State \texttt{$\bm{X'}=\bm{\Sigma^{\dagger}U^HY}$}\;
        \Comment{SVD-based equalization}
    \EndIf
    \State \texttt{$\bm{X'}\in \mathbb{C}^{M\times k} \rightarrow \bm{X'_s}\in \mathbb{R}^{l\times \frac{2MK}{l}}$}
        \Comment{Reshape}
    %\State \texttt{$\bm{{S_{d}}}={concat}(\bm{X'_s},\bm{M})$}\
    \State \texttt{$\bm{{\hat{S}}}=f_{\bm{\phi}}({concat}(\bm{X'_s},\bm{M}))$}\
        \Comment{ViT decoding}
    \end{algorithmic}
    \label{MIMO_algorithm}
    \end{algorithm}
 
\textbf{Siamese layer and positional embedding:} We design a weight-shared Siamese layer, denoted as $\text{Siam}(\cdot)$, consisting of several linear projection layers and GeLU activation functions. To form the input of the Siamese layer $\bm{S_{d}}$, we concatenate $\bm{X^{'}_{s}}$ and $\bm{M}$ as:
\begin{equation}
    \bm{S_{d}}=\texttt{concat}(\bm{X^{'}_{s}},\bm{M})\in \mathbb{R}^{l\times \frac{4Mk}{l}}.
\end{equation} 
$\bm{S_d}$ multiplied by $1$ and $-1$ are fed into several linear projection layers and GeLU functions, as illustrated in Fig. \ref{ViT_structure}. {In doing so, our networks are tasked with handling the positive and the negative noise realizations through two parallel branches, and subsequently, the resultant features are aggregated by a linear layer to obtain the final output.} We expect that these GeLU functions and linear projection layers can learn to truncate excessive noise realizations {to bootstrap the performance\cite{shao2022attentioncode}}. To introduce the positional information in the decoding process, we use the same positional embedding layer as in the Eqn. \eqref{pos_embed} to get the output $\bm{D_{0}}\in \mathbb{R}^{l\times d}$:
\begin{equation}
\bm{D_{0}}=\text{Siam}(\bm{S_d})+P_e(\bm{p}).
\label{siam_embed}
\end{equation}

\textbf{Transformer layer:}
After the Siamese layer and positional embedding, $\bm{D_{0}}$ is passed through $L_t$ transformer layers, where 
\begin{equation}
\bm{D_{l}}=MSA(\bm{D_{l-1}})+MLP(MSA(\bm{D_{l-1}})),
\end{equation}
where $\bm{D_{i}}\in \mathbb{R}^{l\times d}$ is the output of the $i$-th transformer layer at the decoder, and the $MSA$ and $MLP$ blocks share the same structure as those in Eqn. \eqref{trans_layer_eqn}.

%As show in algorithm \ref{decoder_algorithm}, we concatenate $\bm{z'}$ and $\bm{M}$ to construct the input, our decoder reconstructs the input image as $\bm{\hat{x}}=D_{\phi}(\bm{Y})$.
\textbf{Linear Projection:}
Given the output of the $L_t$-th transformer layer $\bm{D_{L_t}}$, we apply a linear projection $\bm{W_{out}}\in \mathbb{R}^{d\times c}$, and then reshape the output into a matrix of size $\mathbb{R}^{h\times w \times 3}$ to reconstruct the input image as: 
\begin{equation}
    \bm{\hat{S}}=\texttt{reshape}(\bm{D_{L_t}}\bm{W_{out}}).
\end{equation}

\subsection{Loss function}
The encoder and decoder are optimized jointly to minimize the loss function 
\textcolor{black}{
\begin{equation}
\mathcal{L}(\bm{\theta},\bm{\phi})=\text{MSE}(\bm{S},\bm{\hat{S}})+\lambda\cdot \text{LPIPS}(\bm{S},\bm{\hat{S}}),
\end{equation}}
\textcolor{black}{\noindent where we introduced an additional LPIPS term\cite{erdemir2022generative} with $\lambda=0.1$, which is set into $0$ when optimizing the PSNR metric.} We train the model to search for the optimal parameters $(\bm{\theta}^*,\bm{\phi}^* )$ with a minimal $\mathcal{L}(\bm{\theta},\bm{\phi})$ as: $(\bm{\theta}^*,\bm{\phi}^* )=\mathop{\arg\min}_{\bm{\theta},\bm{\phi}}\mathbb{E}\big[\mathcal{L}(\bm{\theta},\bm{\phi})\big]$, where the expectation is taken over the image and channel datasets. 
\begin{figure*}[t]
    \centering
    \subfloat[$R=1/24-PSNR$]{
    \label{open_ViT_24} 
    \begin{minipage}[t]{0.24\linewidth}
    \centering
    \includegraphics[scale=0.25]{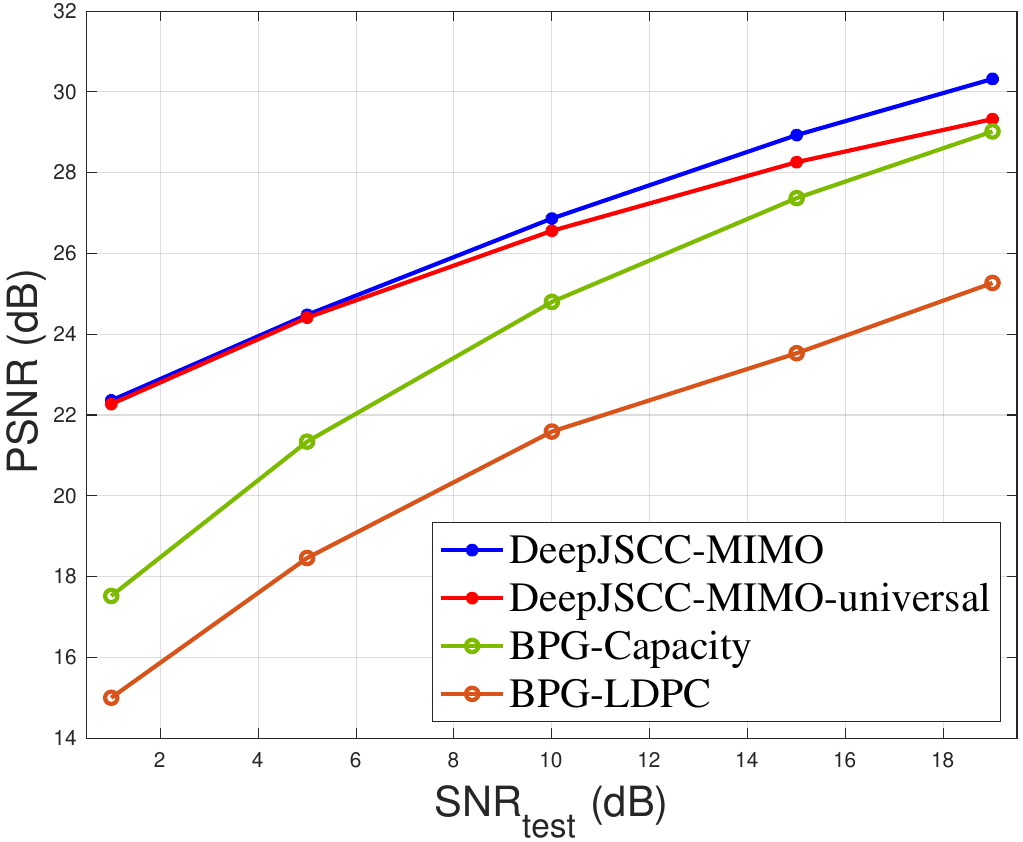}
    \end{minipage}%
    }%
    \subfloat[$R=1/24-LPIPS$]{
    \label{open_ViT_24_lpips} 
    \begin{minipage}[t]{0.24\linewidth}
    \centering
    \includegraphics[scale=0.25]{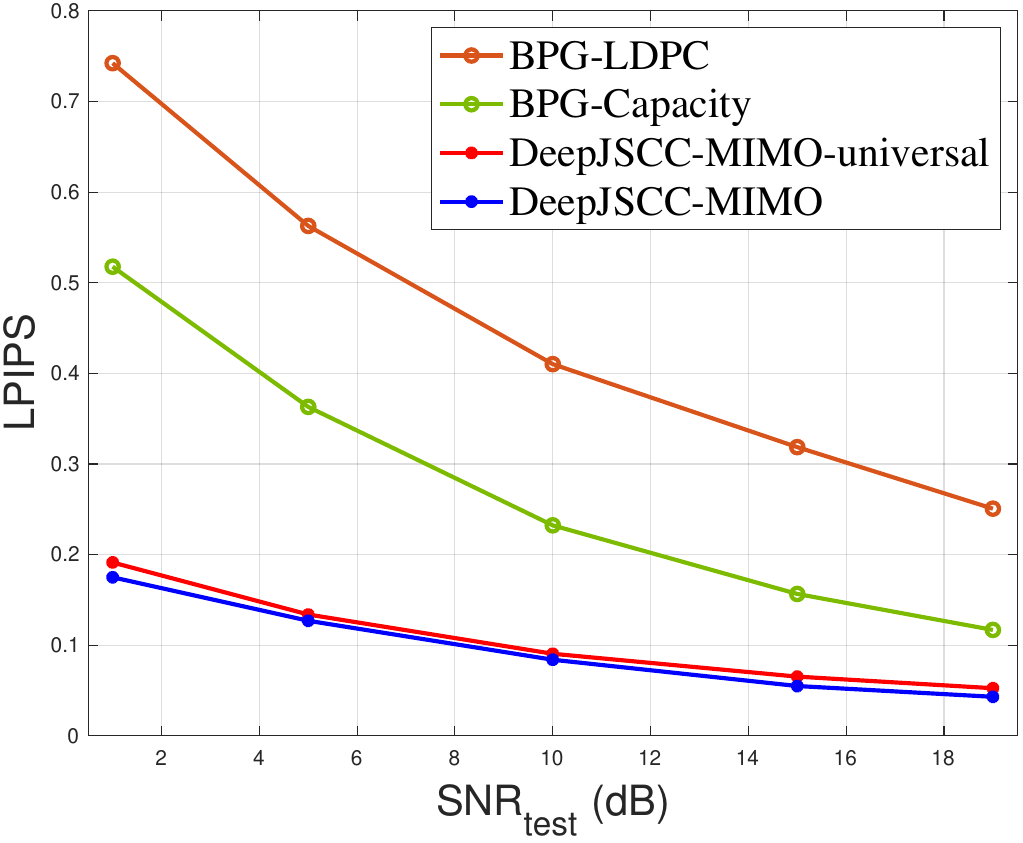}
    %\caption{fig2}
    \end{minipage}%
    }%
    \subfloat[$R=1/12-PSNR$]{
    \label{open_ViT_12} 
    \begin{minipage}[t]{0.24\linewidth}
    \centering
    \includegraphics[scale=0.25]{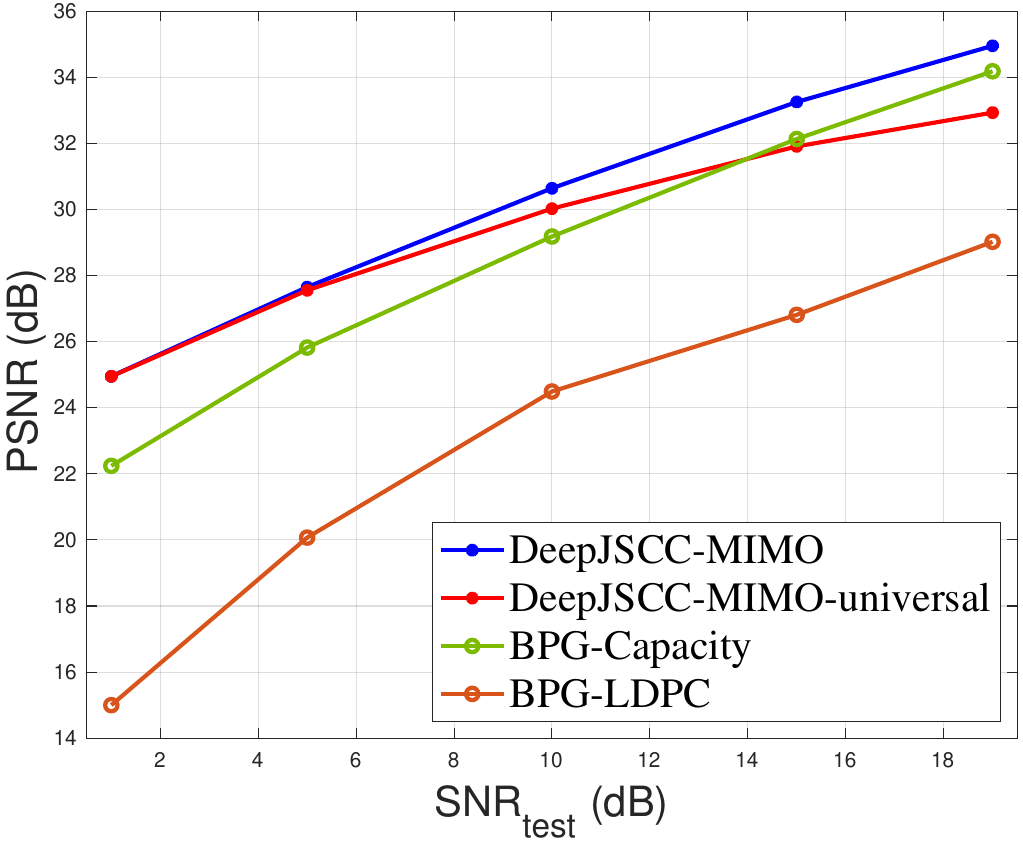}
    \end{minipage}%
    }
    %\hspace{-0.5mm}
    \subfloat[$R=1/12-LPIPS$]{
    \label{open_ViT_12_lpips} 
    \begin{minipage}[t]{0.24\linewidth}
    \centering
    \includegraphics[scale=0.25]{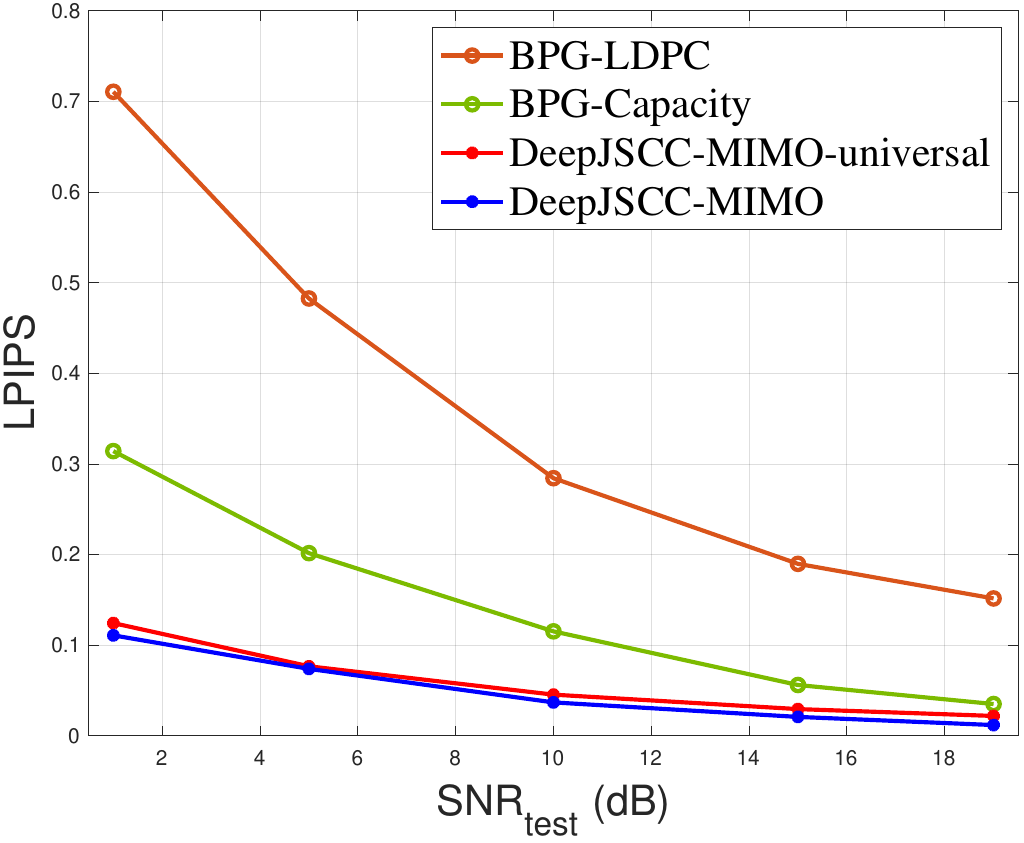}
    %\caption{fig2}
    \end{minipage}%
    }%
    \centering
    \caption{\textcolor{black}{Performance comparisons between the proposed DeepJSCC-MIMO and BPG-Capacity over different SNR values and bandwidth ratios for the MIMO systems with CSIR.}}
    \label{ViT_open_JSCC} 
\end{figure*}

\section{Training and evaluation}
This section conducts a set of numerical experiments to evaluate the performance of our DeepJSCC-MIMO in various bandwidth and SNR scenarios. Unless stated otherwise, we consider a $2\times2$ MIMO system and the CIFAR10 dataset, which has $50000$ images with dimensions $3\times 32\times 32$ (color, height, width) as the training dataset and $10000$ images as the test dataset. We employ both the practical separation-based codes and theoretical bounds (assuming capacity-achieving channel codes) as benchmarks. 
\begin{comment}
\begin{figure*}[t]
  \centering
 \includegraphics[width=0.78\linewidth]{image/vis_open_loop.pdf}
  \caption{Visual comparisons of images from the Kodak dataset transmitted over the open-loop MIMO channel at SNR $5$ dB and $12$ dB with $R=1/24$. The first and second columns are the original image and the original patches of the bounding boxes.}
  \label{vis_open_loop}
\end{figure*}
\end{comment}

\subsection{{Experimental setup}}
For the practical separation-based coding scheme, we use the BPG codec\cite{bpg} as well as the state-of-the-art neural compression algorithms \cite{minnenbt18, vtm,he2022elic, cheng2020image, ballemshj18} as the source coding method. For channel coding, we consider LDPC codes at rates ($1/2,2/3,3/4,5/6$) and various constellation sizes.  \textcolor{black}{A lower LDPC coding rate can yield improved channel coding results but at the cost of a reduced compression rate with notable performance reduction, especially in very low coding rate regimes and short block length scenarios.} In particular, we adopt WiFi (IEEE 802.11n) LDPC code construction, featuring block lengths of $648$, $1296$, and $1944$ bits and 4-QAM, 16-QAM, and 64-QAM constellations. For each channel condition \textcolor{black}{and scenario}, an exhaustive search \textcolor{black}{for all samples} is conducted across all combinations of channel coding rates and constellation orders, each associated with its respective compression rates. The best performances are subsequently identified and employed as the benchmark for the given channel condition \textcolor{black}{and scenario}. \textcolor{black}{Although exploring a larger set of code rate and constellation combinations could potentially enhance performance, the improvement is not expected to be significant and the overall performance is inherently constrained by the capacity-achieving separation-based scheme, as elaborated subsequently.}

For the theoretical bound, we assume capacity-achieving channel codes for transmission, denoted as the BPG-Capacity scheme. For the MIMO system with CSIT, we apply the water-filling algorithm for power allocation  among the parallel channels formed after Eqn. 8. \textcolor{black}{For the open-looop MIMO system with CSIR, we assume that the transmitter knows the instantaneous capacity of each channel block without power allocation, and can select the best compression rate. Note that, in practice, the transmitter does not have access to the CSI, and would probably employ a conservative compression rate to avoid a potential outage.}

% We transmit each image $5$ times over the channel model. 
Our model was implemented in Pytorch with two GTX 3090Ti GPUs. We use a learning rate of $5e^{-5}$ and a batch size of $128$ with an Adam optimizer. Models were trained until the performance on a validation set stopped improving. Considering the model complexity and the performance, we set $p=8$, $l=64$, $c=48$ for image vectorization, and $L_t=8$, $d=256$, and $N_s=8$ for each transformer layer of the ViT. 
\begin{figure*}[t]
  \centering
 \includegraphics[width=0.95\linewidth]{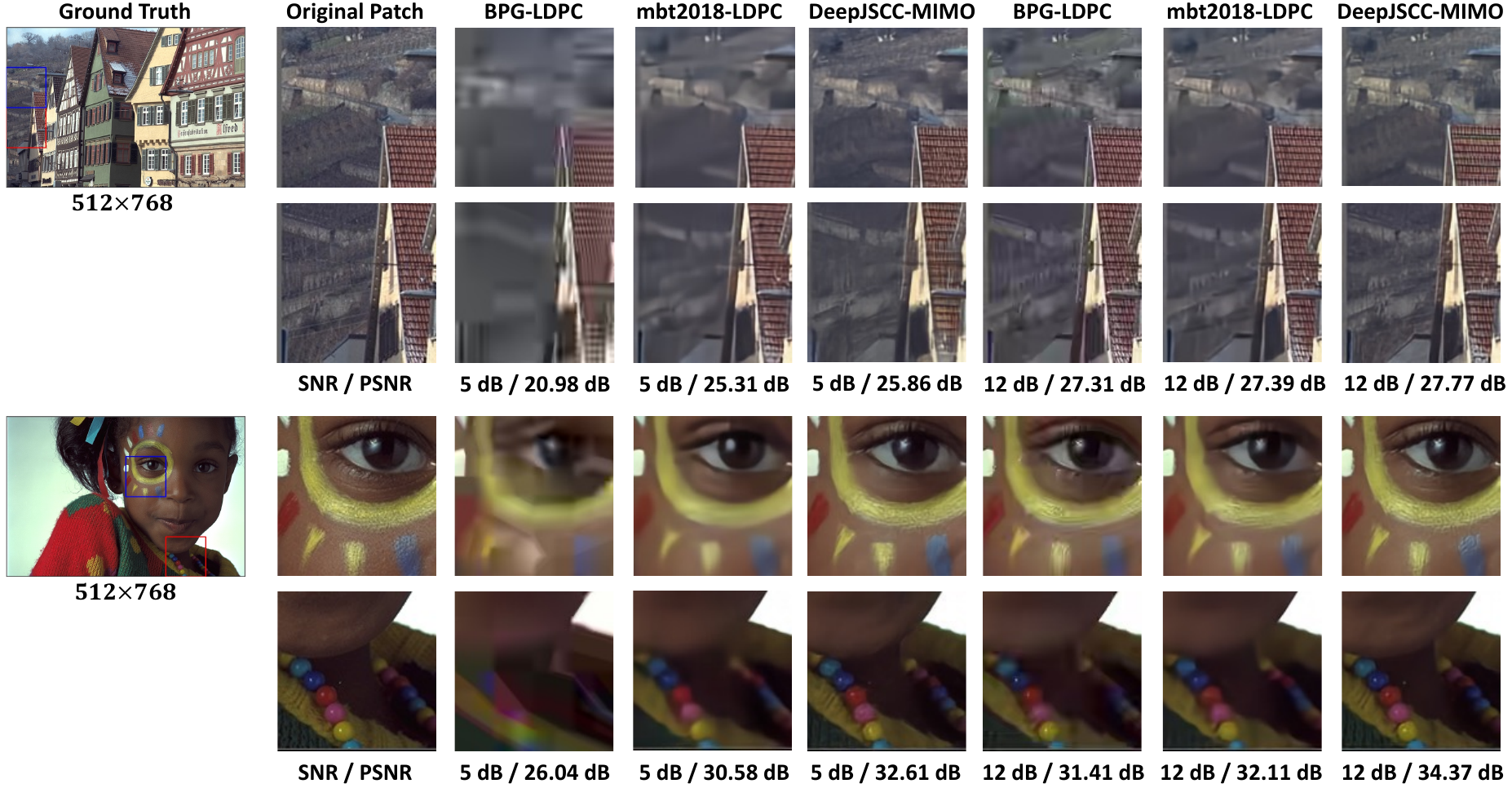}
  \caption{\textcolor{black}{Visual comparisons of images from the Kodak dataset transmitted over the open-loop MIMO channel at SNR $5$ dB and $12$ dB with $R=1/24$. The first and second columns are the original image and the original patches of the bounding boxes.}}
  \label{vis_open_loop}
\end{figure*}

Each element of $\bm{H}$ is sampled from a complex Gaussian process as $H[i,j]\sim \mathcal{CN}(0,1)$, where we set $\sigma_h^2=1$. \textcolor{black}{Results for a more general correlated MIMO channel are presented in \cite{wu2023deep_arxiv}}. To measure the channel quality, we define the channel SNR, denoted by $\mu$, as:
\begin{equation}
    %\mu=10\log_{10} \frac{\mathbb{E}[\bm{\|HX\|}^2_F]}{M\mathbb{E}[\bm{W}_F^2]},
    \mu\triangleq 10\log_{10} \frac{\mathbb{E}_{\bm{H,X}}[\bm{\|HX\|}^2_F]}{\mathbb{E}_{\bm{W}}[\bm{\|W\|}^2_F]}~(\text{dB})=10\log_{10} \frac{M}{\sigma_w^2}.
\end{equation}
\begin{comment}
 \begin{figure}[t]
     \centering
    \includegraphics[scale=0.4]{image/open_loop_rate_1_12_detection.png}
     \caption{Comparisons with performance of different MIMO equalization methods for $R=1/12$}
     \label{exp_MIMO_dl_detection}
 \end{figure}
 \end{comment}
\subsection{Open-Loop MIMO system with CSIR}
We first evaluate DeepJSCC-MIMO in the open-loop MIMO system with CSIR. Specifically, we compare DeepJSCC-MIMO, trained at specific SNRs, with the BPG-Capacity and BPG-LDPC schemes at different SNR values and bandwidth ratios. For the BPG-LDPC scheme, we apply the sphere decoding (SD) algorithm \cite{guo2006algorithm}.

%, where our DeepJSCC-MIMO model is trained at specific SNRs, that is $\textit{SNR}_{\textit{train}}=\textit{SNR}_{\textit{test}}$.

\subsubsection{General performance}
Fig. \ref{open_ViT_24} and Fig. \ref{open_ViT_12} show the PSNR performance of the DeepJSCC-MIMO, BPG-Capacity, and BPG-LDPC schemes over the SNR values from $0$ dB to $20$ dB when the bandwidth ratio is $R=1/24$ and $R=1/12$. We can observe that our DeepJSCC-MIMO can generally outperform the BPG-Capacity and BPG-LDPC schemes in all SNR values. We emphasize that the BPG-Capacity performance shown in these figures is not achievable by practical channel coding schemes. {Thus, it serves as an upper bound on any separation-based scheme that employs BPG for compression of the input image.} We observe that DeepJSCC-MIMO provides significant improvements (at least $1.3$ dB and $5.05$ dB) for $R=1/24$, compared with the BPG-Capacity scheme and the BPG-LDPC scheme, respectively. For $R=1/12$, our DeepJSCC-MIMO can still outperform the BPG-Capacity scheme (at least $0.77$ dB) and the BPG-LDPC scheme (at least $5.93$ dB) in all SNRs. We observe that the gap between DeepJSCC-MIMO and BPG capacity is larger in the lower bandwidth ratio case.

\textcolor{black}{We observe in Figs. \ref{open_ViT_24_lpips} and \ref{open_ViT_12_lpips} that DeepJSCC-MIMO achieves the best perceptual performance across a range of channel conditions and bandwidth ratios. Especially in poor SNR conditions, DeepJSCC-MIMO exhibits a substantial improvement over BPG-LDPC and BPG-Capacity, with a maximum LPIPS discrepancy of $0.59$ and $0.35$ at SNR $1$dB for CIFAR10, respectively.} %A similar trend is observed with the JSCCformer-f (lite), which preserves adaptability to channel variations but suffers a slight performance degradation relative to the original JSCCformer-f model.
\begin{table}[t]
\centering
\caption{\textcolor{black}{Comparison of DeepJSCC-MIMO with separation-based baselines employing different compression algorithms, LDPC code and sphere decoding for the open-loop MIMO system on the Kodak dataset when $SNR=5$ dB.}}
\begin{tabular}{c|c|c|c|c|c}
\textbf{Bandwidth ratio}&{\textbf{$1/24$}} &\textbf{$1/12$}&\textbf{$1/8$} &\textbf{$1/6$}&\textbf{$1/4$}  \\
\hline
{{BPG-LDPC}\cite{bpg}}  &${24.05}$&${25.50}$&${27.02}$ &${28.21}$&${29.70}$\\

{{MBT2018-LDPC\cite{minnenbt18}}} &${24.48}$& ${26.03}$ & ${27.64}$&${28.91}$&${30.39}$\\

{{VTM-LDPC}\cite{vtm}} &${24.81}$& ${26.31}$ & ${27.91}$&${29.15}$&${30.68}$\\

{{ELIC-LDPC}\cite{he2022elic}} &${24.99}$& ${26.54}$ & ${28.16}$&${29.35}$&${30.85}$\\

{{DeepJSCC-MIMO}} &$\bm{29.43}$& $\bm{30.45}$ & $\bm{31.17}$&$\bm{31.56}$&$\bm{31.92}$\\
\end{tabular}
\label{table_dl_open}
\end{table}

\begin{comment}
 \begin{figure}
    \centering
    \includegraphics[scale=0.38]{image/open_loop_ratio_img_snr_5.pdf}
    \caption{Performance comparison of DeepJSCC-MIMO and BPG-LDPC on the Kodak dataset for different bandwidth ratios at SNR=$5$ dB.}
    \label{Vitf_open_ratio}
\end{figure}
\end{comment}

%% (256,192) to (256,tcn)
\begin{comment}
\begin{table}[t]
    \centering
    \caption{\textcolor{black}{Ablation study of different MIMO equalization methods for DeepJSCC-MIMO-universal with $R=1/12$.}}
    \label{table:exp_MIMO_dl_detection}
    \begin{tabular}{c|c|c|c|c|c}
    \textbf{Equalizations} & {1 dB}& {5 dB}& {10 dB}& {15 dB}& {19 dB}\\
    \hline
     {End-to-end w/o ZF}  & 22.65&24.4& 26.16&26.83&27.03\\
     Pure ZF & 24.47&26.98& 29.41&31.43&32.73 \\
     Pure MMSE & 24.61&27.04& 29.48&31.50&32.74\\
     {DL-aided ZF} &$\bm{24.95}$&$\bm{27.56}$&$\bm{30.02}$&$\bm{31.91}$&$\bm{32.93}$ \\    
    \end{tabular}
\end{table}
\end{comment}
\subsubsection{SNR-adaptability}
Note that each point on the DeepJSCC-MIMO curve is obtained by training a separate encoder-decoder pair. We also consider a random SNR training strategy to evaluate the SNR adaptability of the DeepJSCC-MIMO scheme, denoted by DeepJSCC-MIMO-universal. We train the model with random SNR values uniformly sampled from $[0,22]$ dB and test the well-trained model at different SNRs. The comparisons in Fig. \ref{ViT_open_JSCC} show that there is a slight performance degradation compared to training a separate DeepJSCC-MIMO encoder-decoder pair for each SNR (up to $0.99$ dB for $R=1/24$ and $2.02$ dB for $R=1/12$); however, the DeepJSCC-MIMO-universal brings significant advantage in terms of training complexity and storage requirements. We can conclude that the DeepJSCC-MIMO-universal scheme learns to adapt to different SNRs from the random training strategy at the expense of a slight loss in PSNR, which tends to increase with SNR.

%\subsubsection{Ablation experiments over the equalization methods}
%To evaluate the effectiveness of the DL-aided equalization method with the residual block, we repeat the experiments of DeepJSCC-MIMO-universal in Fig. \ref{open_ViT_12} with an MMSE receiver. The experimental results in Table \ref{table:exp_MIMO_dl_detection} show that the residual block improves the performance (up to $0.54$ dB). More interestingly, we observe that the ZF combined with the residual block achieves the best performance, which shows that the residual block helps to compensate for the limitation of ZF and outperforms MMSE.

%. It is because the ZF method converts the MIMO channel into parallel channels, which matches our model's design and allows our model to compute the attention across the parallel channel outputs and the related channel conditions. 

%can improve more in a lower bandwidth ratio, compared with the BPG-Capacity scheme, which shows the advantage of our DeepJSCC-MIMO scheme in the low bandwidth regime.

\begin{figure*}[t]
  \centering
  \begin{tabular}{cccccc}
    Ground Truth & Original Patch& $\sigma_e^2=0$&$\sigma_e^2=0.1$&$\sigma_e^2=0.2$&$\sigma_e^2=1$\\
    \begin{minipage}[b]{0.4\columnwidth}
		\centering
		\raisebox{-.5\height}{\includegraphics[width=\linewidth]{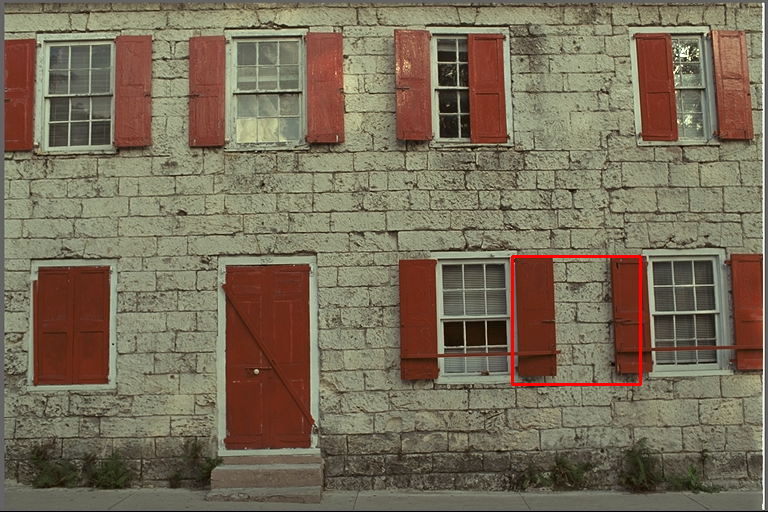}}
	\end{minipage}
	\hspace{-3pt}
    & 
    \begin{minipage}[b]{0.265\columnwidth}
		\centering
		\raisebox{-.5\height}{\includegraphics[width=\linewidth]{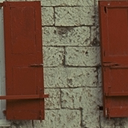}}
	\end{minipage}
	\hspace{-8pt}
	&
        \begin{minipage}[b]{0.265\columnwidth}
		\centering
		\raisebox{-.5\height}{\includegraphics[width=\linewidth]{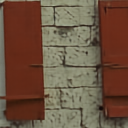}}
	\end{minipage}
	\hspace{-8pt}
    & 
        \begin{minipage}[b]{0.265\columnwidth}
		\centering
		\raisebox{-.5\height}{\includegraphics[width=\linewidth]{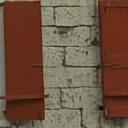}}
	\end{minipage}
	\hspace{-8pt}
	&
	    \begin{minipage}[b]{0.265\columnwidth}
		\centering
		\raisebox{-.5\height}{\includegraphics[width=\linewidth]{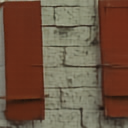}}
	\end{minipage}	
	\hspace{-8pt}
		&
	    \begin{minipage}[b]{0.265\columnwidth}
		\centering
		\raisebox{-.5\height}{\includegraphics[width=\linewidth]{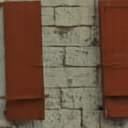}}
	\end{minipage}
    \\ 
    $512\times 768$ & PSNR (dB) & $29.44$ &$28.22$&$27.78$&$26.84$
    \\
            \begin{minipage}[b]{0.4\columnwidth}
		\centering
		\raisebox{-.5\height}{\includegraphics[width=\linewidth]{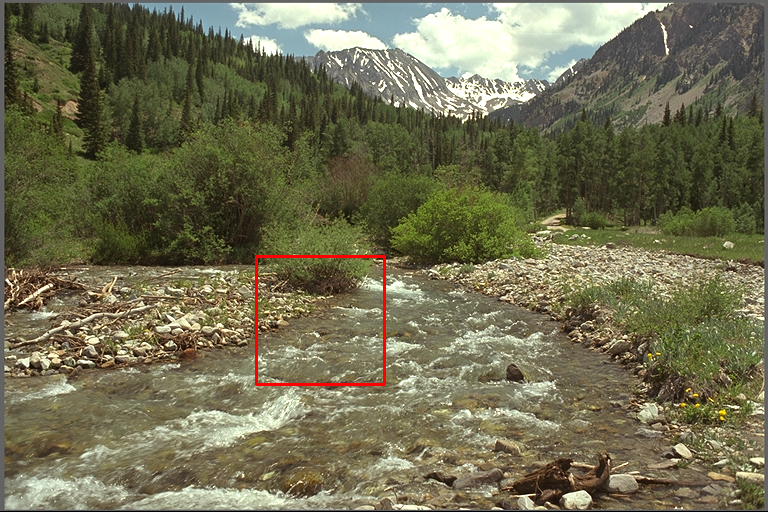}}
	\end{minipage}
	\hspace{-3pt}
    & 
    \begin{minipage}[b]{0.265\columnwidth}
		\centering
		\raisebox{-.5\height}{\includegraphics[width=\linewidth]{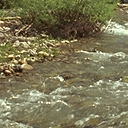}}
	\end{minipage}
	\hspace{-8pt}
	&
        \begin{minipage}[b]{0.265\columnwidth}
		\centering
		\raisebox{-.5\height}{\includegraphics[width=\linewidth]{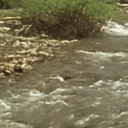}}
	\end{minipage}
	\hspace{-8pt}
    & 
        \begin{minipage}[b]{0.265\columnwidth}
		\centering
		\raisebox{-.5\height}{\includegraphics[width=\linewidth]{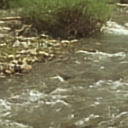}}
	\end{minipage}
	\hspace{-8pt}
	&
	    \begin{minipage}[b]{0.265\columnwidth}
		\centering
		\raisebox{-.5\height}{\includegraphics[width=\linewidth]{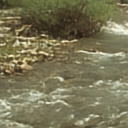}}
	\end{minipage}	
	\hspace{-8pt}
		&
	    \begin{minipage}[b]{0.265\columnwidth}
		\centering
		\raisebox{-.5\height}{\includegraphics[width=\linewidth]{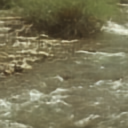}}
	\end{minipage}
    \\ 
    $512\times 768$ & PSNR (dB) & $26.46$ &$25.44$& $25.14$ & $24.35$\\
  \end{tabular}
  \caption{Visual comparisons of images reconstructed by  DeepJSCC-MIMO-universal in the presence of channel estimation errors, where the model is trained on the ImageNet dataset and validated on the Kodak dataset at SNR=$12$ dB and $R=1/24$. The first and second columns are the source image and the image patch from the red bounding box of the original image. The third to sixth columns are the reconstructions of the same patches by DeepJSCC-MIMO universal with different channel estimation errors, respectively. The average PSNR metrics of the entire image are provided at the bottom of each visualized patch.}
  \label{vis_estimation_error}
\end{figure*}

\subsubsection{High resolution datasets and different bandwidth ratios}
\textcolor{black}{To further study the effect of bandwidth ratio and channel SNR on the performance, and evaluate the model's generalizability, we train and test the DeepJSCC-MIMO scheme with various bandwidth ratios and SNRs individually over the Kodak dataset. Specifically, we train the models with randomly cropped $128 \times 128$ patches from the ImageNet dataset, and evaluate the well trained models on the Kodak dataset.}

In Fig. \ref{vis_open_loop}, we present visualization examples of images transmitted by DeepJSCC-MIMO and BPG-LDPC schemes at $5$ dB and $12$ dB when $R=1/24$, {where SNR values and PSNR of the entire image are provided at the bottom of each visualized patch.} It can be observed that reconstructed by DeepJSCC-MIMO are qualitatively better with more detailed high-frequency features. The comparisons here can clearly show the advantages of the DeepJSCC-MIMO in high-resolution datasets. We also conclude that training on a sufficiently large dataset (ImageNet) can allow our DeepJSCC-MIMO scheme to perform well on a never-seen dataset (Kodak) for a wide range of bandwidth ratios.  

We also evaluate the model \textcolor{black}{across various bandwidth ratios and compression algorithms\cite{minnenbt18, vtm,he2022elic, cheng2020image, ballemshj18}} when SNR=$5$ dB over the Kodak dataset, as shown in Table. \ref{table_dl_open}. We want to emphasize that the sphere decoding algorithm in \textcolor{black}{separation-based schemes} is a much better MIMO detection method with higher computation complexity, compared with ZF employed by DeepJSCC-MIMO. \textcolor{black}{We can observe that the superiority of DeepJSCC-MIMO persists across various bandwidth ratios and compression algorithms}, where the gap is even larger at low bandwidth ratios, which shows that DeepJSCC-MIMO outperforms \textcolor{black}{traditional schemes} despite using a weaker detection algorithm.

\subsubsection{Robustness to channel estimation errors}
We take the same approach as in \cite{yoo2006capacity} to evaluate the robustness of MIMO systems to channel estimation errors, where $\bm{H}$ is imperfectly estimated as $\bm{\hat{H}}$ at the receiver. Similar to the work in\cite{hassibi2003much,1614084,yoo2006capacity}, let $\bm{E}_n\triangleq \bm{H}-\bm{\hat{H}}$, where the entries of $\bm{E}_n$ are zero-mean complex Gaussian with variance $\sigma^2_e$. Here $\sigma^2_e$ is the `noise variance' term to capture the channel estimation quality, which can be appropriately selected depending on the channel conditions and channel estimation method.
 \begin{table}[t]
\centering
\caption{\textcolor{black}{Comparisons of DeepJSCC-MIMO-universal and BPG-LDPC schemes with sphere decoding operations under channel estimation errors over the Kodak dataset for $R=1/24$}}
\begin{tabular}{c|c|c|c|c}
{Different $\text{SNR}_{\text{test}}$ \text{(dB)}}&\textbf{$5$}&\textbf{$7$} &\textbf{$9$}&\textbf{$12$}  \\
\hline
{{DeepJSCC-MIMO ($\sigma_e^2=0$)}} & $\bm{29.38}$ & $\bm{30.08}$&$\bm{30.80}$&$\bm{31.62}$\\

{{DeepJSCC-MIMO ($\sigma_e^2=0.2$)}} &${29.18}$ & ${29.86}$&${30.53}$&${31.33}$\\

{{DeepJSCC-MIMO ($\sigma_e^2=1$)}} & ${28.73}$ & ${29.49}$&${30.08}$&${30.78}$\\

{{BPG-LDPC ($\sigma_e^2=0$)}}  &${24.05}$&${27.23}$ &${28.92}$&${30.27}$\\
{{BPG-LDPC ($\sigma_e^2=0.2$)}} &${18.65}$&${21.27}$ &${23.21}$&${24.84}$\\
{{BPG-LDPC ($\sigma_e^2=1$)}}&${15.01}$&${15.37}$ &${15.72}$&${16.03}$\\

\end{tabular}
     \label{MIMO_estimation_error}
\end{table}
\begin{figure*}[t]
    \centering
    \subfloat[$R=1/24-PSNR$]{
    \label{close_ViT_1_12} 
    \begin{minipage}[t]{0.3\linewidth}
    \centering
    \includegraphics[scale=0.3]{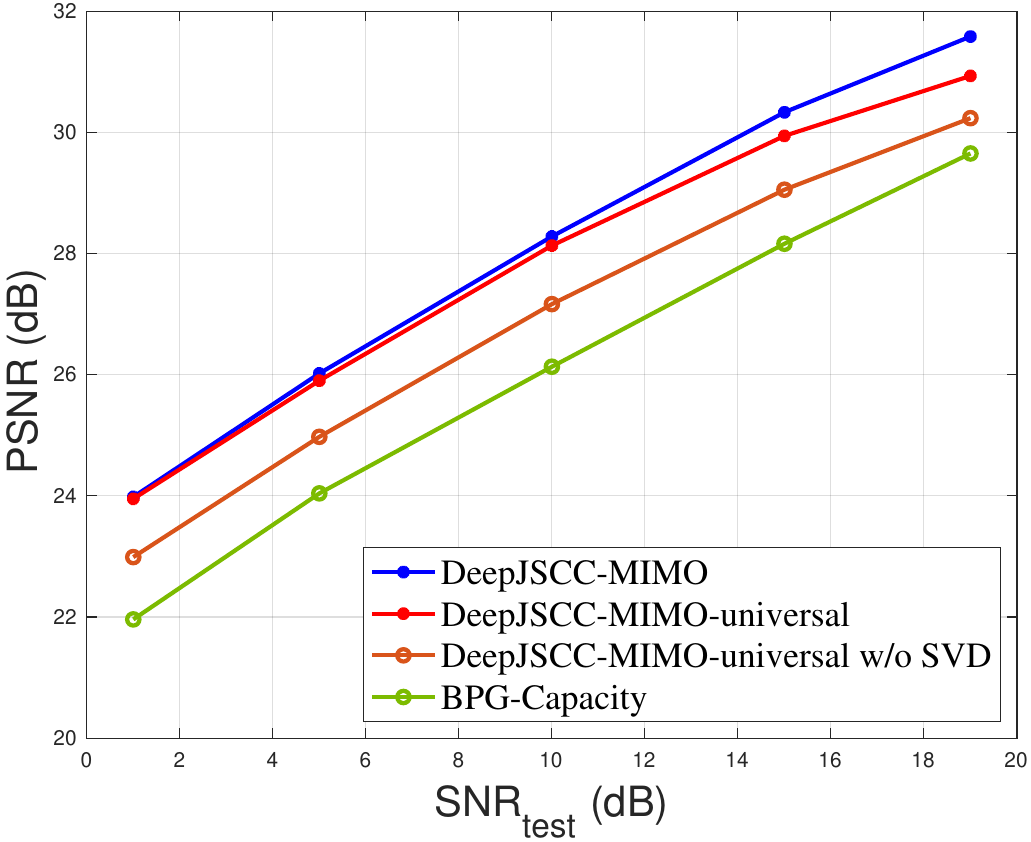}
    \end{minipage}%
    }%
    \hspace{-0.5mm}
    \subfloat[$R=1/12-PSNR$]{
    \label{close_ViT_1_6} 
    \begin{minipage}[t]{0.3\linewidth}
    \centering
    \includegraphics[scale=0.3]{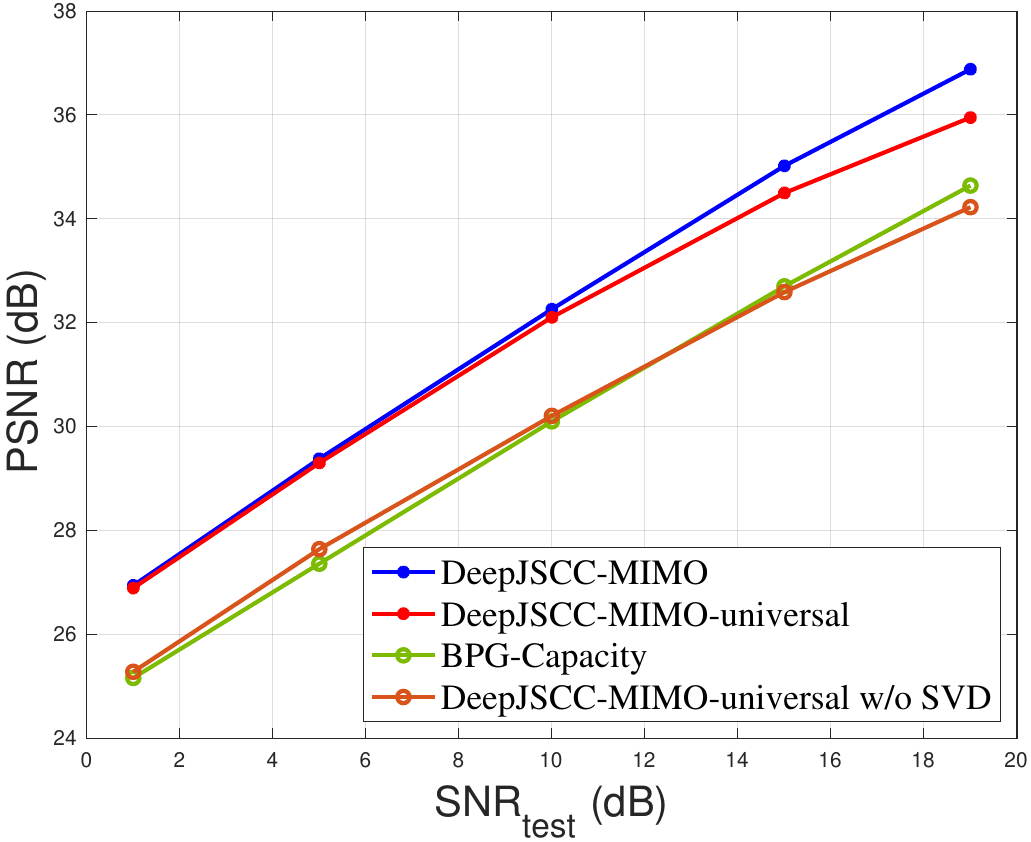}
    %\caption{fig2}
    \end{minipage}%
    }%
    \hspace{-0.5mm}
    \subfloat[$R=1/6-PSNR$]{
    \label{close_ViT_1_3} 
    \begin{minipage}[t]{0.3\linewidth}
    \centering
    \includegraphics[scale=0.3]{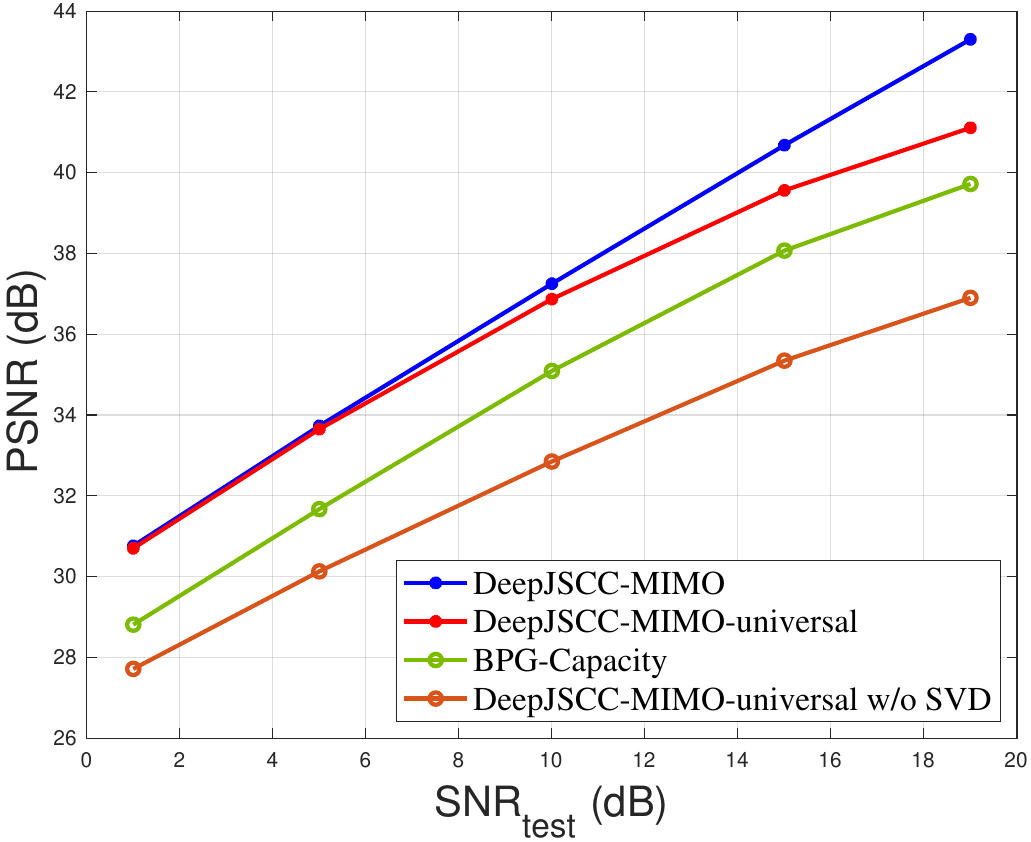}
    %\caption{fig2}
    \end{minipage}%
    }\\
    \subfloat[$R=1/24-LPIPS$]{
    \label{close_ViT_1_24_lpips} 
    \begin{minipage}[t]{0.3\linewidth}
    \centering
    \includegraphics[scale=0.3]{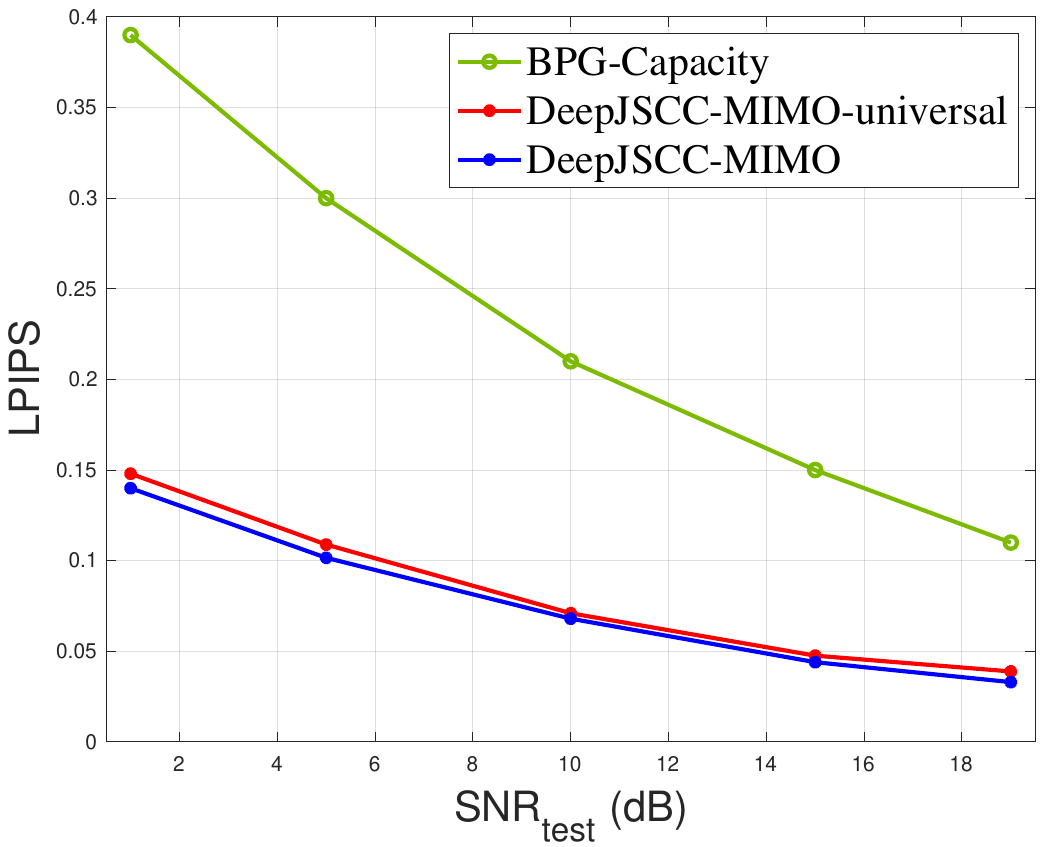}
    \end{minipage}%
    }%
    \hspace{-0.5mm}
    \subfloat[$R=1/12-LPIPS$]{
    \label{close_ViT_1_12_lpips} 
    \begin{minipage}[t]{0.3\linewidth}
    \centering
    \includegraphics[scale=0.3]{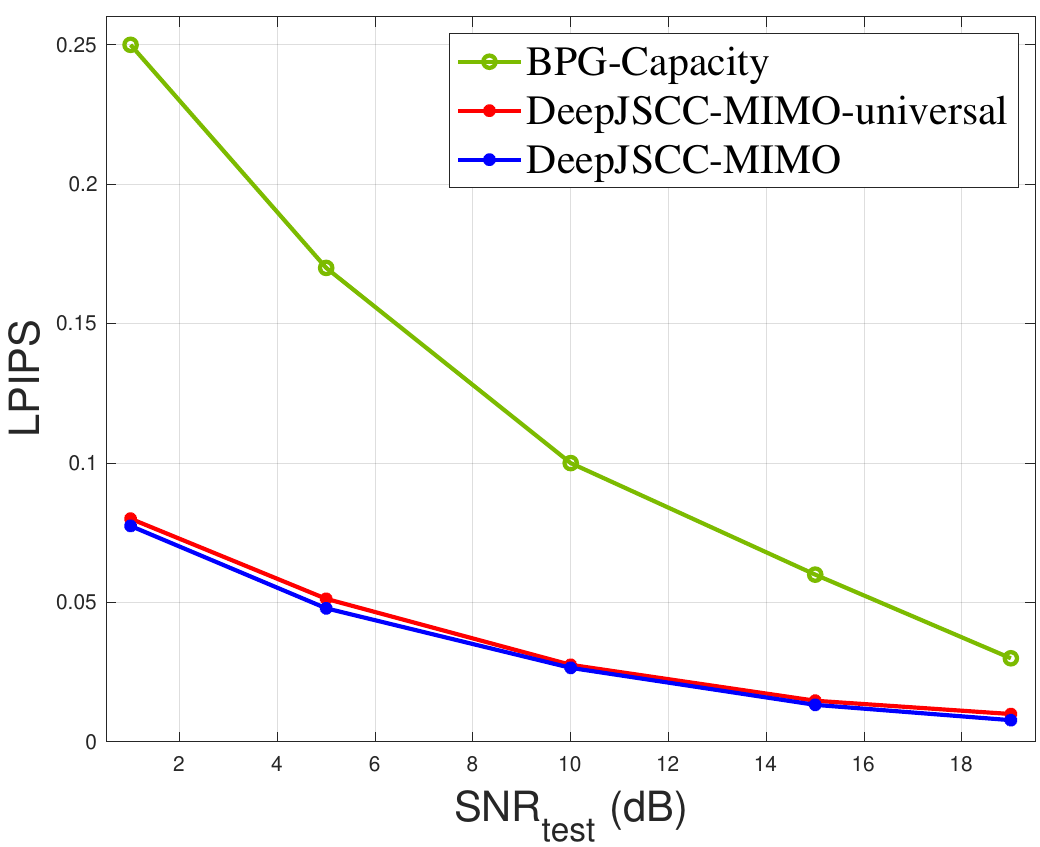}
    %\caption{fig2}
    \end{minipage}%
    }%
    \hspace{-0.5mm}
    \subfloat[$R=1/6-LPIPS$]{
    \label{close_ViT_1_6_lpips} 
    \begin{minipage}[t]{0.3\linewidth}
    \centering
    \includegraphics[scale=0.3]{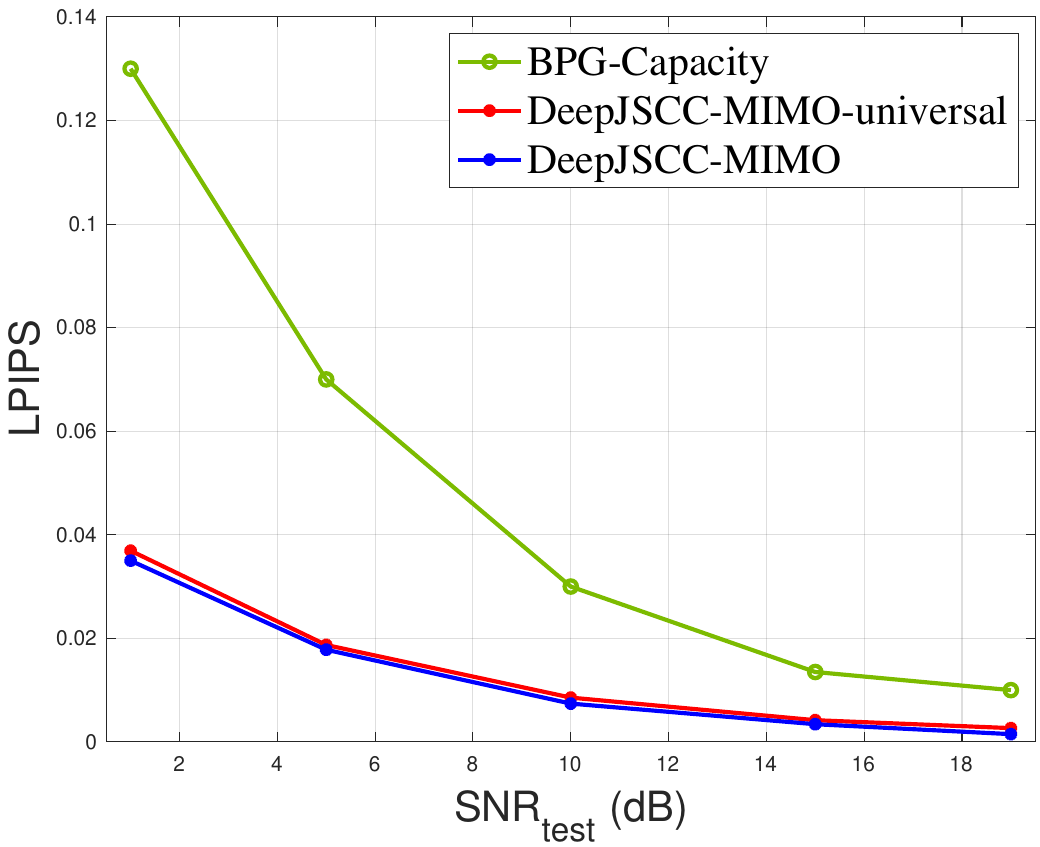}
    %\caption{fig2}
    \end{minipage}%
    }%
    \centering
    \caption{\textcolor{black}{Performance of the proposed DeepJSCC-MIMO model compared with the BPG-Capacity benchmark at different channel SNR and bandwidth ratio scenarios for the closed-loop MIMO.}}
    \label{ViT_close_JSCC} 
\end{figure*} 
We evaluate the DeepJSCC-MIMO-universal and BPG-LDPC schemes with different $\sigma^2_e\in \{0,0.2,1\}$ values over different SNR values on the Kodak dataset with $R=1/24$. For the DeepJSCC-MIMO-universal scheme, we train and test the model both with imperfect $\bm{\hat{H}}$. The experimental results are shown in Table. \ref{MIMO_estimation_error}. As expected, for both schemes, the performance degrades as the channel estimation error increases. 
%the model trained and tested with perfect channel information, i.e., $\sigma^2_e=0$, achieves the best performance. Generally, for both schemes, the performance of models get worse as the error of channel estimation increase. We conclude from these results that a more accurate channel estimation is generally beneficial. 
In specific, for $\sigma^2_e=0.2$, the performance loss of our DeepJSCC-MIMO scheme is up to $0.29$ dB, compared with the performance loss up to $5.96$ dB from the BPG-LDPC scheme. For $\sigma^2_e=1$, the BPG-LDPC scheme starts to fail to work. However, our DeepJSCC-MIMO scheme can still work despite experiencing a limited performance decrease of less than $0.84$ dB, \textcolor{black}{which mainly arises from noisy channel equalization operations, as demonstrated in \cite{wu2023deep_arxiv}.} 

To visualize the performance of DeepJSCC-MIMO-universal scheme in the presence of channel estimation errors, we present the reconstructed images with different channel estimation errors in Fig. \ref{vis_estimation_error}. We can observe that the increasing channel estimation error is reflected in the gradual loss of high-frequency details during transmission. {We want to emphasize that even with a significant channel estimation error ($\sigma_e^2=1$), DeepJSCC-MIMO continues to function, with only limited performance loss}, compared with the significant performance loss with the traditional separation benchmark. We can conclude that the proposed scheme is more robust to channel estimation errors, as it can learn to compensate for it during transmission.
%channel estimation error and the channel adaptability at the same time. 

\subsubsection{\textcolor{black}{Model efficiency and other architectures}}
\textcolor{black}{The proposed DeepJSCC-MIMO scheme is versatile and can be seamlessly integrated with various network architectures.} Table \ref{table_para} presents the model size, floating point operations (FLOPs), \textcolor{black}{PSNR performance, and computational complexity} for alternative DeepJSCC architectures with different \textcolor{black}{block} numbers when transmitting a single sample from the CIFAR10 dataset \textcolor{black}{under $R=1/24$ and SNR=$5$ dB}. \textcolor{black}{$L_t$ represents the number of basic blocks in different DeepJSCC architectures. Specially, each CNN block\cite{wu2022channel,9714510} is characterized by kernel size $k$, feature size $F$, along with input / output channel numbers $C_{in}$ / $C_{out}$. On the other hand, each Swin block within Swin-based DeepJSCC\cite{9791398,yao2022versatile} is characterized by the window size $M_w$, sequence length $l$, and hidden dimension $d$.} We can observe \textcolor{black}{from our experiments on CIFAR10 dataset} that the proposed method maintains the lowest parameter and FLOP counts \textcolor{black}{when achieving the best transmission performance}, reducing the memory consumption and FLOP operations by up to $44.74\%$ and $61.4\%$, respectively. 

 \begin{table}[t]
\caption{\textcolor{black}{The number of parameters (millions), FLOPs (G), PSNR performance (dB), and encoding / decoding time (ms) for different backbones across various basic block numbers $L_t$.}} %Bold figures correspond to the minimal values among different models for each $L_t$ value.}}
\begin{center}
%\resizebox{\textwidth}{!}{
\begin{tabular}{@{\extracolsep{\fill}}|c|c|ccc|}
\hline 
\multicolumn{2}{|c|}{\textbf{Number of basic blocks $L_t$}}&4&8&12\\
\hline 
\multirow{3}{*}{{Parameters}}
&\text{Proposed method}&\textbf{6.61}&\textbf{12.92}&\textbf{19.23}\\
&\text{DeepJSCC-CNN}& 6.89& 20.00&33.11\\
&\text{DeepJSCC-Swin}& 8.89& 15.21&21.52\\
\hline 
\multirow{3}{*}{{FLOPs}}
&\text{Proposed method}&\textbf{0.43}&\textbf{0.83}&\textbf{1.23}\\
&\text{DeepJSCC-CNN}&0.83& 1.67& 2.51\\
 &\text{DeepJSCC-Swin}& 1.22& 2.23&3.24\\
\hline 
\multirow{3}{*}{{Performance}}
 &\text{Proposed method}&\textbf{23.76}&\textbf{24.41}&\textbf{24.32}\\
&\text{DeepJSCC-CNN}&22.83 & 23.54 & 23.04\\ 
&\text{DeepJSCC-Swin}& 22.67& 23.61 & 23.22\\
\hline
\multirow{3}{*}{{Complexity}}
 &\text{Proposed method}&\multicolumn{3}{c|}{$\mathcal{O}(4ld^2+2l^2d)$}\\
&\text{DeepJSCC-CNN}&\multicolumn{3}{c|}{$\mathcal{O}(k^2F^2C_{in}C_{out})$}\\ 
&\text{DeepJSCC-Swin}&\multicolumn{3}{c|}{$\mathcal{O}(4ld^2+2M_w^2ld)$}\\
\hline
\hline 
{{Encoding time} /}
&\multicolumn{1}{c}{\text{BPG-LDPC (CPU)}}&\multicolumn{3}{|c|}{37.25 / 71.28}\\
{decoding time } &\multicolumn{1}{c}{\text{Proposed method (GPU)}}&\multicolumn{3}{|c|}{\textbf{4.82} / \textbf{{5.16}}}\\
\hline 
\end{tabular}
%\footnotesize{Results are calculated through Thop library.}
\label{table_para}
\end{center}
\end{table}

\textcolor{black}{Considering computational complexity, the proposed DeepJSCC method and CNN-based DeepJSCC methods exhibit quadratic computation complexity concerning the sequence length or input size. This underscores the necessity for meticulous design of patch sizes, particularly when dealing with high-resolution images such as 2K images, or alternatively, Swin block can be employed as an optional solution.} It is also noteworthy that the inherently parallelizable architecture of our approach renders it particularly efficient for deployment on GPU-like systems, as opposed to conventional methods typically tailored for optimization on CPUs. The DeepJSCC-MIMO scheme implemented on a GPU is significantly faster compared to the traditional scheme executed on a CPU with the iterative decoding algorithm. 

\textcolor{black}{In summary, the proposed ViT-based solution significantly reduces both the inference complexity and parameter count. Furthermore, it is feasible to retain the total coding time of DeepJSCC-MIMO within $10$ ms, which can be further reduced by encompassing alternative hardware, \textcolor{black}{different architectures, optimized implementations}, and acceleration techniques \cite{wu2023deep_arxiv}.}

\subsection{Closed-Loop MIMO system with CSIT}
In this section, we consider the closed-loop MIMO system with CSI at the transmitter as well as at the receiver. We evaluate the DeepJSCC-MIMO trained at specific SNRs with $\textit{SNR}_{\textit{test}}\in [0,20]$ dB in three bandwidth ratios $R=1/24$, $R=1/12$ and $R=1/6$. 

\subsubsection{General performance}
As before, we first evaluate the DeepJSCC-MIMO model under the setup where the training SNR, denoted by $\textit{SNR}_{\textit{train}}$ matches the test SNR, $\textit{SNR}_{\textit{test}}$. In particular, we set $\textit{SNR}_{\textit{train}}=\textit{SNR}_{\textit{test}}\in \{1,5,10,15,19\}$ dB and the bandwidth ratio $R\in\{1/24,1/12,1/6\}$. The comparisons between DeepJSCC-MIMO and the BPG-Capacity benchmark is shown in Fig. \ref{ViT_close_JSCC}. 

In terms of the PSNR performance, DeepJSCC-MIMO outperforms the separation-based benchmark in all SNR and bandwidth-ratio scenarios. Specifically, we can see that DeepJSCC-MIMO outperforms the benchmark by at least $1.98$ dB and $1.78$ dB for $R=1/24$ and $R=1/12$, respectively. When $R=1/6$, improvements of up to $3.58$ dB can be observed. These significant improvements demonstrate the superiority of the DeepJSCC-MIMO scheme and its capacity in extracting and mapping image features to the available channels in an adaptive fashion. \textcolor{black}{We observe in Figs. \ref{close_ViT_1_24_lpips}, \ref{close_ViT_1_12_lpips} and \ref{close_ViT_1_6_lpips} that DeepJSCC-MIMO achieves the best perceptual performance across a wide range of channel conditions and bandwidth ratios, particularly in the low SNR regimes.}

  \begin{figure}[t]
     \centering
    \includegraphics[scale=0.4]{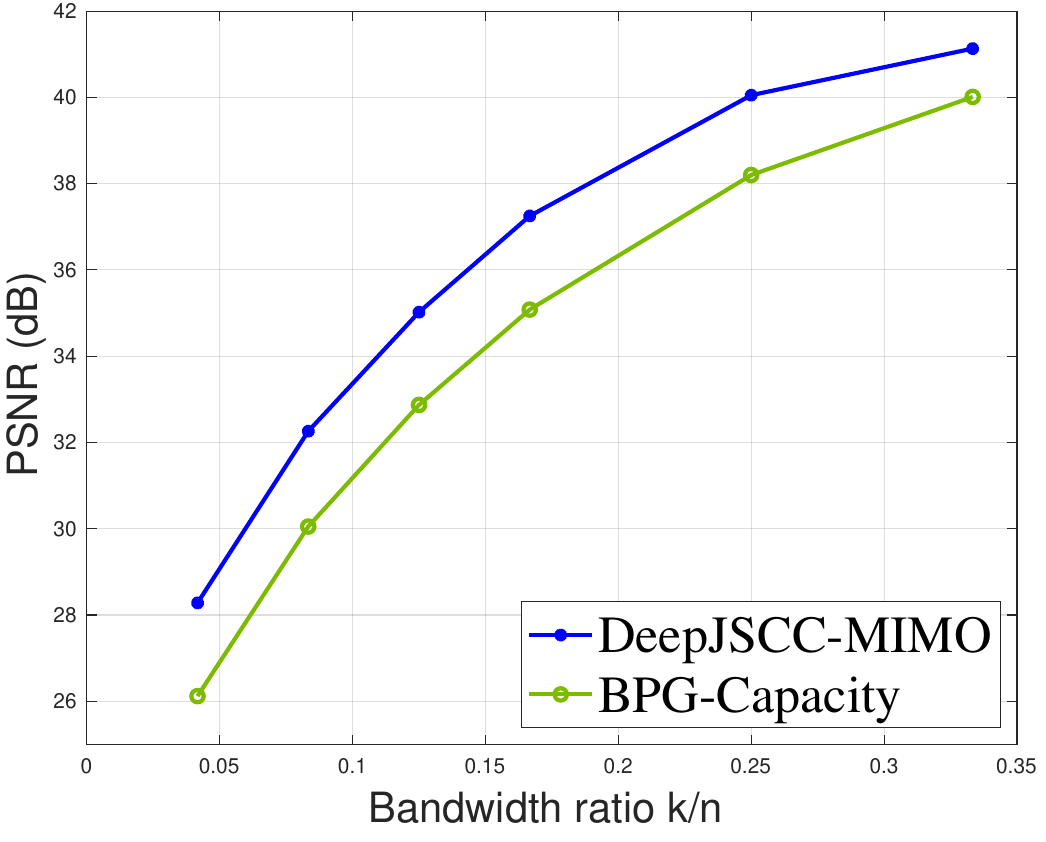}
     \caption{Performance of DeepJSCC-MIMO and BPG-LDPC scheme with respect to the bandwidth ratio at SNR=$10$ dB.}
     \label{close_bd_10}
 \end{figure}
 
\begin{figure*}[t]
\includegraphics[width=0.9\linewidth]{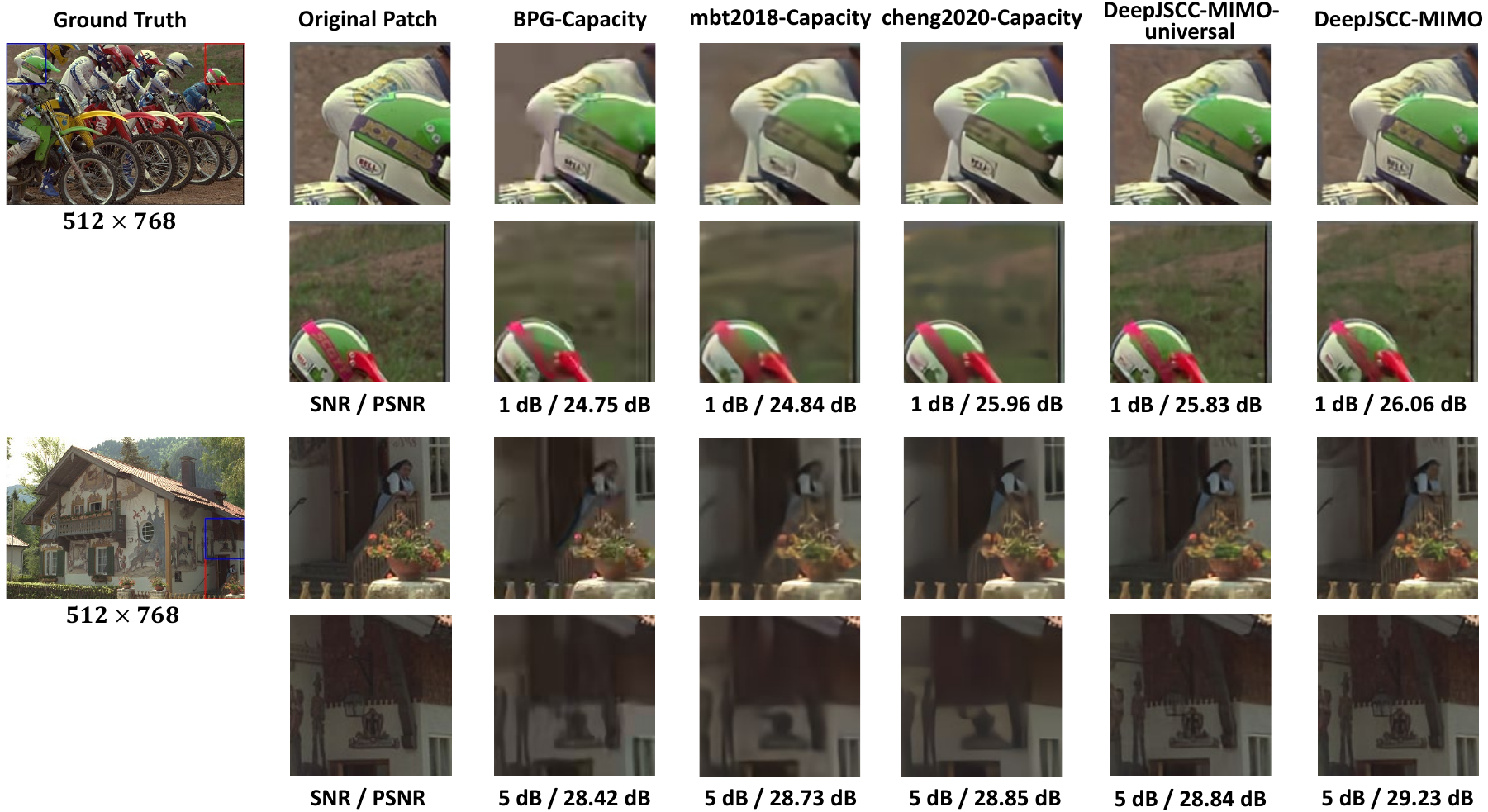}
\centering
\caption{\textcolor{black}{Visual comparison of different schemes on two samples from the Kodak dataset at a bandwidth ratio $R=1/24$ and different SNRs . The first and second columns are the original image and the original patch in the red and blue bounding boxes. The third, fourth, and fifth columns are the patch reconstructed by the BPG-Capacity\cite{bpg}, MBT2018-Capacity \cite{minnenbt18}, and Cheng2020-Capacity \cite{cheng2020image} schemes, respectively. The last two columns demonstrate the patch recovered by the DeepJSCC-MIMO-universal and DeepJSCC-MIMO schemes.} }
\label{vis_close_loop}
\end{figure*}

\subsubsection{SNR-adaptability}
We also consider a random SNR training strategy to evaluate the channel adaptability of DeepJSCC-MIMO. We train the model with random SNR values uniformly sampled from $[0,22]$ dB, denoted by DeepJSCC-MIMO-universal, and test it at different SNRs. The comparisons in Fig. \ref{ViT_close_JSCC} show that there is a slight performance degradation compared to training a separate DeepJSCC-MIMO encoder-decoder pair for each channel SNR (up to $0.65$ dB, $0.93$ dB, and $2.19$ dB for $R=1/24$, $R=1/12$ and $R=1/6$); however, DeepJSCC-MIMO-universal brings significant advantage in terms of training complexity and storage requirements.

We can observe more performance loss in the high SNR regime. However, DeepJSCC-MIMO-universal still significantly outperforms the BPG-Capacity benchmark. We conclude that the DeepJSCC-MIMO-universal scheme can adapt to SNR variations with a slight loss in PSNR, especially in the high-SNR regime.
 
\subsubsection{SVD ablation study}
To evaluate the effectiveness of the SVD strategy in our scheme, we train the models of DeepJSCC-MIMO-universal without SVD decomposition-based precoding in different bandwidth ratios, denoted by DeepJSCC-MIMO-universal w/o SVD. Specifically, we directly feed the channel heatmap $\bm{M}$ into both transceivers and %and apply the zero-forcing equalization in the receiver. 
the performance is shown in Fig. \ref{ViT_close_JSCC}. Compared with the DeepJSCC-MIMO-universal, we can observe that the performance of DeepJSCC-MIMO-universal w/o SVD has a gap of up to $0.97$ dB, $1.91$ dB and $4.21$ dB for $R=1/24$, $R=1/12$, and $R=1/6$, respectively. Although the networks can still learn to communicate over the MIMO channel, the SVD-based model-driven strategy can significantly simplify the training process and improve performance, illustrating the importance of exploiting domain knowledge in designing data-driven communication technologies.
%We observe that DeepJSCC-MIMO can outperform the BPG-Capacity benchmark at all bandwidth ratios for $5$ dB and $10$ dB, with a gain of up to $2.1$ dB and $2.2$ dB in PSNR, respectively. The gap between the two is more significant for low $R$ values. We would like to emphasize that the BPG-Capacity performance shown in these figures serves as an upper bound  for  separate source and channel coding schemes employing BPG compression, and the real gap can be even larger, especially in the short block length regime, i.e., low $R$. Thus, the results here illustrate the clear superiority of DeepJSCC-MIMO for the CIFAR dataset compared to separation-based alternatives. 

\subsubsection{Different bandwidth ratios}
To further evaluate the impact of the bandwidth ratio $R$ on the system performance, we compare the DeepJSCC-MIMO and BPG-Capacity schemes over a wide range of bandwidth ratios in Fig. \ref{close_bd_10}, where the test SNR is set to $10$ dB. We observe that DeepJSCC-MIMO can outperform the BPG-Capacity benchmark at all bandwidth ratios for $10$ dB, with a gain of up to $2.21$ dB in PSNR, respectively. The gap between the two is more significant for low $R$ values. We would like to emphasize that the BPG-Capacity performance shown in these figures serves as an upper bound for separate source and channel coding schemes employing BPG compression, and the real gap can be even larger, especially in the short block length regime, i.e., low $R$. Thus, the results here illustrate the clear superiority of DeepJSCC-MIMO for the CIFAR dataset compared to separation-based alternatives. 

\subsubsection{High resolution datasets}
\begin{figure}[t]
     \centering
    \includegraphics[scale=0.4]{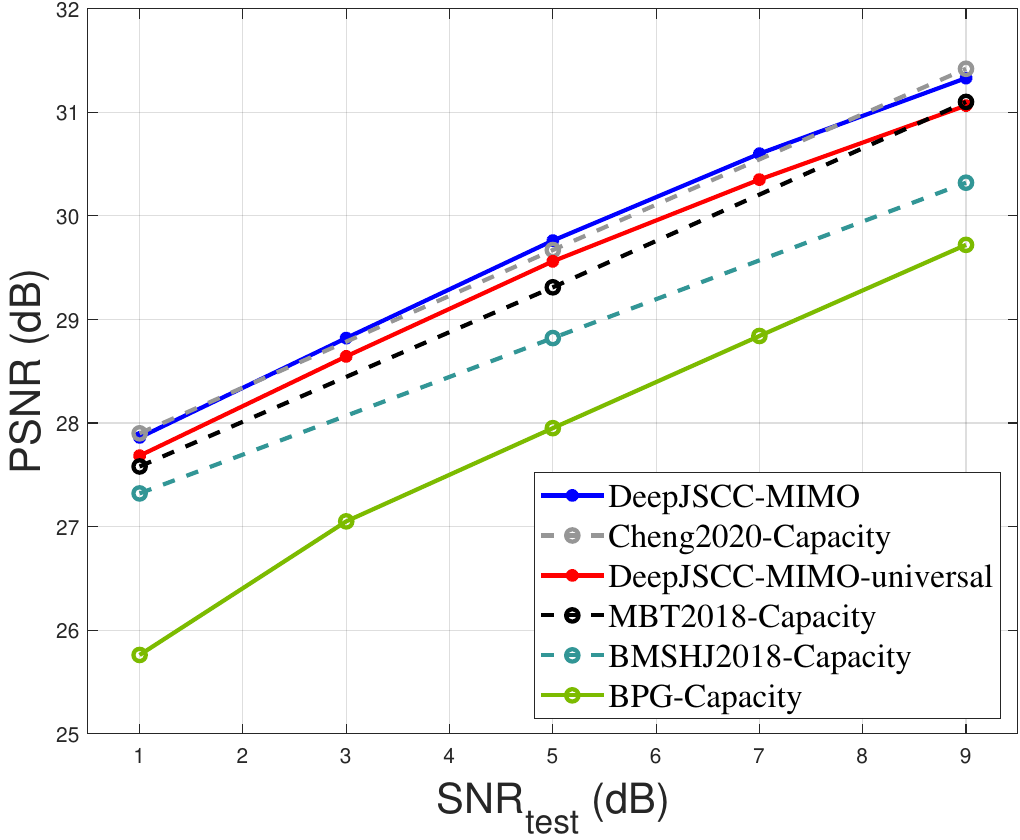}
     %\caption{fig2}
     \caption{\textcolor{black}{Performance of DeepJSCC-MIMO and conventional separation-based scheme in the {CelebA dataset} when $R=1/48$.}}
     \label{vitmimo_close_celeba}
 \end{figure}

To evaluate the model generalizability, we validate the DeepJSCC-MIMO on Kodak and CelebA datasets. Specifically, we train the DeepJSCC-MIMO and DeepJSCC-MIMO-universal schemes with randomly cropped $128 \times 128$ patches from the ImageNet dataset, and evaluate the well-trained models on the Kodak and CelebA datasets. The detailed visualization and the PSNR performance of the transmission, compared with BPG-Capacity, are shown in Fig. \ref{vis_close_loop} and Fig. \ref{vitmimo_close_celeba}, respectively.
 
{Similarly to the CIFAR dataset, from Fig. \ref{vitmimo_close_celeba}}, we can observe that our DeepJSCC-MIMO can outperform BPG-Capacity at all test SNRs and generally provide higher gains at lower SNRs (up to $2.1$ dB) in the CelebA dataset. There is a slight performance degradation of the DeepJSCC-MIMO-universal scheme (up to $0.27$ dB) compared with DeepJSCC-MIMO trained at specific SNRs; however, the DeepJSCC-MIMO-universal can still outperform the BPG-Capacity scheme at all SNRs. Additionally, we consider state-of-the-art learning-based compression algorithms along with capacity-achieving codes and water-filling algorithm, denoted as Cheng2020-Capacity \cite{cheng2020image}, MBT2018-Capacity \cite{minnenbt18}, and BMSHJ2018-Capacity schemes \cite{ballemshj18}. \textcolor{black}{The superior performance of Deep JSCC-MIMO is evident when compared to both the MBT2018-Capacity and BMSHJ2018-Capacity schemes. Furthermore, Deep-JSCC-MIMO achieves similar performance with the Cheng2020-Capacity scheme.} It is also noteworthy that capacity-achieving channel codes are practically not possible, and require accurate channel estimation and sufficiently long block lengths to approximate in practice.
 
\begin{table}[t]
    \centering
    \caption{\textcolor{black}{Ablation study of DeepJSCC-MIMO-universal scheme with/without channel heatmap across various SNRs on the CIFAR10 dataset when $R=1/12$.}}
    \label{table:ab_heatmap}
    \begin{tabular}{|c|c|c|c|c|c|}
    \hline
    \multicolumn{6}{|c|}{\textbf{Open-loop MIMO system with CSIR}}\\
    \hline
     Models& {1 dB}& {5 dB}& {10 dB}& {15 dB}& {19 dB}\\
    %Open-loop MIMO&\textbf{Channel SNR} & {1 dB}& {5 dB}& {10 dB}& {15 dB}& {19 dB}\\    
    \hline
     w/o heatmap & 24.22&26.69& 29.10&30.89&31.86\\
     {with heatmap}&$\bm{24.95}$&$\bm{27.56}$&$\bm{30.02}$&$\bm{31.91}$&$\bm{32.93}$ \\          
              \hline
    \multicolumn{6}{|c|}{\textbf{Closed-loop MIMO system with CSIT}}\\
    \hline
     Models& {1 dB}& {5 dB}& {10 dB}& {15 dB}& {19 dB}\\
    %Open-loop MIMO&\textbf{Channel SNR} & {1 dB}& {5 dB}& {10 dB}& {15 dB}& {19 dB}\\    
    \hline
         w/o heatmap &26.21& 28.71& 31.36&33.41& 34.67\\
     {with heatmap}&
     $\bm{26.89}$&$\bm{29.30}$&$\bm{32.10}$&$\bm{34.50}$&$\bm{35.95}$ \\                        \hline
    \end{tabular}
\end{table}

We also visualize the results of the DeepJSCC-MIMO scheme on the Kadak dataset and compare them with those of the BPG-Capacity scheme in Fig. \ref{vis_close_loop}. We can observe that DeepJSCC-MIMO performs better in the low SNR regime, such as $1$ and $5$ dB, with more detailed high-frequency features. Comparing the performance of DeepJSCC-MIMO-universal and DeepJSCC-MIMO schemes, we can observe that the DeepJSCC-MIMO-universal can achieve comparable performance at the expense of a slight drop in distortion.
 
Considering that the BPG-Capacity performance shown in these figures is an upper bound on the performance of any separation-based scheme that employs the BPG codec, the comparison here clearly shows the benefits of the DeepJSCC-MIMO and DeepJSCC-MIMO-universal schemes in high-resolution datasets. We can also conclude that a sufficiently large dataset (ImageNet) can allow our DeepJSCC-MIMO scheme to perform well on other never-seen datasets (Kodak/CelebA) for a wide range of channel conditions (SNRs). 

\subsubsection{\textcolor{black}{Channel heatmap ablation study}} 
\textcolor{black}{To assess the effectiveness of the channel heatmap, we conduct ablation studies over the DeepJSCC-MIMO-universal scheme, as outlined in Tables. \ref{table:ab_heatmap}. Specifically, we can observe a maximum increase in PSNR of $1.07$ dB and $1.28$ dB for open-loop and closed-loop MIMO systems when $R=1/12$, respectively. Our experimental findings demonstrate that integrating the channel heatmap is essential to enhance the performance of the DeepJSCC-MIMO-universal scheme in both open-loop and closed-loop MIMO systems across various SNRs and bandwidth ratios\cite{wu2023deep_arxiv}.} \textcolor{black}{Similar findings can also be observed in more challenging fast fading channels, where individual channel symbols experience fast fading, resulting in distinct channel heatmap in each time slot, as elaborated in Table. \ref{table:ab_heatmap_fast_fading}.}
\begin{figure}[t]
\centering
\includegraphics[width=0.9\linewidth]{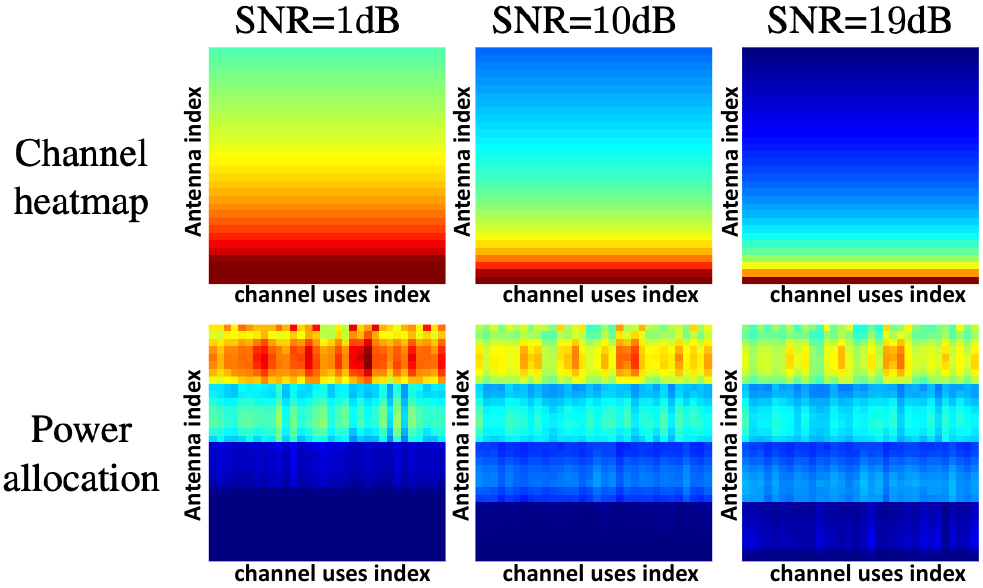}
\caption{{Visualization of the channel heatmap and power allocation across the channel symbols, where antenna number is set as $M=32$ and $R=1/6$}} 
\label{vis_heatmap_close_loop}
\end{figure}

\subsubsection{Visualization of the channel heatmap and power allocation}
We visualize the channel heatmap and the power allocation of the channel input matrix for a closed-loop MIMO system in Fig. \ref{vis_heatmap_close_loop}. Within our channel heatmap, we visualize the equivalent noise power experienced by each channel symbols, with the color red indicating higher noise power, and hence, a poorer channel condition. In the power allocation map, the color red denotes a higher allocation of power to the respective channel symbols. We can observe that for each SNR value, DeepJSCC-MIMO generally assigns more power to antennas with better channel conditions. Specifically, in the low SNR regime, DeepJSCC-MIMO concentrates most of the power on a few select antennas with exceptional quality. In certain extreme cases, DeepJSCC-MIMO refrains from allocating power to the sub-channels characterized by the worst channel conditions. As the SNR increases, we note that DeepJSCC-MIMO demonstrates a tendency to distribute power more evenly across a greater number of antennas, instead of concentrating on a limited set of {antennas} with the worst channel conditions.

\subsubsection{Adaptability to the antenna numbers}
\begin{figure}[t]
     \centering
    \includegraphics[scale=0.4]{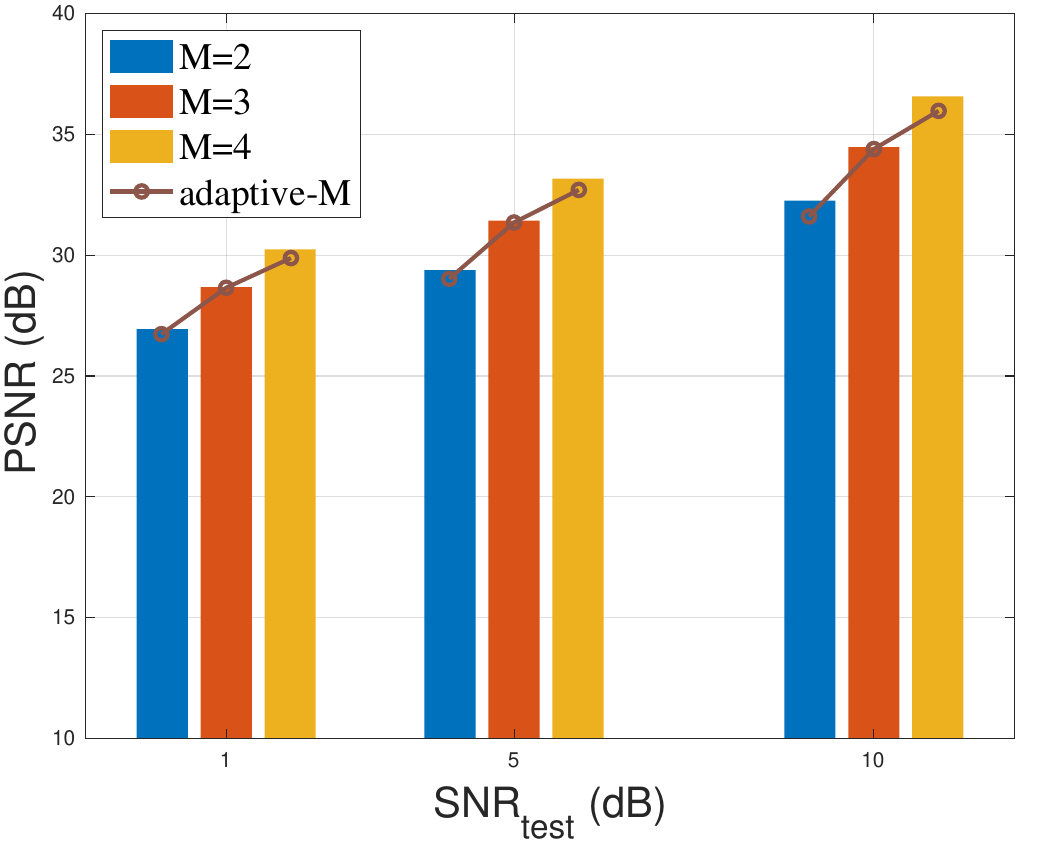}
     \caption{Performance of DeepJSCC-MIMO with different antennas number.}
     \label{at_num}
 \end{figure}

In all the above experiments, we have the same number of $M=2$ antennas at the transmitter and receiver. In this subsection, we investigate the generalizability of the DeepJSCC-MIMO architecture to different antenna numbers. In particular, we repeat the experiment of DeepJSCC-MIMO in Fig. \ref{close_ViT_1_6} for different $M$ values. The performance of our DeepJSCC-MIMO models with different antenna numbers $M\in \{2,3,4\}$ and channel SNRs $SNR_{test}\in \{1,5,10\}$ dB is shown in  Fig. \ref{at_num}. We can observe that DeepJSCC-MIMO can generally achieve better performance with the increase in the number of antennas, thanks to its more flexible power allocation.
\begin{table}[t]
    \centering
    \caption{\textcolor{black}{Ablation study of DeepJSCC-MIMO-universal scheme with/without channel heatmap on the CIFAR10 dataset across various SNRs in the fast fading channel when $R=1/48$.}}
    \label{table:ab_heatmap_fast_fading}
    \begin{tabular}{|c|c|c|c|c|c|}
    \hline
    \multicolumn{6}{|c|}{\textbf{Open-loop MIMO system with CSIR}}\\
    \hline
     Models& {1 dB}& {5 dB}& {10 dB}& {15 dB}& {19 dB}\\
    %Open-loop MIMO&\textbf{Channel SNR} & {1 dB}& {5 dB}& {10 dB}& {15 dB}& {19 dB}\\    
    \hline
    {w/o heatmap}  & ${18.72}$&${20.64}$& ${22.35}$&${23.51}$&${24.04}$\\     with heatmap &$\bm{19.24}$&$\bm{ 21.23}$& $\bm{23.11}$& $\bm{24.47}$& $\bm{25.17}$\\          
              \hline
    \multicolumn{6}{|c|}{\textbf{Closed-loop MIMO system with CSIT}}\\
    \hline
     Models& {1 dB}& {5 dB}& {10 dB}& {15 dB}& {19 dB}\\
    %Open-loop MIMO&\textbf{Channel SNR} & {1 dB}& {5 dB}& {10 dB}& {15 dB}& {19 dB}\\    
    \hline
     {w/o heatmap} & $20.75$& $22.57$& $	24.30$& $	25.50$& $	26.14$\\
     with heatmap &$\bm{21.56}$&$\bm{23.44}$& $\bm{25.23}$& $\bm{26.67}$& $\bm{27.30}$\\    
              \hline
    \end{tabular}
\end{table}
We also consider a random antenna number training strategy with a maximal antenna number setting $M_{max}$, denoted as `adaptive-M'. We train the DeepJSCC-MIMO model with $M$ uniformly sampled from $\{2,\ldots, M_{max}\}$ and test the well-trained model with different $M$ values. Specifically, our encoder generates the channel symbols $\bm{X}\in \mathbb{C}^{M_{max}\times k}$, and only transmits first $\mathbb{C}^{M_i\times k}$ symbols for the $M=M_i$ antenna scenario. At the same time, the receiver pads the received signal into $\mathbb{C}^{M_{max}\times k}$ with zeros for decoding. \textcolor{black}{The representation of CSI heatmaps differs across scenarios with distinct antenna configurations, serving as a guide for effectively encoding the channel symbols. We think these adaptive characteristics are attributed to the transformer's powerful learning and generalization capabilities.} From experiments with $M_{max}=4$, we observe in Fig. \ref{at_num} that the adaptive-M strategy can achieve similar performance compared with models trained for specific $M$ values: the performance loss is at most $0.6$dB. We can conclude that DeepJSCC-MIMO is highly flexible and adaptive and can be applied to different number of antennas without retraining.
 
\section{Conclusion}
We introduced the first DeepJSCC-enabled MIMO communication system designed for wireless image transmission. The proposed DeepJSCC-MIMO model utilizes a vision transformer-based architecture to exploit contextual semantic features and channel conditions through a self-attention mechanism. The results demonstrate significant improvements in transmission quality across a wide range of SNRs and bandwidth scenarios, with enhanced robustness to channel estimation errors. Moreover, DeepJSCC-MIMO is a unified and adaptable design capable of accommodating varying channel conditions and different antenna numbers without the need for retraining.

Moving forward, there are several potential directions for future exploration based on the findings of this paper:
\begin{itemize}
\item JSCC with space-time coding.
In the open-loop scenario, full potential of the spatial and temporal dimensions in MIMO can be harnessed by integrating DeepJSCC with space-time codes \cite{bian2022space}. This synergy can allow for even greater flexibility and performance enhancements, especially in the low-SNR regime.
\item Variable length JSCC. Building upon the adaptability of ViT-based models to varying antenna numbers, future investigations can extend this concept to accommodate variable code lengths, where the channel resources are judiciously exploited depending on the channel state as well as the input signal.
 
%. By incorporating DL techniques such as Gumbel softmax-based mask generation, there is potential to develop improved solutions for varying code-length encoding JSCC schemes over MIMO systems.
\end{itemize}
\bibliographystyle{IEEEtran}
\bibliography{ref}

% Generated by IEEEtran.bst, version: 1.14 (2015/08/26)
\begin{thebibliography}{10}
\providecommand{\url}[1]{#1}
\csname url@samestyle\endcsname
\providecommand{\newblock}{\relax}
\providecommand{\bibinfo}[2]{#2}
\providecommand{\BIBentrySTDinterwordspacing}{\spaceskip=0pt\relax}
\providecommand{\BIBentryALTinterwordstretchfactor}{4}
\providecommand{\BIBentryALTinterwordspacing}{\spaceskip=\fontdimen2\font plus
\BIBentryALTinterwordstretchfactor\fontdimen3\font minus \fontdimen4\font\relax}
\providecommand{\BIBforeignlanguage}[2]{{%
\expandafter\ifx\csname l@#1\endcsname\relax
\typeout{** WARNING: IEEEtran.bst: No hyphenation pattern has been}%
\typeout{** loaded for the language `#1'. Using the pattern for}%
\typeout{** the default language instead.}%
\else
\language=\csname l@#1\endcsname
\fi
#2}}
\providecommand{\BIBdecl}{\relax}
\BIBdecl

\bibitem{bourtsoulatze2019deep}
E.~Bourtsoulatze, D.~Burth~Kurka, and D.~Gündüz, ``Deep joint source-channel coding for wireless image transmission,'' \emph{IEEE Transactions on Cognitive Communications and Networking}, vol.~5, no.~3, pp. 567--579, 2019.

\bibitem{9714510}
M.~Yang, C.~Bian, and H.-S. Kim, ``\textsc{OFDM}-guided deep joint source channel coding for wireless multipath fading channels,'' \emph{IEEE Transactions on Cognitive Communications and Networking}, vol.~8, no.~2, pp. 584--599, 2022.

\bibitem{xu2021wireless}
J.~Xu, B.~Ai, W.~Chen, A.~Yang, P.~Sun, and M.~Rodrigues, ``Wireless image transmission using deep source channel coding with attention modules,'' \emph{IEEE Transactions on Circuits and Systems for Video Technology}, vol.~32, no.~4, pp. 2315--2328, 2022.

\bibitem{9791398}
J.~Dai, S.~Wang, K.~Tan, Z.~Si, X.~Qin, K.~Niu, and P.~Zhang, ``Nonlinear transform source-channel coding for semantic communications,'' \emph{IEEE Journal on Selected Areas in Communications}, vol.~40, no.~8, pp. 2300--2316, 2022.

\bibitem{kurka2021bandwidth}
D.~B. Kurka and D.~Gündüz, ``Bandwidth-agile image transmission with deep joint source-channel coding,'' \emph{IEEE Transactions on Wireless Communications}, vol.~20, no.~12, pp. 8081--8095, 2021.

\bibitem{bian2024process}
C.~Bian, Y.~Shao, H.~Wu, E.~Ozfatura, and D.~Gunduz, ``Process-and-forward: Deep joint source-channel coding over cooperative relay networks,'' \emph{arXiv preprint arXiv:2403.10613}, 2024.

\bibitem{yang2022deep}
M.~Yang and H.-S. Kim, ``Deep joint source-channel coding for wireless image transmission with adaptive rate control,'' in \emph{ICASSP 2022 - 2022 IEEE International Conference on Acoustics, Speech and Signal Processing (ICASSP)}, 2022, pp. 5193--5197.

\bibitem{yoo2023role}
H.~Yoo, L.~Dai, S.~Kim, and C.-B. Chae, ``On the role of \textsc{V}i\textsc{T} and \textsc{CNN} in semantic communications: Analysis and prototype validation,'' \emph{IEEE Access}, vol.~11, pp. 71\,528--71\,541, 2023.

\bibitem{erdemir2022generative}
E.~Erdemir, T.-Y. Tung, P.~L. Dragotti, and D.~Gunduz, ``Generative joint source-channel coding for semantic image transmission,'' \emph{IEEE Journal on Selected Areas in Communications}, vol.~41, no.~8, pp. 2645--2657, 2023.

\bibitem{bian2022deep}
C.~Bian, Y.~Shao, H.~Wu, and D.~Gunduz, ``Deep joint source-channel coding over cooperative relay networks,'' in \emph{2024 IEEE International Conference on Machine Learning for Communication and Networking (ICMLCN)}.\hskip 1em plus 0.5em minus 0.4em\relax IEEE, 2024.

\bibitem{shannon1948mathematical}
C.~E. Shannon, ``A mathematical theory of communication,'' \emph{The Bell system technical journal}, vol.~27, no.~3, pp. 379--423, 1948.

\bibitem{hekland2009shannon}
F.~Hekland, P.~A. Floor, and T.~A. Ramstad, ``Shannon-kotel-nikov mappings in joint source-channel coding,'' \emph{IEEE Transactions on Communications}, vol.~57, no.~1, pp. 94--105, 2009.

\bibitem{vazquez2014analog}
F.~J. Vazquez-Araujo, O.~Fresnedo, L.~Castedo, and J.~Garcia-Frias, ``Analog joint source-channel coding over \textsc{MIMO} channels,'' \emph{EURASIP Journal on Wireless Communications and Networking}, vol. 2014, pp. 1--10, 2014.

\bibitem{aguerri2015joint}
I.~E. Aguerri and D.~G{\"u}nd{\"u}z, ``Joint source-channel coding with time-varying channel and side-information,'' \emph{IEEE Transactions on Information Theory}, vol.~62, no.~2, pp. 736--753, 2015.

\bibitem{vaezi2014distributed}
M.~Vaezi and F.~Labeau, ``Distributed source-channel coding based on real-field \textsc{BCH} codes,'' \emph{IEEE Transactions on Signal Processing}, vol.~62, no.~5, pp. 1171--1184, 2014.

\bibitem{Peng:JSAC:00}
Z.~Peng, Y.-F. Huang, and D.~Costello, ``Turbo codes for image transmission-a joint channel and source decoding approach,'' \emph{IEEE Journal on Selected Areas in Communications}, vol.~18, no.~6, pp. 868--879, 2000.

\bibitem{Pu:TIP:07}
L.~Pu, Z.~Wu, A.~Bilgin, M.~W. Marcellin, and B.~Vasic, ``\textsc{LDPC}-based iterative joint source-channel decoding for \textsc{JPEG2000},'' \emph{IEEE Transactions on Image Processing}, vol.~16, no.~2, pp. 577--581, 2007.

\bibitem{Fresia:SPM:10}
M.~Fresia, F.~Peréz-Cruz, H.~V. Poor, and S.~Verdú, ``Joint source and channel coding,'' \emph{IEEE Signal Processing Magazine}, vol.~27, no.~6, pp. 104--113, 2010.

\bibitem{Golmohammadi:TCOM:22}
A.~Golmohammadi and D.~G.~M. Mitchell, ``Concatenated spatially coupled \textsc{LDPC} codes with sliding window decoding for joint source-channel coding,'' \emph{IEEE Transactions on Communications}, vol.~70, no.~2, pp. 851--864, 2022.

\bibitem{9954060}
Z.~Xuan and K.~Narayanan, ``Low-delay analog joint source-channel coding with deep learning,'' \emph{IEEE Transactions on Communications}, vol.~71, no.~1, pp. 40--51, 2023.

\bibitem{wu2022channel}
H.~Wu, Y.~Shao, K.~Mikolajczyk, and D.~Gündüz, ``Channel-adaptive wireless image transmission with \textsc{OFDM},'' \emph{IEEE Wireless Communications Letters}, vol.~11, no.~11, pp. 2400--2404, 2022.

\bibitem{tung:ICASSP:20}
\BIBentryALTinterwordspacing
T.-Y. Tung, D.~B. Kurka, and D.~G{\"u}nd{\"u}z, ``Deepjscc: The future of wireless video transmission,'' in \emph{IEEE International Conference on Acoustics, Speech and Signal Processing (ICASSP), Demo Session}, 2020. [Online]. Available: \url{https://cmsworkshops.com/ICASSP2020/Papers/ViewPaper.asp?PaperNum=6212}
\BIBentrySTDinterwordspacing

\bibitem{tung2021deepjscc}
T.-Y. Tung, D.~B. Kurka, M.~Jankowski, and D.~Gündüz, ``Deep\textsc{JSCC-Q}: Constellation constrained deep joint source-channel coding,'' \emph{IEEE Journal on Selected Areas in Information Theory}, vol.~3, no.~4, pp. 720--731, 2022.

\bibitem{DTAT}
Y.~Shao and D.~Gunduz, ``Semantic communications with discrete-time analog transmission: A \textsc{PAPR} perspective,'' \emph{IEEE Wireless Communications Letters}, vol.~12, no.~3, pp. 510--514, 2023.

\bibitem{samuel2019learning}
N.~Samuel, T.~Diskin, and A.~Wiesel, ``Learning to detect,'' \emph{IEEE Transactions on Signal Processing}, vol.~67, no.~10, pp. 2554--2564, 2019.

\bibitem{mashhadi2021pruning}
M.~B. Mashhadi and D.~G{\"u}nd{\"u}z, ``Pruning the pilots: Deep learning-based pilot design and channel estimation for \textsc{MIMO-OFDM} systems,'' \emph{IEEE Transactions on Wireless Communications}, vol.~20, no.~10, pp. 6315--6328, 2021.

\bibitem{he2020model}
H.~He, C.-K. Wen, S.~Jin, and G.~Y. Li, ``Model-driven deep learning for \textsc{MIMO} detection,'' \emph{IEEE Transactions on Signal Processing}, vol.~68, pp. 1702--1715, 2020.

\bibitem{wu2022vision}
H.~Wu, Y.~Shao, C.~Bian, K.~Mikolajczyk, and D.~Gündüz, ``Vision transformer for adaptive image transmission over \textsc{MIMO} channels,'' in \emph{ICC 2023 - IEEE International Conference on Communications}, 2023, pp. 3702--3707.

\bibitem{gunduz2008joint}
D.~Gunduz and E.~Erkip, ``Joint source--channel codes for \textsc{MIMO} block-fading channels,'' \emph{IEEE Transactions on Information Theory}, vol.~54, no.~1, pp. 116--134, 2008.

\bibitem{o2017deep}
T.~J. O'Shea, T.~Erpek, and T.~C. Clancy, ``Deep learning based \textsc{MIMO} communications,'' \emph{arXiv preprint arXiv:1707.07980}, 2017.

\bibitem{song2020benchmarking}
J.~Song, C.~Häger, J.~Schröder, T.~J. O’Shea, E.~Agrell, and H.~Wymeersch, ``Benchmarking and interpreting end-to-end learning of \textsc{MIMO} and multi-user communication,'' \emph{IEEE Transactions on Wireless Communications}, vol.~21, no.~9, pp. 7287--7298, 2022.

\bibitem{zhang2022svd}
X.~Zhang, M.~Vaezi, and T.~J. O’Shea, ``\textsc{SVD}-embedded deep autoencoder for \textsc{MIMO} communications,'' in \emph{ICC 2022-IEEE International Conference on Communications}.\hskip 1em plus 0.5em minus 0.4em\relax IEEE, 2022, pp. 5190--5195.

\bibitem{Gunduz:JSAC:23}
D.~Gündüz, Z.~Qin, I.~E. Aguerri, H.~S. Dhillon, Z.~Yang, A.~Yener, K.~K. Wong, and C.-B. Chae, ``Beyond transmitting bits: Context, semantics, and task-oriented communications,'' \emph{IEEE Journal on Selected Areas in Communications}, vol.~41, no.~1, pp. 5--41, 2023.

\bibitem{lo2023collaborative}
W.~F. Lo, N.~Mital, H.~Wu, and D.~Gündüz, ``Collaborative semantic communication for edge inference,'' \emph{IEEE Wireless Communications Letters}, vol.~12, no.~7, pp. 1125--1129, 2023.

\bibitem{shi2023excess}
Y.~Shi, S.~Shao, Y.~Wu, W.~Zhang, X.-G. Xia, and C.~Xiao, ``Excess distortion exponent analysis for semantic-aware \textsc{MIMO} communication systems,'' \emph{IEEE Transactions on Wireless Communications}, vol.~22, no.~9, pp. 5927--5940, 2023.

\bibitem{yao2022versatile}
S.~Yao, S.~Wang, J.~Dai, K.~Niu, and P.~Zhang, ``Versatile semantic coded transmission over \textsc{MIMO} fading channels,'' \emph{arXiv preprint arXiv:2210.16741}, 2022.

\bibitem{baltersee2001achievable}
J.~Baltersee, G.~Fock, and H.~Meyr, ``Achievable rate of \textsc{MIMO} channels with data-aided channel estimation and perfect interleaving,'' \emph{IEEE Journal on Selected Areas in Communications}, vol.~19, no.~12, pp. 2358--2368, 2001.

\bibitem{yoo2006capacity}
T.~Yoo and A.~Goldsmith, ``Capacity and power allocation for fading \textsc{MIMO} channels with channel estimation error,'' \emph{IEEE Transactions on Information Theory}, vol.~52, no.~5, pp. 2203--2214, 2006.

\bibitem{shao2022attentioncode}
Y.~Shao, E.~Ozfatura, A.~G. Perotti, B.~M. Popović, and D.~Gündüz, ``Attentioncode: Ultra-reliable feedback codes for short-packet communications,'' \emph{IEEE Transactions on Communications}, vol.~71, no.~8, pp. 4437--4452, 2023.

\bibitem{ozfatura2022all}
E.~Ozfatura, Y.~Shao, A.~G. Perotti, B.~M. Popović, and D.~Gündüz, ``All you need is feedback: Communication with block attention feedback codes,'' \emph{IEEE Journal on Selected Areas in Information Theory}, vol.~3, no.~3, pp. 587--602, 2022.

\bibitem{wu2023transformer}
H.~Wu, Y.~Shao, E.~Ozfatura, K.~Mikolajczyk, and D.~Gündüz, ``Transformer-aided wireless image transmission with channel feedback,'' \emph{IEEE Transactions on Wireless Communications}, pp. 1--1, 2024.

\bibitem{8578166}
R.~Zhang, P.~Isola, A.~A. Efros, E.~Shechtman, and O.~Wang, ``The unreasonable effectiveness of deep features as a perceptual metric,'' in \emph{IEEE/CVF Conf. on Comp. Vis. and Ptrn. Recog.}, 2018, pp. 586--595.

\bibitem{scutari2009mimo}
G.~Scutari, D.~P. Palomar, and S.~Barbarossa, ``The \textsc{MIMO} iterative waterfilling algorithm,'' \emph{IEEE Transactions on Signal Processing}, vol.~57, no.~5, pp. 1917--1935, 2009.

\bibitem{zheng2021rethinking}
S.~Zheng, J.~Lu, H.~Zhao, X.~Zhu, Z.~Luo, Y.~Wang, Y.~Fu, J.~Feng, T.~Xiang, P.~H. Torr \emph{et~al.}, ``Rethinking semantic segmentation from a sequence-to-sequence perspective with transformers,'' in \emph{Proceedings of the IEEE/CVF conference on computer vision and pattern recognition}, 2021, pp. 6881--6890.

\bibitem{chu2021twins}
X.~Chu, Z.~Tian, Y.~Wang, B.~Zhang, H.~Ren, X.~Wei, H.~Xia, and C.~Shen, ``Twins: Revisiting the design of spatial attention in vision transformers,'' \emph{Advances in Neural Information Processing Systems}, vol.~34, pp. 9355--9366, 2021.

\bibitem{wu2023deep_arxiv}
H.~Wu, Y.~Shao, C.~Bian, K.~Mikolajczyk, and D.~G{\"u}nd{\"u}z, ``Deep joint source-channel coding for adaptive image transmission over \textsc{MIMO} channels,'' \emph{arXiv preprint arXiv:2309.00470}, 2023.

\bibitem{bpg}
``Better portable graphics 0.9.8,'' \emph{URL https://bellard.org/bpg/}, 2018.

\bibitem{minnenbt18}
D.~Minnen, J.~Ball{\'{e}}, and G.~Toderici, ``Joint autoregressive and hierarchical priors for learned image compression,'' in \emph{Neural Information Processing Systems}, 2018, pp. 10\,794--10\,803.

\bibitem{vtm}
``Versatile video coding reference software version 12.1 (vtm-12.1).'' \emph{https://vcgit.hhi.fraunhofer.de/jvet/VVCSoftware\textunderscore VTM}, vol.~6, 2021.

\bibitem{he2022elic}
D.~He, Z.~Yang, W.~Peng, R.~Ma, H.~Qin, and Y.~Wang, ``Elic: Efficient learned image compression with unevenly grouped space-channel contextual adaptive coding,'' in \emph{IEEE/CVF Conf. on Comp. Vis. and Ptrn. Recog.}, 2022.

\bibitem{cheng2020image}
Z.~Cheng, H.~Sun, M.~Takeuchi, and J.~Katto, ``Learned image compression with discretized {G}aussian mixture likelihoods and attention modules,'' in \emph{IEEE/CVF Conf. on Comp. Vis. and Ptrn. Recog.}, 2020.

\bibitem{ballemshj18}
\BIBentryALTinterwordspacing
J.~Ball{\'{e}}, D.~Minnen, S.~Singh, S.~J. Hwang, and N.~Johnston, ``Variational image compression with a scale hyperprior,'' in \emph{6th International Conference on Learning Representations, {ICLR} 2018, Vancouver, BC, Canada, April 30 - May 3, 2018, Conference Track Proceedings}.\hskip 1em plus 0.5em minus 0.4em\relax OpenReview.net, 2018. [Online]. Available: \url{https://openreview.net/forum?id=rkcQFMZRb}
\BIBentrySTDinterwordspacing

\bibitem{guo2006algorithm}
Z.~Guo and P.~Nilsson, ``Algorithm and implementation of the \textsc{K}-best sphere decoding for \textsc{MIMO} detection,'' \emph{IEEE Journal on selected areas in communications}, vol.~24, no.~3, pp. 491--503, 2006.

\bibitem{hassibi2003much}
B.~Hassibi and B.~M. Hochwald, ``How much training is needed in multiple-antenna wireless links?'' \emph{IEEE Transactions on Information Theory}, vol.~49, no.~4, pp. 951--963, 2003.

\bibitem{1614084}
R.~Etkin and D.~Tse, ``Degrees of freedom in some underspread \textsc{MIMO} fading channels,'' \emph{IEEE Transactions on Information Theory}, vol.~52, no.~4, pp. 1576--1608, 2006.

\bibitem{bian2022space}
C.~Bian, Y.~Shao, H.~Wu, and D.~Gündüz, ``Space-time design for deep joint source channel coding of images over \textsc{MIMO} channels,'' in \emph{2023 IEEE 24th International Workshop on Signal Processing Advances in Wireless Communications (SPAWC)}, 2023, pp. 616--620.

\bibitem{oestges2006validity}
C.~Oestges, ``Validity of the kronecker model for \textsc{MIMO} correlated channels,'' in \emph{2006 IEEE 63rd Vehicular Technology Conference}, vol.~6.\hskip 1em plus 0.5em minus 0.4em\relax IEEE, 2006, pp. 2818--2822.

\bibitem{10094735}
K.~Yang, S.~Wang, J.~Dai, K.~Tan, K.~Niu, and P.~Zhang, ``\textsc{WITT}: A wireless image transmission transformer for semantic communications,'' in \emph{ICASSP 2023 - 2023 IEEE International Conference on Acoustics, Speech and Signal Processing (ICASSP)}, 2023, pp. 1--5.

\bibitem{liu2021swin}
Z.~Liu, Y.~Lin, Y.~Cao, H.~Hu, Y.~Wei, Z.~Zhang, S.~Lin, and B.~Guo, ``Swin transformer: Hierarchical vision transformer using shifted windows,'' in \emph{Proceedings of the IEEE/CVF international conference on computer vision}, 2021, pp. 10\,012--10\,022.

\bibitem{Graham_2021_ICCV}
B.~Graham, A.~El-Nouby, H.~Touvron, P.~Stock, A.~Joulin, H.~Jegou, and M.~Douze, ``\textsc{L}e\textsc{V}i\textsc{T}: A vision transformer in convnet's clothing for faster inference,'' in \emph{Proceedings of the IEEE/CVF International Conference on Computer Vision (ICCV)}, October 2021, pp. 12\,259--12\,269.

\bibitem{Dollar2021}
P.~Doll{\'a}r, M.~Singh, and R.~Girshick, ``Fast and accurate model scaling,'' in \emph{Proceedings of the IEEE/CVF Conference on Computer Vision and Pattern Recognition}, 2021, pp. 924--932.

\bibitem{evo_vit}
Y.~Xu, Z.~Zhang, M.~Zhang, K.~Sheng, K.~Li, W.~Dong, L.~Zhang, C.~Xu, and X.~Sun, ``Evo-vit: Slow-fast token evolution for dynamic vision transformer,'' in \emph{Proceedings of the AAAI Conference on Artificial Intelligence}, 2022, pp. 2964--2972.

\bibitem{stevens2021softermax}
J.~R. Stevens, R.~Venkatesan, S.~Dai, B.~Khailany, and A.~Raghunathan, ``Softermax: Hardware/software co-design of an efficient softmax for transformers,'' in \emph{2021 58th ACM/IEEE Design Automation Conference (DAC)}.\hskip 1em plus 0.5em minus 0.4em\relax IEEE, 2021, pp. 469--474.

\bibitem{kurka2020deepjscc}
D.~B. Kurka and D.~G{\"u}nd{\"u}z, ``Deepjscc-f: Deep joint source-channel coding of images with feedback,'' \emph{IEEE Journal on Selected Areas in Information Theory}, vol.~1, no.~1, pp. 178--193, 2020.

\end{thebibliography}
\newpage
\appendix
\begin{comment}
\begin{table*}[h]
    \centering
    \caption{\textcolor{black}{Ablation studies over the model-driven design for the proposed DeepJSCC-MIMO schemes with CSIR.}}
    \label{table:open_ab_heatmap}
    \begin{tabular}{|c|c|c|c|c|c|c|}
    \hline
    \multicolumn{2}{|c|}{\textbf{Open-loop MIMO system}}&\multicolumn{5}{c|}{{Channel SNRs}}\\
    \hline
        Ratio&Models& {1 dB}& {5 dB}& {10 dB}& {15 dB}& {19 dB}\\
    %Open-loop MIMO&\textbf{Channel SNR} & {1 dB}& {5 dB}& {10 dB}& {15 dB}& {19 dB}\\    
    \hline
     \multirow{4}{*}{\textbf{$R=1/24$}}
     &{DeepJSCC-MIMO}  & $\bm{22.27}$&$\bm{24.41}$& $\bm{26.56}$&$\bm{28.26}$&$\bm{29.33}$\\
     & DeepJSCC-MIMO w/o heatmap & 21.83&24.00& 26.07&27.64&28.53 \\
      & DeepJSCC-MIMO-universal & 21.83&24.00& 26.07&27.64&28.53 \\
     & DeepJSCC-MIMO-universal w/o heatmap & 21.83&24.00& 26.07&27.64&28.53 \\
         \hline
     \multirow{4}{*}{\textbf{$R=1/12$}}
     &{DeepJSCC-MIMO}  & $\bm{22.27}$&$\bm{24.41}$& $\bm{26.56}$&$\bm{28.26}$&$\bm{29.33}$\\
     & DeepJSCC-MIMO w/o heatmap & 21.83&24.00& 26.07&27.64&28.53 \\
      & DeepJSCC-MIMO-universal & 21.83&24.00& 26.07&27.64&28.53 \\
     & DeepJSCC-MIMO-universal w/o heatmap & 21.83&24.00& 26.07&27.64&28.53 \\
          \hline
    \end{tabular}
\end{table*}
\end{comment}

\begin{figure*}[t]
    \centering
    \subfloat[$R=1/24$]{
    \label{ab_eq_open_ViT_24} 
    \begin{minipage}[t]{0.45\linewidth}
    \centering
    \includegraphics[scale=0.38]{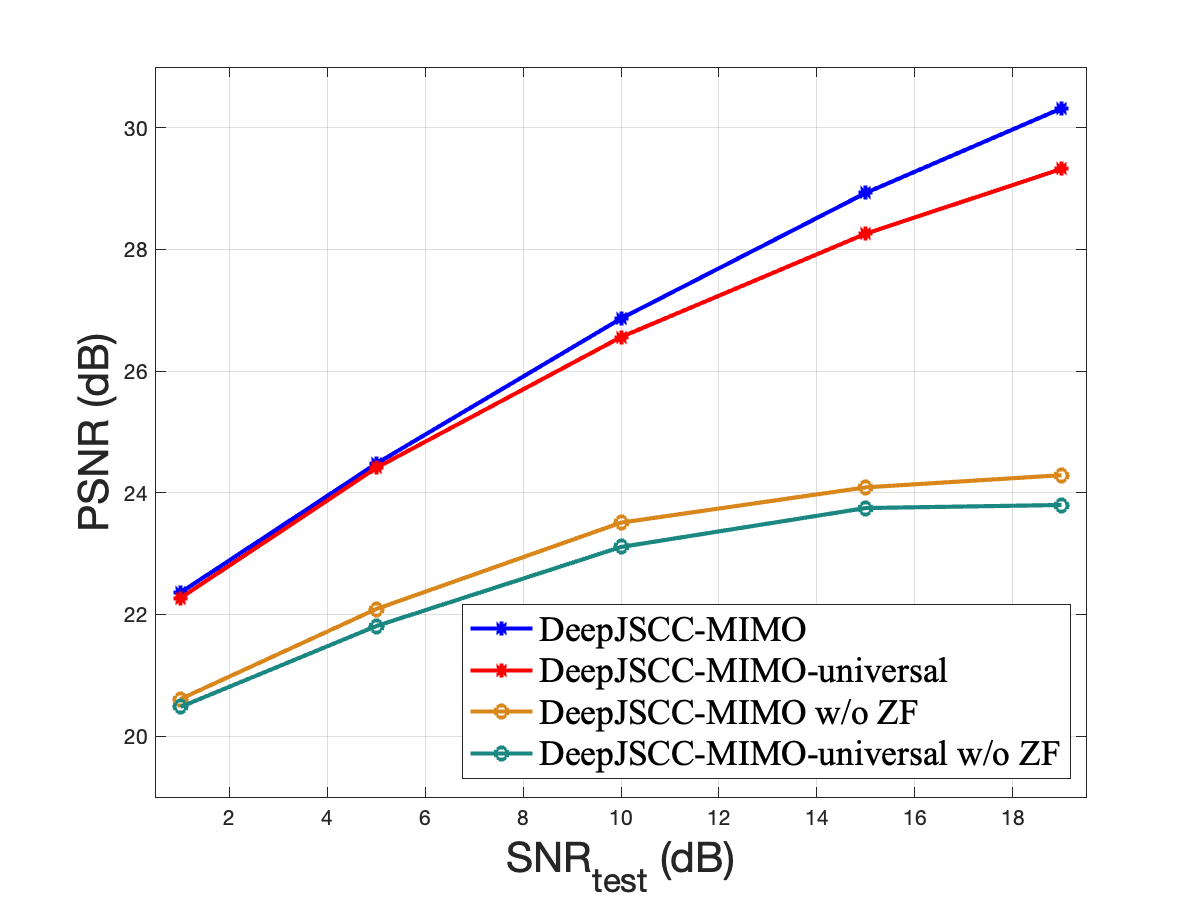}
    \end{minipage}%
    }%
    \hspace{-0.5mm}
    \subfloat[$R=1/12$]{
    \label{ab_eq_open_ViT_12} 
    \begin{minipage}[t]{0.45\linewidth}
    \centering
    \includegraphics[scale=0.38]{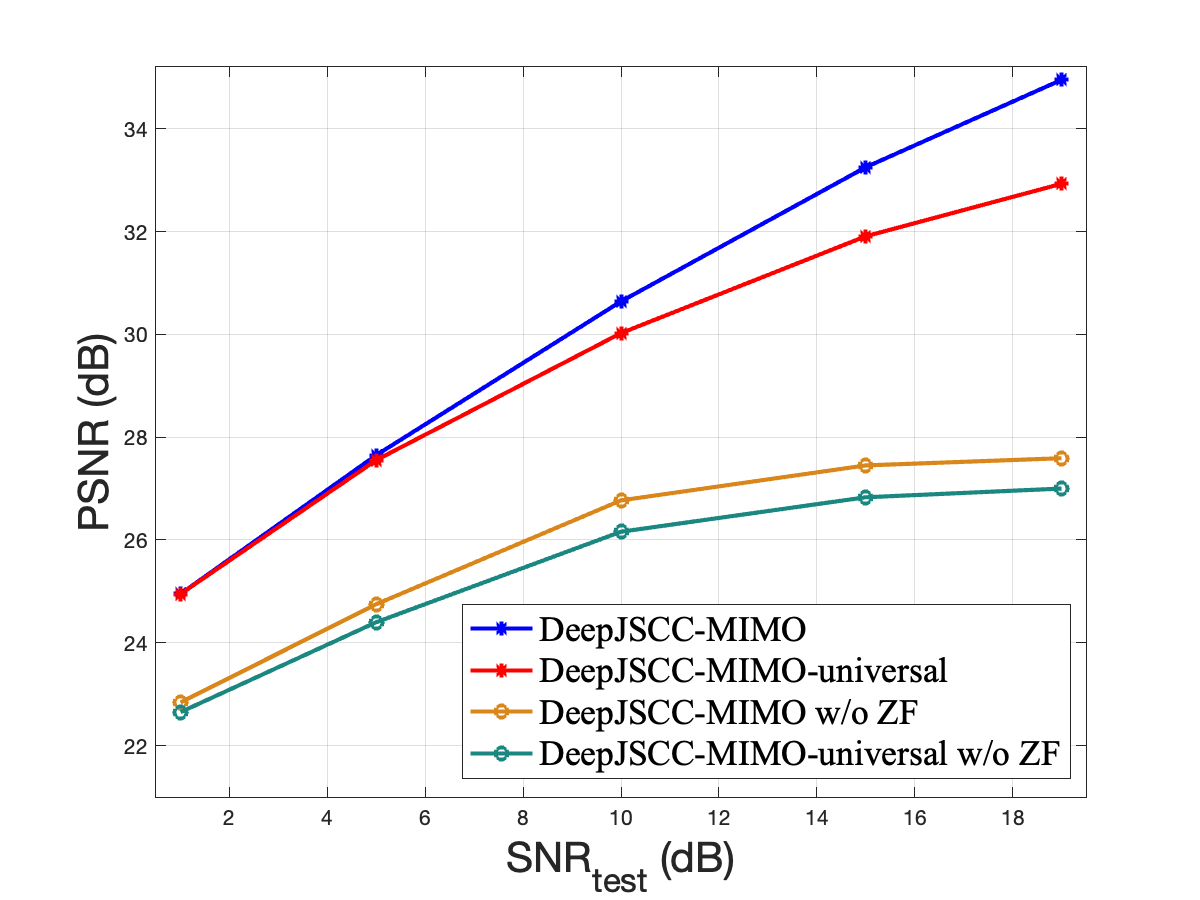}
    %\caption{fig2}
    \end{minipage}%
    }%
    \centering
    \caption{Ablation studies over the model-driven design for the proposed DeepJSCC-MIMO schemes with CSIR.}
    \label{ab_eq_exp} 
\end{figure*}

\begin{table}[h]
    \centering
    \caption{\textcolor{black}{Ablation study of different MIMO equalization methods for DeepJSCC-MIMO-universal with $R=1/12$.}}
    \label{table:exp_MIMO_dl_detection}
    \begin{tabular}{c|c|c|c|c|c}
    \textbf{Equalizations} & {1 dB}& {5 dB}& {10 dB}& {15 dB}& {19 dB}\\
    \hline
     {End-to-end w/o ZF}  & 22.65&24.4& 26.16&26.83&27.03\\
     Pure ZF & 24.47&26.98& 29.41&31.43&32.73 \\
     Pure MMSE & 24.61&27.04& 29.48&31.50&32.74\\
     {DL-aided ZF} &$\bm{24.95}$&$\bm{27.56}$&$\bm{30.02}$&$\bm{31.91}$&$\bm{32.93}$ \\    
    \end{tabular}
\end{table}

\begin{table*}[h]
    \centering
    \caption{\textcolor{black}{Ablation study of DeepJSCC-MIMO-universal with/without channel heatmap over the CIFAR10 dataset with CSIR.}}
    \label{table:open_ab_heatmap}
    \begin{tabular}{|c|c|c|c|c|c|c|}
    \hline
    \multicolumn{2}{|c|}{\textbf{Open-loop MIMO system}}&\multicolumn{5}{c|}{{Channel SNRs}}\\
    \hline
        Ratio&Models& {1 dB}& {5 dB}& {10 dB}& {15 dB}& {19 dB}\\
    %Open-loop MIMO&\textbf{Channel SNR} & {1 dB}& {5 dB}& {10 dB}& {15 dB}& {19 dB}\\    
    \hline
     \multirow{2}{*}{\textbf{$R=1/24$}}
     &{DeepJSCC-MIMO-universal}  & $\bm{22.27}$&$\bm{24.41}$& $\bm{26.56}$&$\bm{28.26}$&$\bm{29.33}$\\
     & DeepJSCC-MIMO-universal w/o heatmap & 21.83&24.00& 26.07&27.64&28.53 \\

         \hline
     \multirow{2}{*}{\textbf{$R=1/12$}}
     &{DeepJSCC-MIMO-universal} &$\bm{24.95}$&$\bm{27.56}$&$\bm{30.02}$&$\bm{31.91}$&$\bm{32.93}$ \\          
     &DeepJSCC-MIMO-universal w/o heatmap & 24.22&26.69& 29.10&30.89&31.86\\
              \hline
    \end{tabular}
\end{table*}
\begin{table*}[h]
    \centering
    \caption{\textcolor{black}{Ablation study of DeepJSCC-MIMO-universal with/without channel heatmap over the CIFAR10 dataset with CSIT.}}
    \label{table:close_ab_heatmap}
    \begin{tabular}{|c|c|c|c|c|c|c|}
        \hline
    \multicolumn{2}{|c|}{\textbf{Closed-loop MIMO system}}&\multicolumn{5}{c|}{{Channel SNRs}}\\
    \hline
    Ratio&Models& {1 dB}& {5 dB}& {10 dB}& {15 dB}& {19 dB}\\
    %\hline
    %Closed-loop MIMO&\textbf{Channel SNR} & {1 dB}& {5 dB}& {10 dB}& {15 dB}& {19 dB}\\    
    \hline
     \multirow{2}{*}{\textbf{$R=1/24$}}
     &{DeepJSCC-MIMO-universal}  & $\bm{23.95}$&$\bm{25.90}$& $\bm{28.13}$&$\bm{29.94}$&$\bm{30.93}$\\
     & DeepJSCC-MIMO-universal w/o heatmap &23.48& 25.63& 27.75& 29.34& 30.24\\
         \hline
     \multirow{2}{*}{\textbf{$R=1/12$}}
     &{DeepJSCC-MIMO-universal}      
     &$\bm{26.89}$&$\bm{29.30}$&$\bm{32.10}$&$\bm{34.50}$&$\bm{35.95}$ \\          
     &DeepJSCC-MIMO-universal w/o heatmap &26.21& 28.71& 31.36&33.41& 34.67\\
       \hline
     \multirow{2}{*}{\textbf{$R=1/6$}}
     &{DeepJSCC-MIMO-universal} 
     &$\bm{30.7}$&$\bm{33.65}$&$\bm{36.87}$&$\bm{39.56}$&$\bm{41.11}$ \\          
     &DeepJSCC-MIMO-universal w/o heatmap&29.19& 32.03&35.02& 37.31&38.63\\
         \hline
    \end{tabular}
\end{table*}
\subsection{Ablation study over the model-driven design}
\textcolor{black}{To evaluate the effectiveness of the DL-aided equalization method with the residual block, we repeat the experiments of DeepJSCC-MIMO-universal in Fig. \ref{open_ViT_12} with an MMSE receiver. The experimental results in Table \ref{table:exp_MIMO_dl_detection} show that the residual block improves the performance (up to $0.54$ dB). More interestingly, we observe that the ZF combined with the residual block achieves the best performance, which shows that the residual block helps to compensate for the limitation of ZF and outperforms MMSE.}

To provide a more comprehensive study of this issue, we conducted additional ablation studies over this model-driven approach for $R=1/24$ and $R=1/12$. Specifically, we removed the ZF operations and trained DeepJSCC-MIMO models in an end-to-end fashion, denoted by ``DeepJSCC-MIMO w/o ZF'' and ``DeepJSCC-MIMO-universal w/o ZF''. The experimental results are presented in Fig. \ref{ab_eq_open_ViT_24} and Fig. \ref{ab_eq_open_ViT_12} for $R=1/24$ and $R=1/12$, respectively. It is evident from the figures that the performance of DeepJSCC-MIMO models experiences a notable decline in the absence of ZF operations. Furthermore, the DeepJSCC-MIMO-universal models exhibit a diminished capacity for channel adaptability, as the channel heatmap tailored for our model-driven approach proves ineffective in conveying channel-related information.
\subsection{Ablation study of channel heatmap}
In order to assess the effectiveness of the channel heatmap, we have conducted ablation studies over the DeepJSCC-MIMO-universal scheme and DeepJSCC-MIMO scheme without channel heatmap. \textcolor{black}{The ablation results over the DeepJSCC-MIMO-universal scheme are presented in Tables. \ref{table:open_ab_heatmap} and \ref{table:close_ab_heatmap} for open-loop and closed-loop MIMO systems, respectively.} The experiments demonstrate that integrating the channel heatmap is essential to enhance the performance of the DeepJSCC-MIMO-universal scheme \textcolor{black}{in both open-loop and closed-loop MIMO systems.} Additional experiments for the DeepJSCC-MIMO scheme are conducted in the open-loop MIMO scenario. As depicted in Table. \ref{table:ab_heatmap_ratio}, despite the model being optimized for a particular channel condition, the performance can be further improved through the utilization of the channel heatmap.

\begin{table}[h]
%saved at hpc
\centering
\caption{{Ablation study of DeepJSCC-MIMO with/without channel heatmap over the Kodak dataset and SNR=$5$dB with CSIR.}}
\begin{tabular}{|c|c|c|c|c|c|}
\hline
{Proposed schemes}&\multicolumn{5}{c|}{Bandwidth ratio $R$}  \\
\hline
{DeepJSCC-MIMO}&{\textbf{$1/24$}} &\textbf{$1/12$}&\textbf{$1/8$} &\textbf{$1/6$}&\textbf{$1/4$}  \\
\hline
{{with heatmap}} &$\bm{29.43}$& $\bm{30.45}$ & $\bm{31.17}$&$\bm{31.56}$&$\bm{31.92}$\\
\hline
{{without heatmap}} &${28.88}$& ${29.73}$ & ${30.36}$&${30.84}$&${31.02}$\\
\hline
\end{tabular}
\label{table:ab_heatmap_ratio}
\end{table}

\textcolor{black}{Furthermore, we have conducted an additional ablation study of the channel heatmap when considering channel estimation errors, as detailed in Table. \ref{MIMO_estimation_error_wo_heat}. When jointly considering the proposed schemes in the presence or absence of a noisy channel heatmap, we can observe that the primary performance degradation arises from the imperfect MIMO equalization, where the performance degradation is up to $0.44$ dB and $1.03$ dB for $\sigma_e^2=0.2$ and $\sigma_e^2=1$, respectively. The additional imperfect channel heatmap can slightly improve the performance, which is up to $0.17$ dB and $0.19$ dB for $\sigma_e^2=0.2$ and $\sigma_e^2=1$, respectively. This observation can be attributed to the potential benefits of imperfect channel condition information in facilitating robust DeepJSCC transmission. Nevertheless, given the limited performance improvement, the effect of the channel heatmap remains relatively minor under the channel estimation errors. We can then conclude that the primary performance degradation of DeepJSCC-MIMO scheme under channel estimation errors arises from the imperfect MIMO equalization.}
%\textcolor{black}{The rationale behind this phenomenon lies in the potential advantage of incorporating additional noisy information of channel conditions to facilitate the robust DeepJSCC transmission. Nevertheless, given the limited overall degradation resulting from channel estimation errors in the DeepJSCC-MIMO scheme, the effect of the channel heatmap remains relatively minor under the channel estimation errors.}

In summary, the channel heatmap serves two objectives: a) it provides CSI for the transmitter/receiver, enabling the model to adapt to diverse channel conditions; b) it simplifies the learning process for the transformer model by presenting the channel condition that each channel symbol faces after specific equalization operations for the attention computation. As a direct result, it yields substantial performance improvements for the DeepJSCC-MIMO scheme as seen from the ablation studies above.

The integration of the above model-driven designs form our DeepJSCC-MIMO approach, each of them contributing to the competitive transmission performance while maintaining an acceptable computational complexity.  It is important to note that each constituent of this integrated approach exerts a distinct influence. Notably, the equalization operation emerges as a pivotal element in the MIMO channel, and its absence would render the proposed DeepJSCC scheme ineffective, similar to its role in traditional systems. While the SVD, heatmap design, and various other architectural choices substantially improve the performance, we can still surpass the performance of traditional separation-based benchmarks even in their absence.

\subsection{Additional experiments over realistic correlated MIMO channel models}
To investigate a more realistic correlated MIMO channels, we have added supplementary experiments over the MIMO channels for both open-loop and closed-loop MIMO systems. Specifically, following \cite{yao2022versatile} and \cite{shi2023excess}, we employed the transmitter SNR and the Kronecker channel model\cite{oestges2006validity}, given as:
\begin{equation}
    \bm{H}=(\bm{R_H^R})^{1/2}\bm{G}(\bm{R_H^T})^{1/2},
\end{equation}
where the channel covariance is approximated by the Kronecker product of the covariance matrices from the transmit and receive side. $\bm{R_H^T}=\begin{bmatrix}
1 & 0.3 \\
0.3 & 1
\end{bmatrix}$, $\bm{R_H^R}=\begin{bmatrix}
1 & 0.5 \\
0.5 & 1
\end{bmatrix}$, and each element from $\bm{G}$ is sampled from a standard Gaussian complex distribution. 

\begin{figure*}[t]
    \centering
    \subfloat[$R=1/24$]{
    \label{ab_eq_open_ViT_12_k} 
    \begin{minipage}[t]{0.47\linewidth}
    \centering
    \includegraphics[scale=0.38]{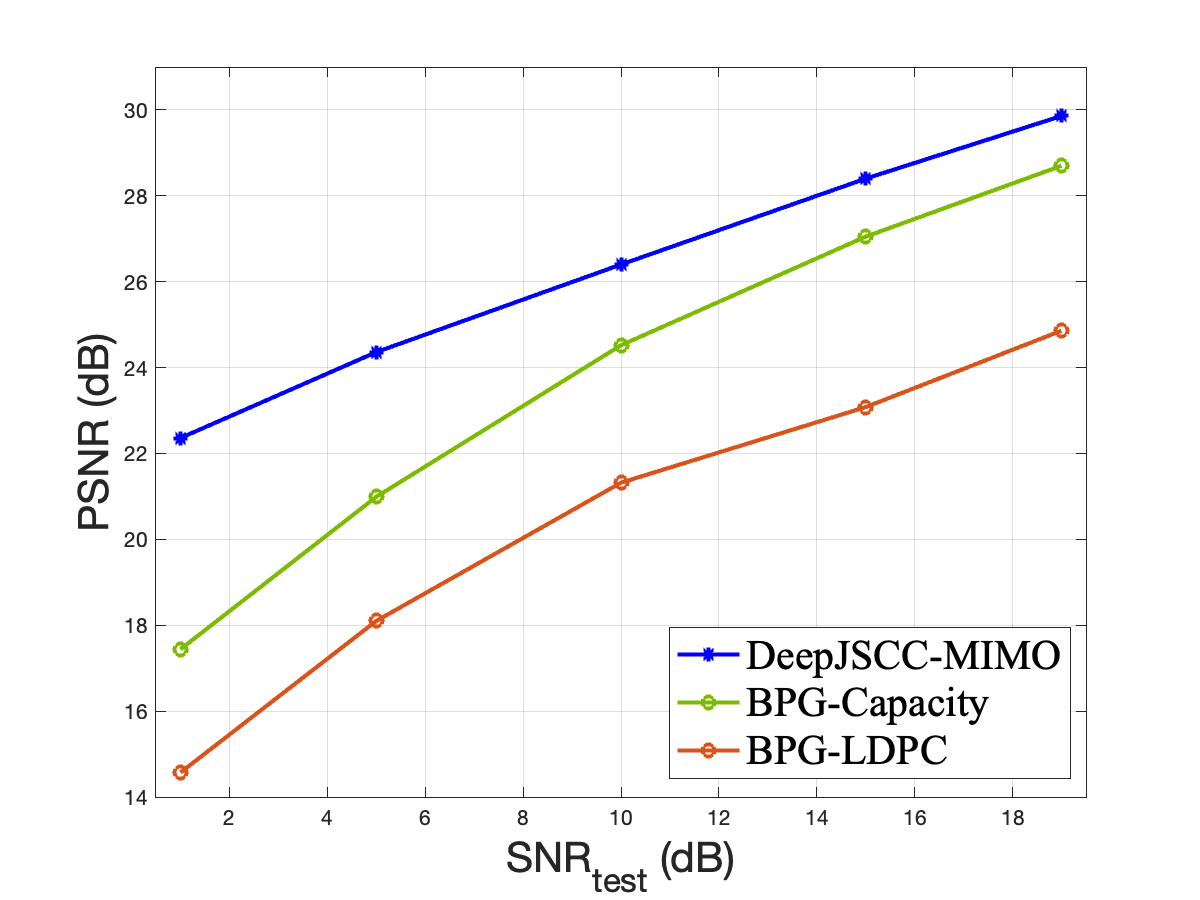}
    \end{minipage}%
    }%
    \hspace{-0.5mm}
    \subfloat[$R=1/12$]{
    \label{ab_eq_open_ViT_6_k} 
    \begin{minipage}[t]{0.47\linewidth}
    \centering
    \includegraphics[scale=0.38]{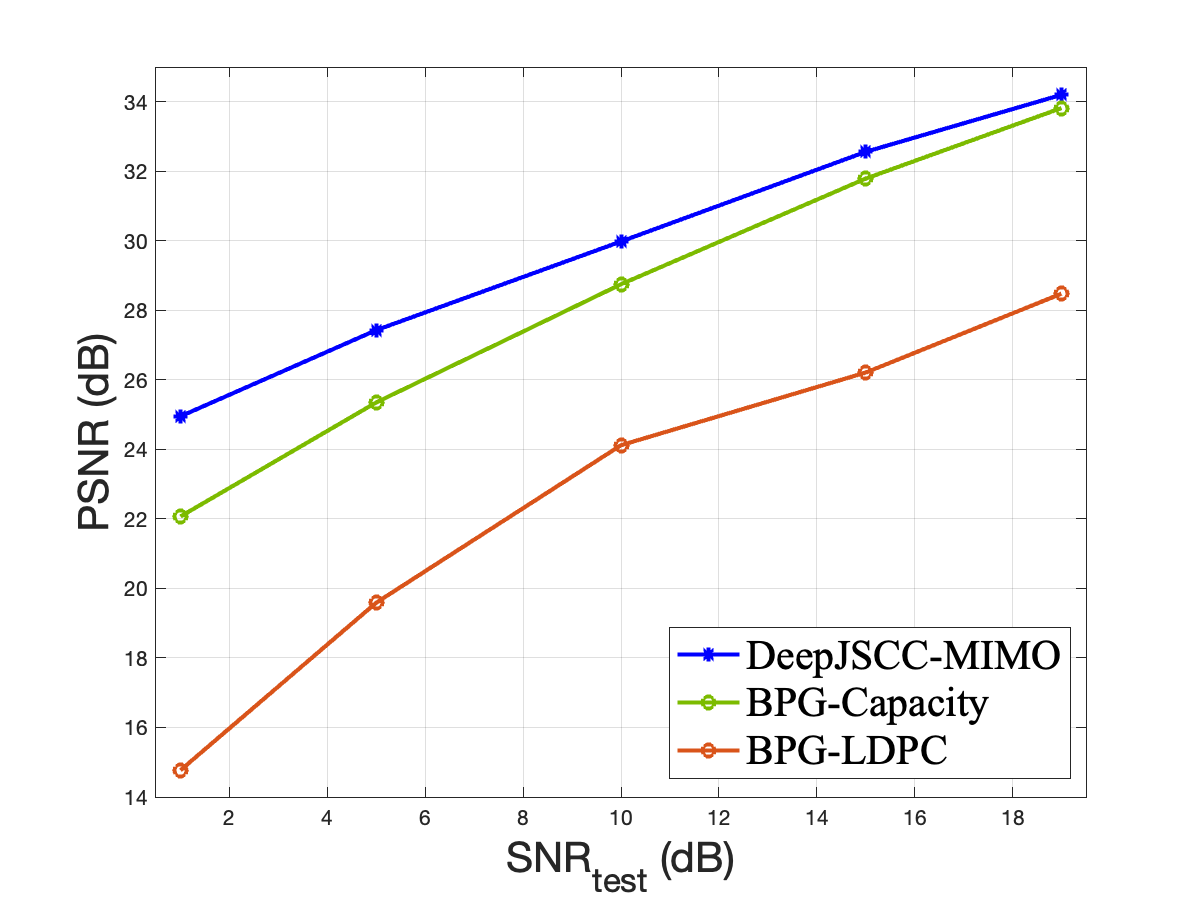}
    %\caption{fig2}
    \end{minipage}%
    }%
    \centering
    \caption{{Performance comparisons between the proposed DeepJSCC-MIMO and BPG-Capacity over different SNR values and bandwidth ratios for the open-loop Kronecker MIMO channels with CSIR.}}
    \label{ab_open_exp_k} 
\end{figure*}

\begin{figure*}[h]
    \centering
    \subfloat[$R=1/24$]{
    \label{close_ViT_24_k} 
    \begin{minipage}[t]{0.47\linewidth}
    \centering
    \includegraphics[scale=0.38]{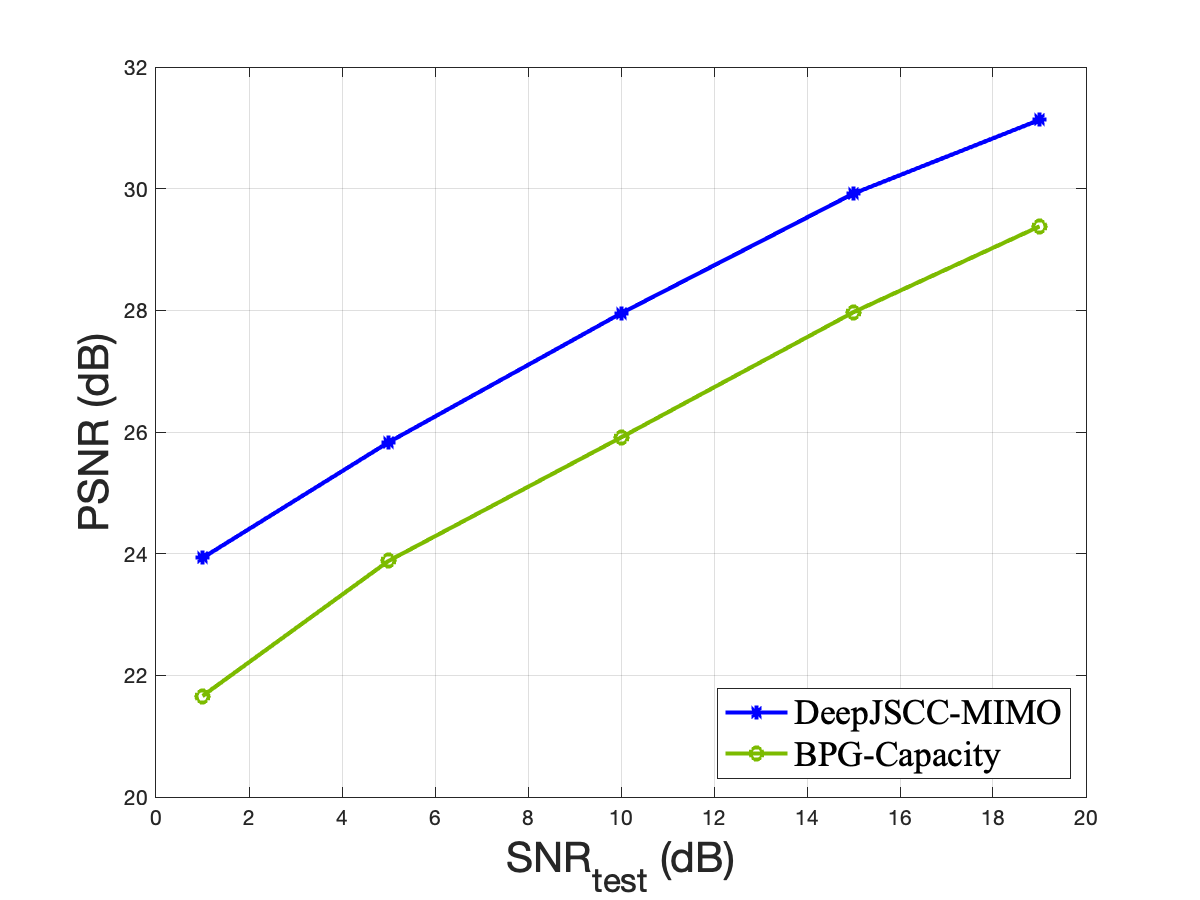}
    \end{minipage}%
    }%
    \hspace{-0.5mm}
    \subfloat[$R=1/12$]{
    \label{close_ViT_12_k} 
    \begin{minipage}[t]{0.47\linewidth}
    \centering
    \includegraphics[scale=0.38]{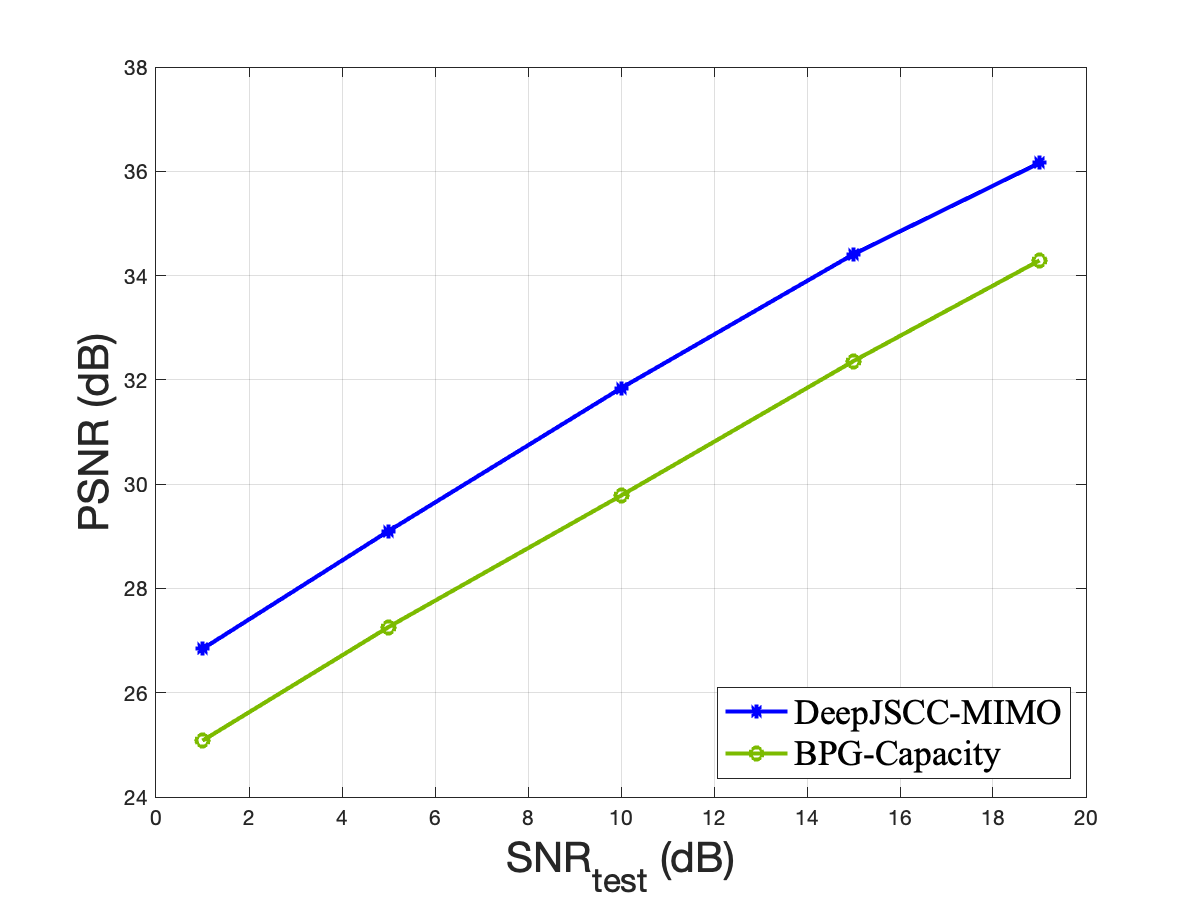}
    %\caption{fig2}
    \end{minipage}%
    }%
    \centering
    \caption{{Performance comparisons between the proposed DeepJSCC-MIMO and BPG-Capacity over different SNR values and bandwidth ratios for the closed-loop Kronecker MIMO channels with CSIT.}}
    \label{ab_close_exp_k} 
\end{figure*}

The experimental results of DeepJSCC-MIMO in the context of spatially correlated channel models for both open-loop MIMO and closed-loop MIMO systems are presented in Figs. \ref{ab_open_exp_k} and \ref{ab_close_exp_k}. It is evident from the figures that the superiority of DeepJSCC-MIMO continues of hold in this channel scenario as well.

\begin{table}[t]
\centering
\caption{\textcolor{black}{Ablation study of channel heatmap for the DeepJSCC-MIMO-universal scheme under channel estimation errors over the Kodak dataset for $R=1/24$ with CSIR.}}
\begin{tabular}{|c|c|c|c|c|c|}
\hline
\multicolumn{2}{|c|}{DeepJSCC-MIMO-universal} &
\multicolumn{4}{c|}{Different $\text{SNR}_{\text{test}}$}\\
\hline
{$\sigma_e^2=0$}&{{with heatmap}} & $\bm{29.38}$ & $\bm{30.08}$&$\bm{30.80}$&$\bm{31.62}$\\
\hline
\multirow{2}{*}{$\sigma_e^2=0.2$}&
{{with heatmap}} &${29.18}$ & ${29.86}$&${30.53}$&${31.33}$\\
&{{w/o heatmap}} &${29.05}$ & ${29.74}$&${30.36}$&${31.18}$\\
\hline
\multirow{2}{*}{$\sigma_e^2=1$}&{{with heatmap}} & ${28.73}$ & ${29.49}$&${30.08}$&${30.78}$\\
&{{w/o heatmap}} &28.68&29.47&29.93&30.59\\
\hline
\end{tabular}
     \label{MIMO_estimation_error_wo_heat}
\end{table}

\begin{table*}[h]
\caption{{The number of parameters and FLOPs for different models across different layers ($L_t$), where DeepJSCC (CNN) represents the CNN-based DeepJSCC architecture introduced in \cite{wu2022channel, 9714510}. Bold figures correspond to the minimal values among different models for each $L_t$, which is consistently achieved by the proposed ViT-based architecture in this paper.}}
\begin{center}
%\resizebox{\textwidth}{!}{
\begin{tabular*}{1\textwidth}{@{\extracolsep{\fill}}|c|c|ccccccc|}
\hline 
\multicolumn{2}{|c|}{\textbf{Layer number $L_t$}}&4&6&8&10&12&14&16 \\
\hline 
\multirow{3}{*}{\textbf{No. of Parameters} (millions)}
&\text{DeepJSCC-MIMO (ours)}&\textbf{6.61}&\textbf{9.76}&\textbf{12.92}&\textbf{16.07}&\textbf{19.23}&{\textbf{22.38}}&\textbf{25.54}\\
&\text{DeepJSCC (CNN)}& 6.89& 13.44& 20.00& 26.56& 33.11& 39.67&46.22\\
&\text{DeepJSCC (Swin)\cite{10094735}}& 8.89& 12.05& 15.21& 18.36& 21.52& 24.68&27.83\\

\hline 
\multirow{3}{*}{\textbf{FLOPs} (G)}
&\text{DeepJSCC-MIMO (ours)}&\textbf{0.43}& \textbf{0.63}&\textbf{0.83}&\textbf{1.03}&\textbf{1.23}&\textbf{1.43}&\textbf{1.64}\\
&\text{DeepJSCC (CNN)}&0.83& 1.25& 1.67& 2.08& 2.51& 2.93&3.35\\
&\text{DeepJSCC (Swin)\cite{10094735}}& 1.22& 1.73& 2.23& 2.74& 3.24& 3.74&4.25\\
\hline
%\multirow{3}{*}{\textbf{{Coding time}} (ms)}&\text{DeepJSCC-MIMO (ViT)}&6.26& 8.67&10.77&12.36&13.94&{16.81}\\
%&\text{DeepJSCC (CNN)}& 1.41& 1.8243& 2.34& 2.78& 3.27& 3.89\\
%&\text{BPG-LDPC (CPU)}& 0& 0& 0& 0& 0& 0\\

%\multicolumn{7}{c}{12.93}\\

%12.93&\textbf{12.93}&\textbf{12.93}&\textbf{12.93}&\textbf{12.93}&\textbf{12.93}&{\textbf{12.93}}\\
\hline 
\end{tabular*}
%}
\vspace{-5pt}
\label{table_para_ab}
\end{center}
\end{table*}

\subsection{More detailed study over the model efficiency}
Meanwhile, the particular structure of our proposed ViT-based methodology significantly limits both the inference complexity and the parameter count. In the proposed DeepJSCC-MIMO architecture, we have carefully balanced the performance with model efficiency, maintaining acceptable computational and storage costs. The ViT-based model proposed herein demonstrates state-of-the-art performance results while requiring limited computational and memory resources. 

In order to better evaluate the complexity of the proposed model, we provide a comparative analysis with an alternative DeepJSCC architecture using convolutional neural networks (CNN) \cite{wu2022channel, 9714510} or Swin transformers \cite{liu2021swin, 10094735} in Table \ref{table_para_ab}. Specifically, we compare the number of floating point operations (FLOPs) as a representative of the time complexity, and the number of model parameters for the memory complexity, calculated from thop library. We can observe that the proposed DeepJSCC-MIMO scheme requires the least number of computational operations and parameters compared to all other approaches under the same network layer configuration. 

We carried out another ablation study whose results are shown in Table \ref{table_para_ab}. In particular, we conducted experiments over different backbones across different layer numbers, where all models are equipped with ZF precoding. We can observe that our design achieves the best performance with $L_t=8$. %, so we finally set $L_t=8$. 

Additionally, performance ablation study is conducted to assess the performance of different layer settings, as detailed in Table \ref{table_para_ab}. We perform experiments utilizing various backbones and layer numbers, with all models incorporating DL-aided equalization operations. The results indicate that our design attains optimal performance when $L_t=8$, leading us to establish $L_t=8$ as the final configuration.

In summary, the proposed DeepJSCC-MIMO scheme demonstrates both practicality and efficiency, making it a promising alternative for emerging semantic communication systems. It is worth highlighting that there is potential for further improvements over our models' inference speed. This can be achieved through alternative hardware options, different architectural implementations, or acceleration techniques for vision transformers (ViTs), such as structure compression \cite{Graham_2021_ICCV}, fast scaling \cite{Dollar2021}, slow-fast evolution \cite{evo_vit}, efficient softmax \cite{stevens2021softermax} or employing a state-of-the-art inference acceleration library (such as NVIDIA FasterTransformer v4.0).

\subsection{Implementation details}
This section presents the implementation details. %Subsequently, we provide additional analysis and explanation, incorporating supplementary experiments and references to similar observations in the related literature.

\textbf{Part A. Implementation details:} 

For source coding, we adopt BPG version 0.9.8 using the binary file with the following command:
\begin{equation}
\begin{array}{r}
\textit{bpgenc -f 444 -q (quantization level)}\\
\textit{-o (out filepath) (in filepath)}.  
\end{array}  
\end{equation}

The 0.9.8 version is extensively used in the literature on image compression and achieves the best performance in our experiments in CelebA and Koadak datasets. For the CelebA dataset, we adopt YCgCo color space to yeild a better performance.

%$$\textit{bpgenc -f 444 -c ycgco -q (quantization level) -o (out filepath) (in filepath)}$$

For channel coding and modulation, the detailed implementation of LDPC is given in Section IV.

\textbf{Part B. Analysis:} 
%The large performance gain of ``DeepJSCC-MIMO'' over ``BPG+LDPC'' can be attributed to the absence of CSI at the transmitter (i.e., the CSIR scenario), the limited bandwidth regime, and the low-SNR regime.
In general, DeepJSCC demonstrates significant advantages in scenarios characterized by low bandwidth ratios and low SNRs \cite{erdemir2022generative, kurka2020deepjscc}. Moreover, \cite{song2020benchmarking, o2017deep} underscores the advantages of DL-based methods compared with traditional MIMO schemes, particularly in CSIR scenarios. Further investigations tailored for the open-loop MIMO image transmission problem, as shown in \cite{yao2022versatile}, exhibit similar findings, revealing significant performance gains ranging from $7$ dB to $9$ dB at low SNRs (i.e., SNR = $4-6$ dB).

We have also performed additional experiments to validate these observations. We first compare the findings of low SNR regimes under the same setting of an open-loop MIMO system with $R=1/24$ over various SNRs. We can see that even for the BPG-Capacity scheme with the instantaneous channel capacity, the DeepJSCC-MIMO can still outperform this bound over $3$dB, clearly demonstrating the benefits of the DeepJSCC-MIMO pipeline. When it comes to higher SNRs, the performance gap becomes smaller. Providing more details for the reviewer: for SNR=$5$dB, the average available total capacity for each image is around $420$ bits, resulting in a high compression ratio with relatively poor performance. In conclusion, it is apparent that the compression algorithm's efficacy is limited by the available bandwidth and channel SNR, constituting the primary factor contributing to the significant performance gap.

We have conducted supplementary experiments at a higher SNR value of $10$dB. The results are presented in Table \ref{table_dl_open_10}. Observations reveal that DeepJSCC-MIMO can consistently outperform the BPG-LDPC benchmark even in the higher SNR regime. However, compared with Table \ref{table_dl_open}, there is a discernible trend indicating a gradual reduction in the performance gap.
\begin{table}[t]
\centering
\caption{{Ablation study over DeepJSCC-MIMO scheme and BPG-LDPC benchmark for the open-loop MIMO system at $SNR=10 dB$ on Kodak dataset.}}
\begin{tabular}{c|c|c|c|c}
\textbf{Different ratios $R$}&{\textbf{$1/24$}} &\textbf{$1/16$}&\textbf{$1/12$}&\textbf{$1/8$}  \\
\hline
{{BPG-LDPC}\cite{bpg}}  &${29.45}$&${30.51}$&${31.69}$ &${34.04}$\\
{{DeepJSCC-MIMO}} &$\bm{32.32}$& $\bm{33.71}$ & $\bm{34.39}$&$\bm{35.21}$\\
\end{tabular}
\label{table_dl_open_10}
\end{table}

\end{document}